\documentclass[prd,12pt]{article}
\usepackage[margin=1.0in]{geometry}

\usepackage{amsmath,amssymb,graphicx,multirow,xspace,slashed}
\usepackage[usenames,dvipsnames]{xcolor}
\usepackage[colorlinks=true,urlcolor=blue,anchorcolor=blue,citecolor=red,filecolor=blue,linkcolor=blue,menucolor=blue,pagecolor=blue]{hyperref}
\usepackage[compress,numbers]{natbib}
\usepackage{subfigure, placeins}
\usepackage{booktabs}
\usepackage{blindtext, rotating}
\usepackage{afterpage}
\usepackage{enumitem}
\usepackage{ marvosym }
\usepackage{authblk} 
\usepackage{verbatim}
\usepackage{soul} 
\usepackage[normalem]{ulem}
\usepackage{pifont}
\usepackage[usestackEOL]{stackengine}
\usepackage{subfiles}
\newcommand\sq{\framebox(10,10){}\kern\fboxrule}

\setstackgap{S}{\fboxrule}

\usepackage[utf8]{inputenc}

\usepackage{floatrow}
\newfloatcommand{capbtabbox}{table}[][\FBwidth]

\usepackage[font=footnotesize,labelfont=bf]{caption}

\usepackage{lineno}

\allowdisplaybreaks

\usepackage{bm}

\newcommand{\be}{\begin{equation}}
\newcommand{\ee}{\end{equation}}
\newcommand{\ba}{\begin{align}}
\newcommand{\ea}{\end{align}}
\usepackage{titlesec}

\newcommand\BRmutoe{{\mathrm{BR}\left( \mu \rightarrow e \gamma  \right)}}
\newcommand\BRmutoeee{{\mathrm{BR}\left( \mu \rightarrow 3e  \right)}}
\newcommand\RmutoeN{{\mathrm{R}\left( \mu N\rightarrow e N  \right)}}

\title{
\vspace{-1.5cm}
\hspace{4.9in}
\small{MIT-CTP/5294}
\vspace{1.5cm}
\vspace{-0cm} \bf \Large
Spontaneous CP Violation and Horizontal Symmetry in the MSSM: Toward Lepton Flavor Naturalness
\vspace{0.5cm}}

\author{Daniel Aloni$^a$, Pouya Asadi$^b$, Yuichiro Nakai$^{c}$, \\
\vspace{-0.5cm}
Matthew Reece$^{d}$, and Motoo Suzuki$^{c}$  \\
\vspace{0.5cm}

{\small $^{a}$ Department of Physics, Boston University,} \\
\vspace{-0.2cm}
{\small 590 Commonwealth Ave., Boston, MA, 02215, USA} \\

{\small $^{b}$ Center for Theoretical Physics, Massachusetts Institute of Technology,} \\
\vspace{-0.2cm}
{\small Cambridge, MA, 02139, USA} \\

{\small $^{c}$ Tsung-Dao Lee Institute and School of Physics and Astronomy,} \\
\vspace{-0.2cm}
{\small  Shanghai Jiao Tong University, 800 Dongchuan Road, Shanghai, 200240, China} \\

{\small $^{d}$ Department of Physics, Harvard University, Cambridge, MA, 02138, USA}}

\date{}

\begin{document}

\maketitle

\vspace{-0.5cm}

\begin{abstract}

We study the contributions of supersymmetric models with a $U(1)$ horizontal symmetry and only spontaneous CP breaking to various lepton flavor observables, such as $\mu \to e\gamma$ and the electron electric dipole moment.
We show that both a horizontal symmetry and a lack of explicit CP violation can alleviate the existing bounds from such observables.
The undetermined $\mathcal{O}(1)$ coefficients in such mass matrix models muddle the interpretation of the bounds from various flavor observables.
To overcome this, we define a new fine-tuning measure for different observables in such setups. This allows us to study how naturally the observed IR flavor observables can emerge from a given mass matrix model.
We use our flavor-naturalness measure in study of our supersymmetric models and quantify the degree of fine tuning required by the bounds from various lepton flavor observables at each mass scale of sleptons, neutralinos, and charginos.

\end{abstract}

\newpage
\tableofcontents

\section{Introduction}

The origin of fundamental particle masses is a central issue in modern particle physics.
The Standard Model (SM) partly addresses this issue by the higgs mechanism.
However, it leaves two major questions unanswered. 
The first question is why the three generations of quarks and leptons couple to the higgs field so differently. 
The couplings range from $\mathcal{O}(10^{-5})$ to $\mathcal{O}(1)$, and
it seems unlikely that they are determined completely randomly.
The second question is on the stability of the scale of the electroweak symmetry breaking (EWSB)
under quantum corrections, commonly known as the naturalness problem or the hierarchy problem
\cite{Weinberg:1975gm,Gildener:1976ai,Susskind:1978ms,tHooft:1979rat}. 
A significant fine tuning is required for the higgs mass-squared parameter
to realize the electroweak scale much smaller than UV mass scales such as the Planck scale. 
These two questions have driven numerous theoretical and experimental studies of physics beyond the SM (see Refs.~\cite{Martin:1997ns,Csaki:2018muy} for reviews).

The stability of the electroweak scale is elegantly ensured by supersymmetry (SUSY).
In SUSY models, a large quantum correction from a SM particle to the higgs mass-squared parameter is canceled with
that of its corresponding superpartner.
If SUSY is realized in nature, it must be broken; however, SUSY breaking can radiatively drive the higgs mass-squared parameter negative so that the electroweak symmetry is dynamically broken.
While the discovery of the higgs boson with the mass of around 125 GeV at the Large Hadron Collider (LHC) indicates that
the superpartner of the top quark is significantly heavier than the weak scale (see Ref.~\cite{Draper:2016pys} for a recent review, and Refs.~\cite{Haber:1990aw,Okada:1990vk,Barbieri:1990ja} for early references),
to avoid a severe tuning,
the mass scale of supersymmetric particles should be within just a few orders of magnitude
above the electroweak scale
(for a recent study of tuning of the minimal supersymmetric SM (MSSM) or other SUSY extensions of the SM,
see Ref.~\cite{Buckley:2016kvr}). 

Nonetheless, a random breaking of SUSY generates significant flavor- and CP-violating processes, in contradiction with precise experimental data.
One possible solution to this problem is to engineer a mechanism to realize flavor independent masses of superparticles.
Gauge mediation is the most popular example (see, e.g., Refs.~\cite{Giudice:1998bp,Kitano:2010fa} for reviews), but it predicts a light gravitino which generally leads to cosmological problems such as overclosure of the universe \cite{Pagels:1981ke, Weinberg:1982zq, Khlopov:1984pf, Moroi:1993mb}. 
Gravity mediation with a cosmologically harmless gravitino of $O(10) \, \rm TeV$ appears to be more natural, but it is not equipped with a mechanism to suppress flavor and CP violating processes. 

An intriguing approach to both the SM Yukawa hierarchy and the SUSY flavor problem
is provided by a $U(1)$ horizontal symmetry and its spontaneous breaking \cite{Froggatt:1978nt}. 
Depending on $U(1)$ charge assignments of quark and lepton supermultiplets, their Yukawa couplings are suppressed
by some power of a small parameter $\epsilon \equiv \langle S \rangle / \Lambda$ where $\Lambda$ is some UV mass scale
and $\langle S \rangle$ is a vacuum expectation value (vev) of a flavon field $S$ charged under the $U(1)$.
Appropriate charge assignments can lead to the observed SM fermion mass and mixing pattern.
Since the same symmetry is applied to SUSY breaking masses of superparticles,
the mass matrices of quarks and leptons and their superpartners can be approximately diagonal in the same basis. This \textit{alignment} can suppress dangerous flavor violating processes induced by supersymmetric particles
\cite{Leurer:1992wg,Nir:1993mx,Leurer:1993gy,Ibanez:1994ig,Grossman:1995hk}. The $U(1)$ symmetry can arise as an accidental symmetry enforced by discrete gauge symmetries, or as a low-energy remnant of an anomalous gauged $U(1)$ obtaining a mass from the 4d Green-Schwarz mechanism~\cite{Witten:1984dg}. These options allow very good approximate global symmetries, while being compatible with the expectation that exact global symmetries do not exist in gravitational theories~\cite{Zeldovich:1976vq,Banks:2010zn,Harlow:2018jwu}.

The quark sector with three generations contains a CP violating complex phase
called the Cabibbo-Kobayashi-Maskawa (CKM) phase.
Recently, a nonzero phase has been hinted at in the lepton sector as well~\cite{Abe:2019vii,Kolupaeva:2020pug,Abe:2021gky}.
The existence of CP violation (CPV) in nature and the SUSY CP problem are reconciled
by the idea of that CP is a gauge symmetry~\cite{Dine:1992ya, Choi:1992xp}, which is broken spontaneously~\cite{Nir:1996am}. 
Multiple flavon fields can realize this idea: their vevs are complex and break CP spontaneously.
In Ref.~\cite{Nir:1996am} this mechanism gives rise to the correct CKM phase, while
suppressing the neutron electric dipole moment (EDM) through CP violating squark mass matrices.
Taking account of a hinted phase in the Pontecorvo–Maki–Nakagawa–Sakata (PMNS) matrix and
the stringent upper bound on the electron EDM (eEDM) \cite{Andreev:2018ayy},\footnote{See also Refs.~\cite{Nakai:2016atk,Cesarotti:2018huy} for interpretations of these bounds on SUSY models' parameter space.} it is natural to apply the same idea to the lepton sector.

The $\mu \to e \gamma$ process is a well-known observable
which gives a strong constraint on new physics with lepton flavor violation (LFV). 
The current most stringent limit on this process comes from the MEG experiment \cite{TheMEG:2016wtm}.
Future experiments such as the upgraded MEG II experiment~\cite{Baldini:2018nnn}
will increase the sensitivity by an order of magnitude.
Moreover, the $\mu \to e$ conversion process has been currently searched for
and its future limit is expected to be dramatically improved.
Thus, it is timely to assess up-to-date constraints from CPV and LFV observables
on supersymmetric models with a $U(1)$ horizontal symmetry and discuss future prospects.

In this paper, we consider a $U(1)$ horizontal symmetry and its spontaneous breaking
to realize the correct SM fermion masses and mixings, including CP violating phases, in the SUSY framework.
We focus on the lepton sector.
We do not study the squarks and gluinos here, since they will not contribute to the observables of interest at leading order. A detailed study of the quark sector observables in our setup is left for future works.

To accommodate a nonzero phase in the PMNS matrix without inducing sizable EDMs,
the idea of spontaneous CP violation~\cite{Nir:1996am} is extended to the lepton sector.
We explore current constraints from CPV and LFV observables
on models with two flavons and discuss future prospects of the constraints.
The allowed mass scale of supersymmetric particles is lowered in the models with two flavons
compared to the case with anarchic soft masses containing uncontrolled complex phases.\footnote{By ``anarchic,'' we mean completely uncontrolled by any flavor symmetry, as opposed to the well-known neutrino anarchy scenario~\cite{Hall:1999sn} in which the left-handed leptons have a random mixing matrix but the right-handed leptons' mass matrix is structured.}

The assessment of constraints on models with horizontal symmetries is generally obscured
by uncertainties originating from undetermined $\mathcal{O}(1)$ numbers in mass matrices.
Horizontal symmetries only tell us the scaling of mass matrix entries with the parameter $\epsilon$,
while undetermined $\mathcal{O}(1)$ numbers multiplied in each entry
can have profound effects on SUSY contributions to CP and flavor violating processes
as well as the obtained SM fermion masses and mixings.
We improve this situation by proposing a new fine-tuning measure to assess the CP and flavor constraints
that captures the effects of undetermined $\mathcal{O}(1)$ numbers.
The new measure makes it possible to estimate constraints on models with horizontal symmetries
in a more numerically precise way.
This technique is generic and not limited to the present SUSY models.
Similar tuning measures can be defined for other SUSY or non-SUSY models that explain the SM flavor structure.

The rest of the paper is organized as follows.
In Sec.~\ref{sec:setup}, we summarize the SM flavor structure, as well as the current and projected future bounds from measurements of eEDM and LFV processes. 
The severity of the SUSY CP and flavor problems is illustrated with the case of anarchic sfermion masses. 
In Sec.~\ref{sec:minimal} we introduce a supersymmetric model with a horizontal symmetry and spontaneous CPV and discuss the effect on suppressing disastrous eEDM and LFV processes. We also introduce a new notion of naturalness for these observables and use it to quantify their bounds on a handful of sample charge assignments. 
Section~\ref{sec:conclusion} is devoted to conclusions and discussions. We also include further details in three appendices. 
We review the contribution of generic UV models to the relevant dipole operator in the IR in App.~\ref{app:dipole-calc} and list the diagrams contributing to this operator in a generic MSSM model in App.~\ref{app:all_contributions}. We also elaborate on how we generate random numbers in the mass matrices in our numerical scans in App.~\ref{app:matrices}

\section{Flavor \& SUSY}
\label{sec:setup}

In this section, we review the existing literature on various ingredients in our study. We start by reviewing the lepton flavor structure in the SM and the current and projected future bounds on the leptonic CPV and LFV observables that we study in this paper.
The SUSY flavor problem and contributions to CPV and LFV observables in the MSSM are then summarized.
We will see that the current bounds on these observables severely constrain the SUSY parameter space without further structure or severe tuning; we will systematically study new structures that may open up the parameter space in the next section.

\subsection{The SM flavor structure}
\label{subsec:SM_flavor}

There are clear hierarchies in the flavor parameters of the SM. Within the strong sector, the quark masses span five orders of magnitude. Moreover, the diagonal elements of the CKM matrix are of  order one, while off-diagonal elements are suppressed. 
Those hierarchies suggest the existence of a structure with a fundamental small parameter $\lambda\sim0.2$, originally introduced by Wolfenstein~\cite{Wolfenstein:1983yz} for the CKM matrix elements, and later applied to the quark masses hierarchies~\cite{Leurer:1993gy}.

Within the lepton sector, the charged lepton masses are hierarchical. The mass ratios can be parameterized by the Wolfenstein parameter $\lambda$ as
\begin{align}
	m_\mu/m_\tau\sim \lambda^2\ , \quad m_e/m_\tau\sim \lambda^5-\lambda^6\ .
\end{align}

The absolute scale of neutrino masses is still unknown, but a hierarchy is observed in the ratio of mass differences
\begin{align}
	\frac{\Delta m_{21}^2}{\Delta m_{32}^2} \sim \lambda^2 ~(\lambda^{-2})\ ,
\end{align}
for the normal (inverted) hierarchy~\cite{Esteban:2020cvm}. 
In contrast to the CKM matrix, the PMNS matrix of the lepton sector does not show any clear structure or hierarchy~\cite{Esteban:2020cvm}. 
The PMNS matrix has one (three) CPV phase in the case of Dirac (Majorana) neutrinos. Only the Dirac phase $\delta_{CP}$ can be measured by oscillation experiments. There is some tension between measurements of this phase. The  NO$\nu$A~\cite{Kolupaeva:2020pug} experiment favors $\delta_{CP}<\pi$, still allowing a CP-conserving phase, while T2K~\cite{Abe:2021gky} favors $\delta_{CP}>\pi$ and disfavors CP-violation at the level of $2\sigma$.

\subsection{Lepton flavor observables}
\label{subsec:lepton_observables}

There are many IR observables that can probe deviations from the SM flavor structure outlined above. In Table~\ref{tabs:obslist} we show a handful of such observables, including both current constraints and those that are expected in the next decade.

\begin{table}
\scalebox{1.0}{\resizebox{\columnwidth}{!}{
\begin{tabular}{|c|c|c|c|c|}
\hline 
Observable & Current bound & Current $\Lambda$ (TeV)  & Future reach & Future $\Lambda$ (TeV) \\ 
\hline 
eEDM & $1.1 \times  10^{-29}$ $e$\,cm & $1.0 \times 10^3$ & $\sim 10^{-32}$ $e$\,cm & $3.3 \times 10^4$
\\ 
\hline 
$\BRmutoe$ & $4.2 \times 10^{-13}$ & $57.$ & $\sim 10^{-14}$ & $1.5 \times 10^2$  \\ 
\hline 
$\RmutoeN$ & $7 \times 10^{-13}$ (Au) & $12.$ & $\sim 10^{-17}$ (Al) & $1.8 \times 10^2$ \\ 
\hline 
$\BRmutoeee$ & $10^{-12}$ & $13.$ & $\sim 10^{-16}$ & $1.3 \times 10^2$ \\ 
\hline 
\end{tabular} 
}}
\caption{
The current and future $90\%$ confidence level bounds on the LFV processes and the electron EDM that we study in this work. We have also converted them each to a representative mass scale $\Lambda$ by setting the coefficient of the appropriate dipole operator in Eq.~\eqref{eq:L-dipole} to $c_{ij} = \frac{eg^2}{16\pi^2} \frac{m_\mu}{\Lambda^2}$. The actual mass sensitivity, of course, depends on the details of the lepton flavor model. Note that the scale $\Lambda$ probed by EDMs and MDMs scales like the inverse square root of the experimental sensitivity, whereas for LFV observables it scales like the inverse one-fourth power. 
}
\label{tabs:obslist}
\end{table}

The current strongest constraint on the eEDM comes from the ACME experiment, using the thorium monoxide (ThO) molecule~\cite{Andreev:2018ayy}. Proposed future experiments are expected to improve the experimental sensitivity by a few orders of magnitude in the near future; see, for instance, Refs.~\cite{Kozyryev:2017cwq,Vutha:2017pej,Aggarwal:2018pru,Ho:2020ucd,Hutzler:2020lmj,Fitch:2021pfs}. 
In terms of a simple parametrization of the scale probed by an operator, the eEDM is the most sensitive lepton-sector observable, as indicated by the scale $\Lambda$ quoted in Table~\ref{tabs:obslist}. However, we will see that model-dependent details can significantly modify this conclusion.

A number of stringent constraints arise from charged lepton flavor violation observables probing $\mu \to e$ transitions. A useful, recent summary prepared as an input to the European Strategy for particle physics may be found in Ref.~\cite{Baldini:2018uhj}; see, especially, its Fig.~1 for the sensitivity of several planned experiments. The current most stringent bound on the LFV observable $\BRmutoe$ comes from the MEG experiment \cite{TheMEG:2016wtm}. Approximately an order of magnitude improvement is projected for this observable after three years of data-taking at the upgraded MEG II experiment~\cite{Baldini:2018nnn}. The conversion of a muon to an electron in the Coulomb field of a nucleus has been constrained by the SINDRUM~II experiment~\cite{Bertl:2006up}. 
Orders of magnitude increase in sensitivity are expected in the near future as well from experiments such as Mu2e~\cite{Kutschke:2011ux} or COMET~\cite{Adamov:2018vin,Shoukavy:2019ydh}. The SINDRUM experiment also provides the current most stringent limit on the $\BRmutoeee$ process~\cite{Bellgardt:1987du}.
A new search for this process at the Mu3e experiment will improve the limit by a few orders of magnitude~\cite{Blondel:2013ia,Berger:2014vba}. All three charged LFV processes ($\mu \to e\gamma$, $\mu \to 3e$, and $\mu N \to eN$) are sensitive to similar dipole operators (see Eqs.~\eqref{eq:mu2e-def} -- \eqref{eq:mu3e-dipole-only} below). Taking into account only these operators, the current most stringent constraints arise from $\mu \to e\gamma$, but will be surpassed by $\mu N \to e N$ and $\mu \to 3e$ in the future. Because $\mu N \to e N$ can receive additional important contributions from box diagrams in portions of the MSSM parameter space, we will focus on the dipole operator contributions to $\mu \to 3e$ as our benchmark for sensitive future charged LFV constraints.

There are other similar observables, e.g., $\mathrm{BR} \left( \tau \rightarrow \mu \gamma \right)$, that will be studied in the near future. In this paper we neglect these observables since the bounds on them are far less constraining than the ones studied in Table~\ref{tabs:obslist} in the foreseeable future, for the class of flavor models that we consider.

The ongoing experimental efforts in measuring eEDM and LFV observables will provide us with a trove of new data that can be used to probe various new physics models. The leading new physics contribution to all the observables in Table~\ref{tabs:obslist} arises from the dipole operators in the IR,
\begin{equation}
    \mathcal{L} \supset c^R_{ij} \bar{f}_i \sigma^{\mu\nu} P_R f_j F_{\mu\nu} + \mathrm{h.c.} ,
    \label{eq:L-dipole}
\end{equation}
where $i,j$ indices refer to SM fermions, $P_R=(1+\gamma^5)/2$, $F_{\mu\nu}$ is the $U(1)_{em}$ field strength, and $\sigma^{\mu\nu}=i/2 \left[ \gamma^\mu,\gamma^\nu \right]$.\footnote{It should be noted that to make this operator invariant under $SU(2)_L \times U(1)_Y$ of the SM (without introducing new sources of EW symmetry breaking), a higgs vev insertion is required. This higgs insertion is absorbed into the Wilson coefficient $c^R_{ij}$.} 
Notice that the operator with $P_L$ is included in the $\mathrm{h.c.}$ part of Eq.~\eqref{eq:L-dipole} since $(\bar{f}_i \sigma^{\mu\nu} P_L f_j)^\dagger=\bar{f}_j \sigma^{\mu\nu} P_R f_i$. Thus,
\begin{equation}
    c^L_{ij}=\left( c^R_{ji}\right)^*,
    \label{eq:WC-PL}
\end{equation}
with $c^L_{ij}$ being the Wilson coefficient of the operator $\bar{f}_i \sigma^{\mu\nu} P_L f_j F_{\mu\nu}$.

\begin{figure}
    \centering
    \resizebox{0.5\columnwidth}{!}{
    \includegraphics{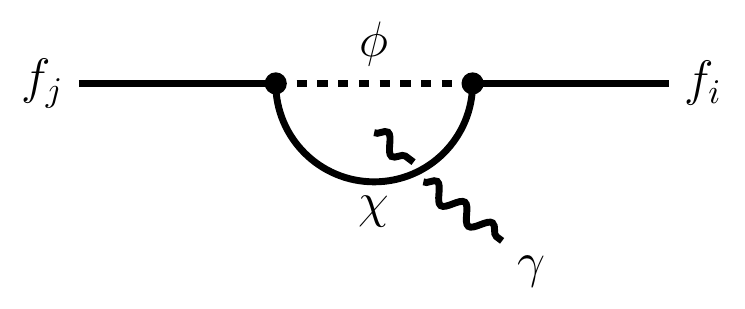}
    }
    \caption{The 1-loop diagram generating the dipole operator of Eq.~\eqref{eq:L-dipole} from a generic UV theory with new heavy scalars ($\phi$) and fermions ($\chi$). We present the contribution of generic scalars and fermions in the loop in App.~\ref{app:dipole-calc}. The photon can be emitted from either of the particles running in the loop, provided that they are electrically charged.  }
    \label{fig:dipole-diag}
\end{figure}

Any generic new physics model with new fermions or bosons can contribute to this operator through the diagram of Fig.~\ref{fig:dipole-diag}. In App.~\ref{app:dipole-calc} we review the contribution of generic new heavy scalars and fermions that can run in the loop to this IR operator. We explicitly check that our calculation is in agreement with the previous calculations in the literature \cite{Hisano:1995cp,Ellis:2008zy,Crivellin:2018qmi}. Discussions of the renormalization group evolution of such operators may be found in Refs.~\cite{Alonso:2013hga,Aebischer:2021uvt}, but we do not require this level of precision and so do not include them in our calculations.

In terms of these dipole operators, the eEDM is given by 
\begin{equation}
    d_e = 2\, \mathrm{Im}\, c^R_{ee}.
    \label{eq:eEDM-def}
\end{equation}
In addition to the eEDM observable, Eq.~\eqref{eq:L-dipole} contributes to the LFV observables such as $\BRmutoe$: 
\begin{equation}
    \BRmutoe = \frac{48 \pi^2}{G_F^2 m_\mu^2 } \left(   |c^R_{e\mu}|^2+|c^R_{\mu e}|^2    \right),
    \label{eq:mu2e-def}
\end{equation}
where $m_\mu$ is the muon mass and $G_F$ is the Fermi constant. 
The dipole operator in Eq.~\eqref{eq:L-dipole} can also give rise to conversion of $\mu$ to $e$ in a nucleus. When this dipole operator is the only source of this conversion, we have~\cite{Ellis:2016yje}
\begin{equation}
    \RmutoeN \approx \frac{\alpha_{em}}{\mathcal{O}(1)} \BRmutoe , 
    \label{eq:conversion-dipole-only}
\end{equation}
where the $\mathcal{O}(1)$ prefactor depends on the nucleus~\cite{Ellis:2016yje,Kitano:2002mt}. However, this observable can get contributions from other penguin and box operators as well. 
The dipole operator of Eq.~\eqref{eq:L-dipole} can also give rise to an enhanced $\BRmutoeee$. This contribution dominates the four-fermion diagrams contribution and gives rise to \cite{Ellis:2016yje}
\begin{equation}
    \BRmutoeee \approx \frac{\alpha_{em}}{3\pi} \left( 2  \log \frac{m_\mu}{m_e} - \frac{11}4  \right) \BRmutoe .
    \label{eq:mu3e-dipole-only}
\end{equation}
The contribution from all  sub-dominant four-fermion operators to this observable can be found in Ref.~\cite{Ellis:2016yje} as well. Below, we will consider only the dipole-operator contributions to $\mu \to e\gamma$ and $\mu \to 3e$.

\subsection{SUSY breaking and $R$-symmetry}
\label{subsec:susy_structure}

In this paper, we will assume a rather generic supersymmetric extension of the Standard Model. The breaking of SUSY can be parametrized through a chiral superfield $X$ whose auxiliary field $F_X$ acquires a SUSY-breaking vev, $F_X/\Lambda \sim m_{\rm soft}$. For simplicity, we will assume that $X$ couples in the most generic way possible through $\Lambda$-suppressed operators,  unless they are forbidden by additional symmetries. If the cutoff is taken to be the Planck scale, this is the scenario generally known as gravity mediation.

Any supersymmetric extension of the SM must confront two dramatic problems: the stability of the proton, and the mass scale of the higgs doublets (i.e., the $\mu$-problem). We invoke a standard solution to both of these problems, namely a $\mathbb{Z}_{4R}$ $R$-symmetry under which the MSSM matter superfields $Q, ~\bar{u}, ~\bar{d}, ~L$, and $\bar{e}$ carry charge $+1$ and the higgs superfields $H_u, H_d$ carry charge $0$~\cite{Lee:2010gv} (see also Refs.~\cite{Kurosawa:2001iq,Babu:2003qh} for similar scenarios). We assume that the SUSY-breaking field $X$ and the flavon-sector fields that we will introduce below also carry zero charge under the $R$-symmetry. As a result, Yukawa terms like $\int \mathrm{d}^2\theta\, y_u H_u Q \bar{u}$ are allowed, but the dangerous $\mu$-term $\int \mathrm{d}^2\theta\, \mu H_u H_d$ is forbidden. Forbidding a large superpotential $\mu$-term allows for the Giudice-Masiero solution of generating the $\mu$ and $b_\mu$  terms at the scale of soft SUSY breaking masses via the K\"ahler potential operators $\int \mathrm{d}^4\theta\, \frac{X^\dagger}{\Lambda} H_u H_d$ and $\int \mathrm{d}^4\theta\, \frac{X^\dagger X}{\Lambda^2} H_u H_d$, respectively \cite{Giudice:1988yz}. The $\mathbb{Z}_{4R}$ symmetry forbids not only the renormalizable baryon- and lepton-number violating $R$-parity odd operators $H_u L$, $L L \bar{e}$, $\bar{u} \bar{d} \bar{d}$, and $Q L \bar{d}$ from the superpotential, but also the dangerous nonrenormalizable superpotential operators $QQQL$ and $\bar{u}\bar{u}\bar{d}\bar{e}$ that can induce proton decay. It allows unsuppressed, holomorphic contributions to the $A$-terms via the operators $\int \mathrm{d}^2\theta \frac{X}{\Lambda} H_d L \bar{e}$, and to gaugino masses via the operators $\int \mathrm{d}^2\theta \frac{X}{\Lambda} {\cal W}_\alpha {\cal W}^\alpha$.

In the future, it may be interesting to explore models with additional structure in the SUSY-breaking sector, especially those with mildly split SUSY that offer one of the simplest explanations for the observed higgs boson mass. For example, if $X$ carries a charge under some symmetry, then gaugino masses and $A$-terms will be suppressed because holomorphic operators with a single $X$ insertion will be forbidden. In this case, anomaly mediation could supply small gaugino mass terms \cite{Randall:1998uk, Giudice:1998xp}, leading to a mildly split spectrum \cite{Wells:2003tf, ArkaniHamed:2004fb, Ibe:2011aa, Ibe:2012hu, Arvanitaki:2012ps, ArkaniHamed:2012gw}. Similar phenomenology arises if the imaginary part of $X$ is a periodic axion field, such that $X$ comes equipped with a discrete shift symmetry. K\"ahler potential soft terms involving $X + X^\dagger$ are still allowed, as is the gaugino mass term $\int \mathrm{d}^2\theta X {\cal W}_\alpha {\cal W}^\alpha$, provided that the coupling is adjusted to match the periodicity of $\mathrm{Im}\, X$ to that of the gauge theory's $\theta$-angle. This symmetry suppresses $A$-terms, but also tends to suppress gaugino masses by a loop factor relative to scalar masses (see, e.g., Refs.~\cite{Baryakhtar:2013wy}). Such structure is relatively common in scenarios with SUSY breaking mediated by moduli arising from extra dimensions, which sometimes have further structure modifying the form of the EFT~\cite{Binetruy:1985ap,Binetruy:1987xj,Conlon:2005ki,Blumenhagen:2009gk,Aparicio:2014wxa,Reece:2015qbf}. For simplicity, in this paper we do not consider this range of possibilities. Instead, we will consider mediation only through a neutral SUSY-breaking field $X$, with couplings that are unsuppressed save for the additional spurions needed to account for flavor and CP symmetries, as we discuss below.

\subsection{The SUSY CP and flavor problem}
\label{subsec:SUSY_CP_flavor_problem}

Supersymmetric extensions of the SM as described in Sec.~\ref{subsec:susy_structure} are generically severely constrained by flavor and CP violating observables. 
We intend to study the bounds on such models from the observables of Table~\ref{tabs:obslist}. 
In such models sleptons, neutralinos, and charginos can run in the loop of Fig.~\ref{fig:dipole-diag} and contribute to the observables from Table~\ref{tabs:obslist}. This can introduce stringent bounds on these particles' masses.\footnote{The LFV bounds can be avoided with gauge-mediated SUSY breaking, but at the cost of other problems such as the gravitino problem \cite{Pagels:1981ke, Weinberg:1982zq, Khlopov:1984pf, Moroi:1993mb}. As a result, we focus on the gravity-mediated SUSY breaking models, which lack an intrinsic mechanism for suppressing the dangerous contributions to the observables of Table~\ref{tabs:obslist}.} In this section, we review the basics of lepton flavor physics in the MSSM and argue that, without further structure, the bounds from the observables in Table~\ref{tabs:obslist} severely constrain its parameter space.

In the MSSM, the scalar (the heavy fermion) in the loop in Fig.~\ref{fig:dipole-diag} is replaced by various sleptons (charginos or neutralinos) to generate the dipole operators of Eq.~\eqref{eq:L-dipole} in the IR. Once the couplings between the various sleptons, charginos/neutralinos, and the SM fermions are determined, we can simply use the results of App.~\ref{app:dipole-calc} to find the MSSM contribution to the dipole operator.

The chargino and neutralino mass eigenvalues are dominated by the gaugino masses and $\mu$; see, for instance, Ref.~\cite{Hisano:1995cp} for their mass matrices.
We work in a basis where these mass matrices are diagonalized. 
We assume the mixing between different mass eigenstates, which is proportional to the ratio of SM gauge boson masses to these mass eigenvalues, is small. 
This suggests that the gauge eigenstates and the mass eigenstates have substantial overlap, allowing us to refer to the mass eigenstates with their dominant component in the gauge basis, i.e., gaugino or higgsino. 

While the MSSM model has many degrees of freedom, our qualitative conclusions are independent of the numerical relations between most of them. As a result, we use a few working assumptions about the MSSM spectrum to simplify our analysis. We list some of these assumptions below. 

The gaugino masses are dominated by their soft masses $M_1$ and $M_2$. It is usually expected that these soft masses are comparable. For simplicity, we assume the bino and wino soft masses are related as $M_2/M_1=2$, as is often the case in models with gaugino mass unification; changing this ratio does not affect our final qualitative conclusions. 
In the SM fermion mass basis, the charged slepton mass matrices are parameterized as 
\begin{equation}
\tilde{M}_{\tilde{f}}^2 = \tilde{m}^2 \left(	\begin{matrix}
\delta^{LL} & \delta^{LR} \\
(\delta^{LR})^\dagger & \delta^{RR}
\end{matrix}	\right),
\label{eq:slepton-mass-mtx}
\end{equation}
while the sneutrino mass matrices are simply $\tilde{M}_{\tilde{\nu}}^2 = \tilde{m}^2 \delta^{LL}$. Here $\tilde{m}$ is a shared mass scale for all the sleptons. The dominant part of the diagonal blocks $\delta^{LL/RR}$ come from the SUSY breaking slepton masses, all of which are assumed to be of the order $\tilde{m}$, while 
\begin{equation}
    \delta^{LR} \approx - \frac{t_\beta}{\sqrt{2} \tilde{m}^2} \mu M_f + \frac{\langle H_d \rangle }{\sqrt{2}} \frac{A}{\tilde{m}^2},
    \label{eq:deltaLR}
\end{equation}
where $M_f$ is the SM leptons' $3\times 3$ mass matrix, $t_\beta= \langle H_u \rangle / \langle H_d \rangle $ is the ratio of the two higgs vevs, $\mu$ is the usual mass scale in the MSSM higgs sector, and $A$ is the leptons' $A$-terms matrix.\footnote{Generically the mass scale of the $LL$ block differs from that of the $RR$, at least due to different running from the scale of SUSY breaking. We neglect this $\mathcal{O}(1)$ effect for the sake of simplicity of the numerical study. Taking this effect into account typically eases the constraints from experiments as it decreases the $L-R$ slepton mixing.
}
 We also assume $|\mu | = \tilde{m}$ and $A= \tilde{m} \delta^A$, which suggests the higgsino masses and the scale of $A$-terms are comparable to the sleptons mass scale. Had we neglected the $A$-terms, in the SM fermion mass basis, $\delta^{LR}$ would have been flavor-diagonal. 
All the $\delta$ matrices ($\delta^{LL},~\delta^{RR},~\delta^{LR},~\delta^{A}$) have dimensionless entries whose magnitude is determined by the details of the UV model. 

Notice that unlike the charginos and neutralinos, in our working basis the sfermion mass matrix is not diagonal. The off-diagonal entries can give rise to LFV effects. The contributions of all sfermions and charginos/neutralinos to the dipole operators using the mass insertion approximation are included in App.~\ref{app:all_contributions}. The diagrams in the appendix are only meant to provide us with an intuition about the contribution of different states. In our numerical calculations in the next section, we properly diagonalize these mass matrices and calculate the dipole operators exactly.

Without any suppression of flavor-violating couplings, i.e., with arbitrary mixing pattern among sfermions in Eq.~\eqref{eq:slepton-mass-mtx}, the  observables of Table~\ref{tabs:obslist} put very strong bounds on the mass scale of SUSY particles. 
The eEDM and $\BRmutoe$ in this scenario depend on the UV parameters as (see, e.g.,~\cite{Paradisi:2005fk,Altmannshofer:2013lfa})
\begin{align}
    \label{eq:anarchy_scaling}
    \frac{|d_e|}{e} & \sim 
    10^{-25}\,\left(\frac{5}{\tan\beta}\right)\left(\frac{10\,{\rm TeV}}{\tilde m}\right)^3 \left(\frac{M_1}{1\,{\rm TeV}}\right)
    \,{\rm cm}\, , \\
    \BRmutoe & \sim
10^{-9}\,\left(\frac{5}{\tan\beta}\right)^2\left(\frac{10\,{\rm TeV}}{\tilde m}\right)^6 \left(\frac{M_1}{1\,{\rm TeV}}\right)^2  \, . \nonumber 
\end{align}
Here, we assume all slepton masses are the same order as $\tilde m$ and that $\tilde m\sim |\mu|\gg M_2 \sim M_1$. Off-diagonal elements of the $\delta^A$ matrices are taken to be $\delta^A\sim 1$ so that the $A$-term contributions are dominant for both the eEDM and $\BRmutoe$. The CP phases in $\delta$s are also assumed to be maximal, contributing to the eEDM.

\section{Horizontal Symmetry and Spontaneous CPV}
\label{sec:minimal}

It is interesting to consider simple mechanisms that can suppress LFV and CPV effects. 
In the present section, we will see that introducing a new horizontal symmetry and forbidding explicit CPV can suppress these observables in the MSSM (see Eq.~\eqref{eq:scaling_horizontal} and Fig.~\ref{fig:horizontal+SCPV}), thus opening up the parameter space for lighter supersymmetric particles.

\subsection{A $U(1)$ horizontal symmetry}
\label{subsec:model}

A simple mechanism for suppressing the LFV effects in SUSY models is \textit{alignment} \cite{Nir:1993mx,Grossman:1995hk}. In this mechanism, the slepton mass matrices are diagonal in the SM fermion mass basis, eliminating any sources of LFV. 
The most straightforward way to achieve this is by augmenting the model with a $U(1)$ horizontal symmetry \cite{Froggatt:1978nt}, i.e., a $U(1)$-augmented MSSM, under which superfields of different generation have different charges. As shown in Ref.~\cite{Leurer:1992wg}, such a horizontal symmetry can not be exact. Generically, it is assumed that this symmetry is broken spontaneously via the vev of some ``flavon". 
Although we do not expect UV-complete theories to have exact global symmetries (spontaneously broken or not), a very good approximate $U(1)$ symmetry could arise, protected by discrete gauge symmetries or perhaps as a remnant of an anomalous gauged $U(1)$ obtaining a Green-Schwarz mass.

By charging the MSSM fields under this symmetry, we not only can generate the SM pattern of leptons' masses and mixings, but we can also suppress the disastrous LFV contributions from slepton loops. (The contribution of different superpartners are still captured by the diagrams in App.~\ref{app:all_contributions}.) 
In such a setup, the eEDM and $\BRmutoe$ observables depend on the UV parameters as~\cite{Altmannshofer:2013lfa} 
\begin{align}
    \label{eq:scaling_horizontal}
    \frac{|d_e|}{e} & \sim 10^{-29}
\left(\frac{10\,{\rm TeV}}{\tilde m}\right)^3 \left(\frac{M_1}{1\,{\rm TeV}}\right)\left(\frac{\tan\beta}{5}\right)\,{\rm cm}\, , \\
    \BRmutoe & \sim  10^{-12}
\left(\frac{10\,{\rm TeV}}{\tilde m}\right)^6 \left(\frac{M_2}{1\,{\rm TeV}}\right)^2\left(\frac{\tan\beta}{5}\right)^2 \, , \nonumber 
\end{align}
where we use a shared slepton mass scale $\tilde m$ with $\tilde m \sim |\mu|\gg M_2 \sim M_1$. 
Off-diagonal elements of the soft mass squared matrices are suppressed by a horizontal symmetry as well, and the $A$-terms are neglected. We assume the fermion charges under this symmetry have the right values implied by the right SM fermion masses and lepton mixings (see Eq.~\eqref{eq:charges_conditions} below). The off-diagonal elements $\delta^{LL,RR}$ used here in the Yukawa diagonal basis are given in Eq.~\eqref{eq:scen2_texture} below, and the CP phases are assumed to be maximal.

Comparing these equations to Eq.~\eqref{eq:anarchy_scaling} indicates that, generically, with the right horizontal symmetry charges the disastrous contributions to the eEDM and $\BRmutoe$ can be suppressed compared to a structureless setup with random CPV entries in the mass matrices. 
In light of this, in what follows we introduce a general setup in which we augment the MSSM with a horizontal $U(1)$ symmetry. We will study the bounds on this setup from the observables listed in Table~\ref{tabs:obslist}.

Let us start by assuming we have a single flavon field $S_1$. We can normalize all the charges such that this flavon's charge under the horizontal symmetry is $[ S_1 ] = -1$ and that its vev's phase is rotated away using an appropriate field redefinition. For simplicity, we assume all the lepton superfields have integer charges under this symmetry too. Let us also assume the small expansion parameter $\langle S_1\rangle/\Lambda \equiv \epsilon_1$, where $\Lambda$ is a UV cut-off where the details of the $S_1$ potential become relevant, and should not be confused with the SUSY breaking scale $\Lambda_{\rm SUSY}$. We also assume that $\langle S_1 \rangle$ is generated through supersymmetric dynamics, so that the flavons do not themselves mediate SUSY-breaking (and conversely, we will assume that SUSY-breaking spurions do not provide additional breaking of flavor symmetries). 
Conventionally, we assume $\epsilon_1 \sim \lambda$.\footnote{Notice that we can capture the effect of smaller $\epsilon_1$s by increasing the absolute value of the lepton superfield charges under the $U(1)$ horizontal symmetry.}

In addition to the horizontal symmetry that can suppress disastrous flavor-violating effects, it was argued in Ref.~\cite{Nir:1996am} that by prohibiting explicit CPV effects in the UV theory, light squarks could evade the stringent bounds from the neutron EDM. In such a theory, CP is assumed to  be a  gauge symmetry that is only broken spontaneously and via the interference of two flavon vevs, one of which has a physical phase. We now extend the idea of spontaneous CPV to the leptonic sector of the MSSM.
In our setup we introduce another flavon field, $S_2$, with $[ S_2 ] = -c_2$ and $\langle S_2\rangle/\Lambda \equiv \epsilon_2 \sim \lambda^{n_2}$, with $c_2$ and $n_2$ integer numbers greater than 1. We assume $S_2$ gets a complex vev and spontaneously breaks CP.

Let us study in more details how spontaneous CPV can suppress different CPV effects in the MSSM. In such a setup, the only source of CPV is the interference between terms with different powers of the two flavons. For instance, if a holomorphic operator made of different superfields has the horizontal charge $[O]=c_O \geqslant 0$, the coefficients multiplying it can have the following flavon spurion insertions:
\begin{equation}
\label{eq:interferences1}
\epsilon_1^{c_O- \xi c_2 } \epsilon_2^{\xi} \sim \lambda^{c_O + \xi (n_2 - c_2) },~\mathrm{for}~c_O - \xi c_2 , \xi \in \mathbb{Z}^{\geqslant 0}.
\end{equation}
An effective holomorphic superpotential term can be generated by inserting vevs in non-holomorphic K\"ahler potential terms, as in the Giudice-Masiero mechanism. In this case, the resulting term is suppressed by both SUSY-breaking spurions and flavor spurions. 
For non-holomorphic insertion of $\epsilon_1$ or $\epsilon_2$, we have 
\begin{eqnarray}
\label{eq:interferences2}
\epsilon_1^{|c_O- \xi c_2|} \epsilon_2^{\xi} \sim \lambda^{|c_O- \xi c_2| + \xi n_2 },~\mathrm{if}~\xi \in \mathbb{Z}^{\geqslant 0}, \\
\epsilon_1^{|c_O- \xi c_2|} (\epsilon_2^{*})^{|\xi|} \sim \lambda^{|c_O- \xi c_2| + |\xi| n_2 },~\mathrm{if}~\xi \in  \mathbb{Z}^{\leqslant 0}. \nonumber
\end{eqnarray}
In writing these equations we used the fact that $\epsilon_1=\epsilon_1^{*}$. It can be shown that the least $\lambda$ suppression happens for terms where both flavon spurions $\epsilon_{1,2}$ have non-negative powers, i.e., the holomorphic condition in Eq.~\eqref{eq:interferences1} is satisfied. 
Moreover, from Eq.~\eqref{eq:interferences1} we find that when $c_2=n_2$, different interfering terms have the same $\lambda$ scaling, thus their interference can give rise to $\mathcal{O}(1)$ phases. This is the maximally violating CP scenario, as in cases with $c_2\neq n_2$ the interfering phase is suppresed by powers of $\lambda$. 
Since we still need to generate the observed CPV in the CKM and PMNS matrices, we focus on the case with $c_2=n_2$ and assume $\mathrm{Arg}(\epsilon_2) = \pi/4$, which maximizes the amount of CPV in our setup. 

Nonetheless, even when $c_2=n_2$, spontaneous CPV can suppress various disastrous contributions to the eEDM. In particular, we consider cases where the higgses are uncharged under the horizontal symmetry. Thus, the $\mu$ term multiplies a neutral operator ($c_O=0$) and, according to Eq.~\eqref{eq:interferences1}, the holomorphic contribution to $\mu$ is real. The leading CPV 
contribution to $\mu$ from non-holomorphic K$\ddot{\mathrm{a}}$hler potential terms is suppressed by $\epsilon_1^{c_2}\epsilon_2^{*} \sim \lambda^{c_2+n_2}$ (see Eq.~\eqref{eq:interferences2}), implying 
\begin{equation}
    \label{eq:muphase}
    \mathrm{Arg}(\mu) \sim \mathrm{Arg} \left( 1+ i \lambda^{c_2+n_2} \right) \sim \lambda^{c_2+n_2},
\end{equation}
which suppresses its contribution to the eEDM. It should be noted that this suppression crucially depends on the two higgses being neutral under the horizontal symmetry. We use this assumption in our model too.

With neutral higgses, the SM charged lepton mass matrix is then given by
\begin{equation}
    M^f_{ij} = \langle H_d \rangle C^{f,ij}_{a_1,a_2} \epsilon_1^{a_1} \epsilon_2^{a_2}~:~a_1+a_2 c_2 = [L_i]+[\bar{e}_j]~\& ~ a_i \geqslant 0,
    \label{eq:mass_mtx_fermions}
\end{equation}
while the neutrinos' Majorana mass matrix is
\begin{equation}
    M^\nu_{ij} = \frac{\langle H_u \rangle^2}{M_\nu} C^{\nu,ij}_{a_1,a_2} \epsilon_1^{a_1} \epsilon_2^{a_2}~:~a_1+a_2 c_2 = [L_i]+[L_j]~\& ~ a_i \geqslant 0.
    \label{eq:mass_mtx_neutrinos}
\end{equation}
Here $\langle H_u \rangle$ is the up-type higgs vev, $i,j$ are SM fermion generation indices, $C^{f/\nu,ij}_{a_1,a_2}$ are $\mathcal{O}(1)$ numbers, and $M_\nu$ is an undetermined UV scale where new heavy neutrinos appear~\cite{Minkowski:1977sc,Yanagida:1979as,GellMann:1980vs,Glashow:1979nm}. Notice that all $C^{ij}_{a_1,a_2}$s are real; the only source of CPV is $\epsilon_2$. The hermiticity of the Lagrangian forces the undetermined numbers in the neutrino mass matrix to follow $C_{a_1,a_2}^{\nu,ij}=C_{a_1,a_2}^{\nu,ji}$.\footnote{In writing these equations we have neglected possible kinetic mixing terms among different superfields and start from a basis where the kinetic terms are canonical. As discussed in appendix B of Ref.~\cite{Leurer:1993gy}, including these terms will merely give rise to some corrections to the undetermined $\mathcal{O}(1)$ coefficients in the mass matrices, i.e., slightly changes their distribution.}

We then perform a rotation to the mass basis of the SM fermions. Similar to Ref.~\cite{Froggatt:1978nt,Grossman:1995hk}, we can show that the PMNS matrix entries scale like 
\begin{equation}
    \label{eq:vpmns_FN}
    |V^{\mathrm{PMNS}}_{ij}| \sim \epsilon_1^{|[L_i]-[L_j]|}.
\end{equation}
The large mixing angles among the LH leptons \cite{Zyla:2020zbs} suggests that various $[L_i]$ should be close to each other. Comparing the fermion mass and mixing matrices with the observed values suggests the following relations between different fermions' charges: 
\begin{align}
    \label{eq:charges_conditions}
    [L_1] - 1  & \approx  [ L_2 ] \approx [ L_3 ], \nonumber \\
    [L_1] + [\bar{e}_1] & \sim  8+x_\beta , \nonumber \\
    [ L_2 ] + [\bar{e}_2] & \sim  5+x_\beta , \\
    [ L_3 ] + [\bar{e}_3] & \sim  3+x_\beta , \nonumber 
\end{align}
where $x_\beta = \log_{0.2} t_\beta$. The first line is a result of the observed neutrino mixing pattern $(V^{\mathrm{PMNS}})$, while the next lines are due to the observed charged lepton masses. Notice that thanks to undetermined $\mathcal{O}(1)$ numbers in the original mass matrices, we can still obtain the observed masses and mixing patterns even if we slightly deviate from these relations.

The holomorphy of the SUSY models does not allow terms with negative powers of $\epsilon_1$ or $\epsilon_2$ to appear in the fermion mass matrices; this is enforced by the $a_i \geqslant 0$  condition above. The SUSY-breaking contributions to the sfermion mass matrices, on the other hand, can include non-holomorphic terms. The charged sfermion mass matrices are given by
\begin{eqnarray}
    \label{eq:conditions}
    \delta^{LL}_{ij} &=& C^{L,ij}_{a_1,a_2} \epsilon_1^{a_1} \epsilon_2^{(*)|a_2|}~:~a_1 = |[L_i]-[L_j]-a_2 c_2|, \\
    \delta^{RR}_{ij} &=& C^{R,ij}_{a_1,a_2} \epsilon_1^{a_1} \epsilon_2^{(*)|a_2|}~:~a_1 = |[\bar{e}_j]-[\bar{e}_i]-a_2 c_2|, \nonumber
\end{eqnarray}
where the complex conjugate superscript ($*$) appears when $a_2 \leqslant 0$. Meanwhile, $\delta^{LR}$ is given by 
\begin{equation}
    \label{eq:deltaLRwithC}
    \delta^{LR} = - \frac{t_\beta}{\sqrt{2} \tilde{m}^2} \mu M_f +  \frac{\langle H_d \rangle }{\sqrt{2} \tilde{m}} \delta^A,
\end{equation}
where the last term captures the effect of $A$-terms from Eq.~\eqref{eq:deltaLR} and 
\begin{equation}
    \delta^A_{ij} = C_{a_1,a_2}^{A,ij} \epsilon_1^{a_1} \epsilon_2^{a_2}~:~a_1+a_2 c_2 = [L_i]+[\bar{e}_j]~\& ~ a_i \geqslant 0.
    \label{eq:deltaA_def}
\end{equation}
Recall that we assume all the $A$-terms have a shared scale, which is the same as the shared slepton mass scale $\tilde{m}$. 
We also assume $|\mu|=\tilde{m}$. 
As mentioned before, we also assume $M_1=0.5 M_2$. Thus, apart from the horizontal charges of the superfields,  the only free parameters in our setup are $M_{1,2}$, $\tilde{m}$, and $\tan \beta$.

With a hierarchical choice of charges for superfields of different generations, the horizontal symmetry pushes us near the alignment limit for the sfermion mass matrix. Similarly, spontaneous CPV suppresses the eEDM contributions. This allows us to lower the scale of new SUSY partners in our setup, without violating the stringent LFV or eEDM bounds included in Table~\ref{tabs:obslist}.

The $C^{ij}$ parameters in all the mass matrices originate from undetermined UV dynamics. Depending on their values, the model prediction for different IR observables can change significantly. As a result, in our study, we treat these numbers as random $\mathcal{O}(1)$ coefficients. For any fixed set of charges, we generate all the mass matrices many times and calculate the contribution of each trial to different IR observables. 
Further details on how we generate these mass matrices and the distribution of these random coefficients are included in App.~\ref{app:matrices}, 
where we also show the resulting distribution of the slepton masses and in particular show that $\tilde{m}$ is indeed the average slepton mass scale. 

\begin{figure}
\centering
\resizebox{ \columnwidth}{!}{
\includegraphics[scale=1]{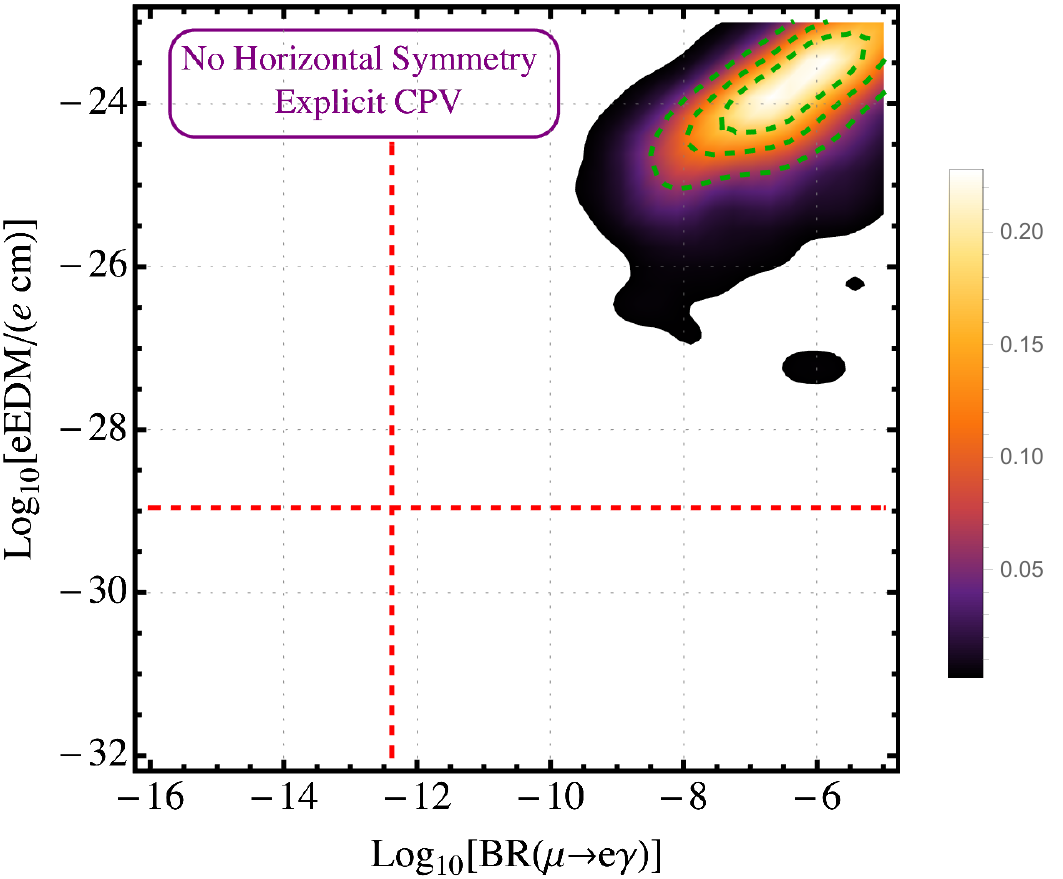}
\includegraphics[scale=1]{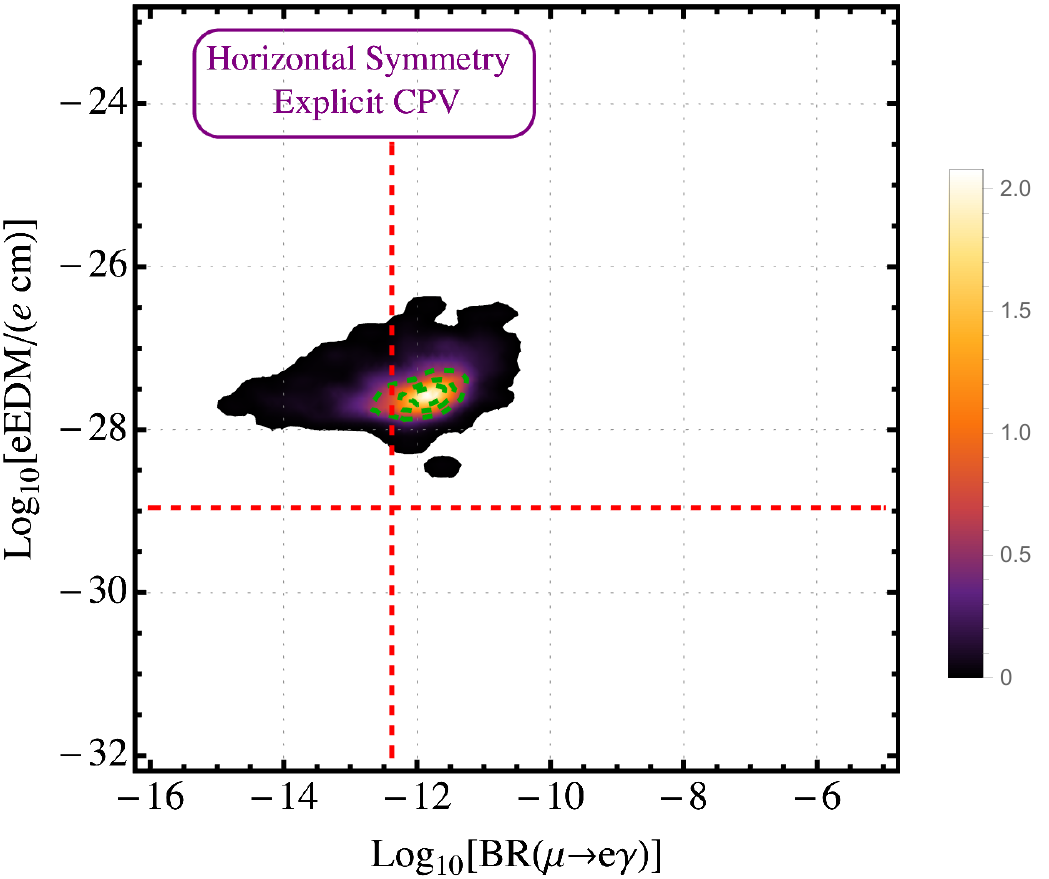}
}\\
\resizebox{ \columnwidth}{!}{
\includegraphics[scale=1]{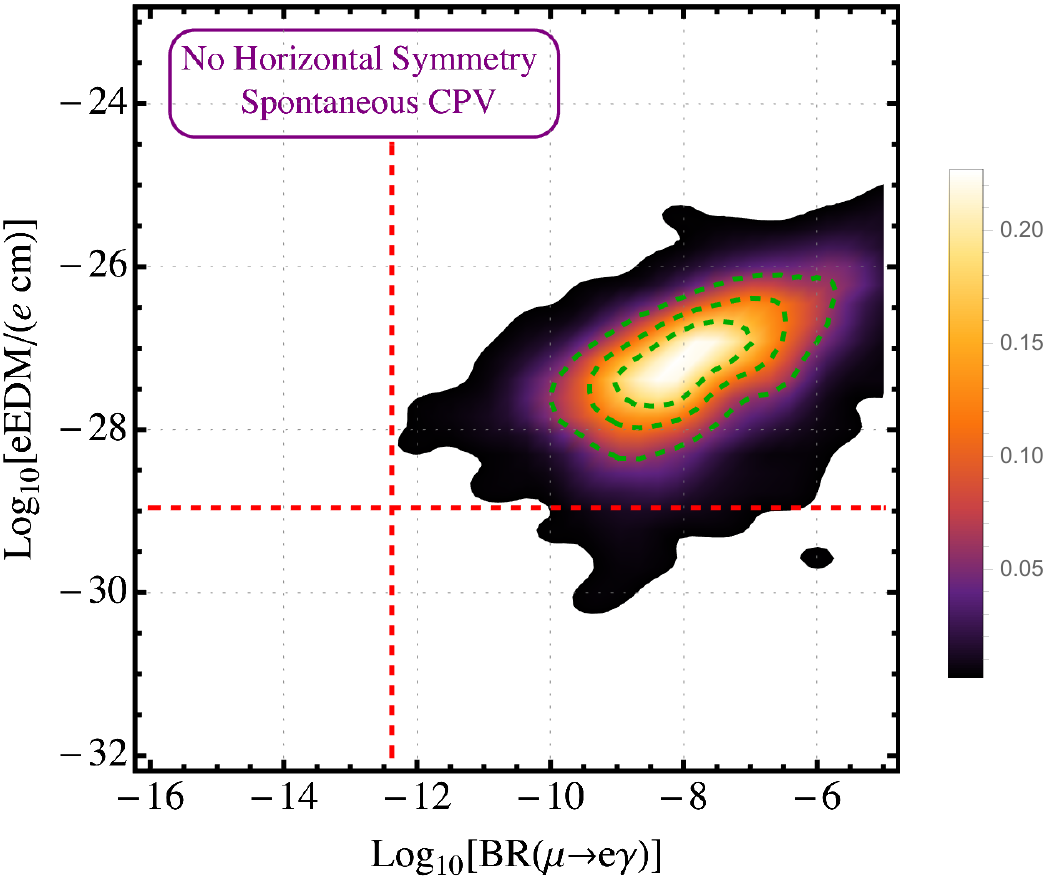}
\includegraphics[scale=1]{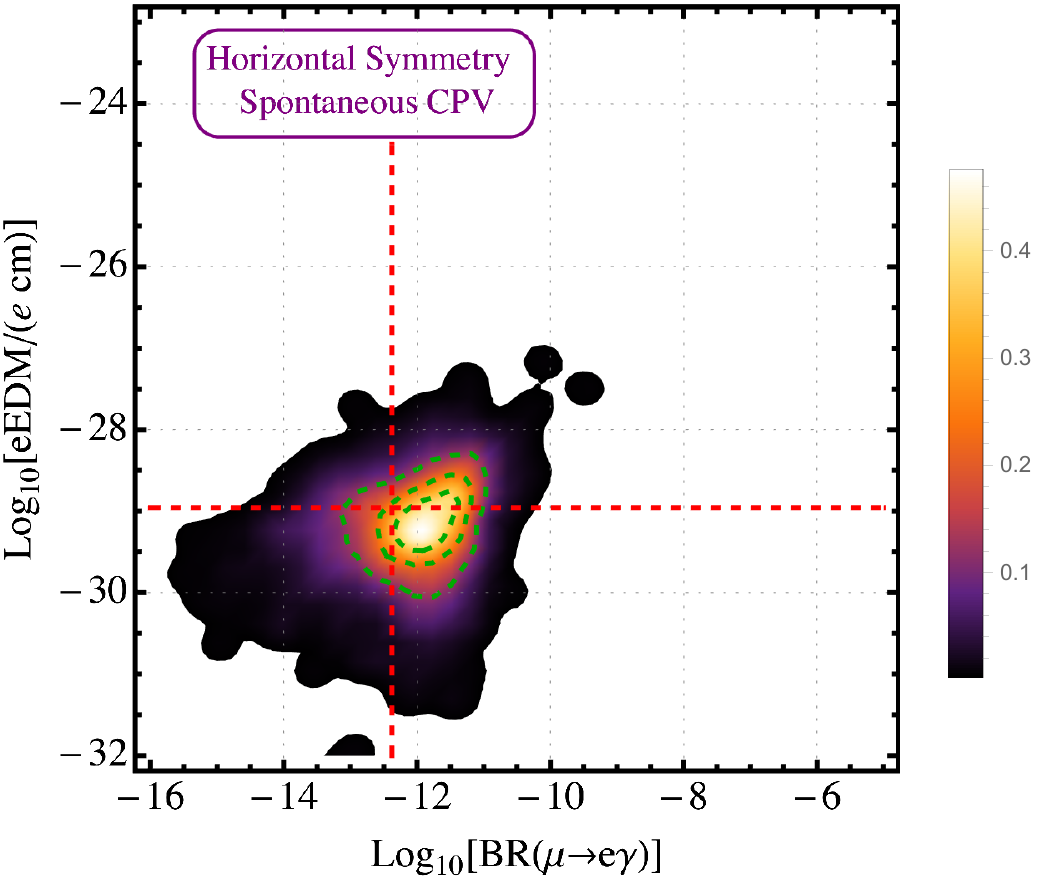}
}
\caption{The distribution of the prediction of four different extensions of the MSSM for the eEDM and $\BRmutoe$. Dashed red lines indicate the current experimental upper bounds. We fix $\tilde{m}=10M_1=10M_2=10~$TeV and $t_\beta =5$ in all the scenarios. In each scenario, we fix the lepton masses to their SM values (despite these being quite unlikely in the anarchic cases) and generate the slepton mass matrices 1000 times with random $\mathcal{O}(1)$ in each mass matrix entry (see App.~\ref{app:matrices} for how these random numbers are generated) and find a few orders of magnitude change in the observables depending on these random numbers. In each figure, the dashed green contours divide the space into four different regions with equal number of total random trials. The four scenarios we consider are (i) generic MSSM without any horizontal symmetry and with explicit CPV (\textbf{top left}), (ii) MSSM augmented with a horizontal $U(1)$ symmetry and with explicit CPV (\textbf{top right}), (iii) generic MSSM without any horizontal symmetry but with spontaneous CPV (\textbf{bottom left}), and (iv) MSSM augmented with a $U(1)$ horizontal symmetry with spontaneous CPV (\textbf{bottom right}). See the text for further details on each scenario. We observe that both the horizontal symmetry and spontaneous CPV can suppress the stringently constrained observables $\BRmutoe$ and eEDM; this opens up parameter space for lighter superpartners. Similar conclusions can be drawn using other LFV observables from Table~\ref{tabs:obslist} instead of $\BRmutoe$.}
\label{fig:horizontal+SCPV}
\end{figure}

To illustrate the effect of these undetermined coefficients, in Fig.~\ref{fig:horizontal+SCPV}, for fixed values of the MSSM mass parameters, we show the contribution to the eEDM and $\BRmutoe$ in four different extensions of the MSSM. 
In each panel of the figure, we start in the SM fermion mass basis and generate the UV mass matrices for the sleptons 1000 times with different random $\mathcal{O}(1)$ coefficients reflecting our agnosticism about the UV model parameters. 
More details on the distribution of these random numbers are included in App.~\ref{app:matrices}. 
In the figure, we also fix $\tilde{m}=10M_1=10M_2=10~$TeV\footnote{In these plots we deviate from our assumption of $M_1=0.5M_2$ so as to compare the results directly with Eqs.~\eqref{eq:anarchy_scaling}-\eqref{eq:scaling_horizontal}; we go back to $M_1=0.5M_2$ assumption for the analysis of our models in the rest of the paper. } 
and $t_\beta =5$ in all the scenarios. The four scenarios we consider are summarized below. 
\begin{itemize}
    \item In the top left plot of Fig.~\ref{fig:horizontal+SCPV} we consider a case with random complex numbers in all the slepton mass matrix entries and with no flavon spurion suppression. This is essentially a structureless (anarchic) MSSM model with explicit CPV. We assume $\mathrm{Arg}(\mu) = \pi/4$. 
    Such a structureless mass matrix, where each entry is an $\mathcal{O}(1)$ number times a shared mass scale, does not typically give rise to mass eigenvalues that follow the SM hierarchical pattern of masses.\footnote{For that to happen, the  random numbers in the charged leptons mass matrix had to roughly be spread in the range $10^{-5}-10^{-2}$. Assuming these random numbers are drawn from similar distributions (say a uniform distribution between $-1$ and 1), the probability of this outcome is exceedingly small.} 
By starting from the SM fermion mass basis, we are neglecting this shortcoming of these structureless models in favor of a better comparison of their contribution to the eEDM and $\BRmutoe$ observables to the scenarios with horizontal symmetries.
    \item In the top right plot of Fig.~\ref{fig:horizontal+SCPV} we consider a generic case with a horizontal $U(1)$ symmetry and sample superfield charges $\{[L_1],~[L_2],~[L_3],~[\bar{e}_1]~[\bar{e}_2]~[\bar{e}_3] \}=\{4,~3,~3,~3,~0,~ -2\}$ (with the higgses still uncharged under the horizontal symmetry). These charges obey Eq.~\eqref{eq:charges_conditions} that are implied by the observed SM lepton masses and mixing. Using Eqs.~\eqref{eq:conditions}-\eqref{eq:deltaA_def} (without any $\epsilon_2$ and $\epsilon_1\sim \lambda$), for these charges we find 
    \begin{equation}
    \tilde{M}_f^2 \sim \left(10 ~\mathrm{TeV} \right)^2~\times ~\left(\begin{matrix}
    1 & \lambda^1 & \lambda^1 & \lambda^7 & \lambda^4 & \lambda^2 \\
    \lambda^1 & 1 & 1 & \lambda^6 & \lambda^3 & \lambda^1 \\
    \lambda^1 & 1 & 1 & \lambda^6 & \lambda^3 & \lambda^1 \\
    \lambda^7 & \lambda^6 & \lambda^6 & 1 & \lambda^3 & \lambda^5 \\
    \lambda^4 & \lambda^3 & \lambda^3 & \lambda^3 & 1 & \lambda^2 \\
    \lambda^2 & \lambda^1 & \lambda^1 & \lambda^5 & \lambda^2 & 1 \\
    \end{matrix}
    \right),
    \label{eq:scen2_texture}
    \end{equation}
    times random complex numbers in each entry. 
    Again we assume $\mathrm{Arg}(\mu) = \pi/4$.
    \item In the bottom left plot of Fig.~\ref{fig:horizontal+SCPV} we consider an anarchy scenario where none of the superfields are charged under a horizontal symmetry. However, we still introduce two flavons with $[S_1]=-1$ and $[S_2]=-2$ and the only source of CPV is the vev of the second flavon that has a phase of $\pi/4$. CPV is introduced only via the interference of spurion terms in the non-holomorphic soft slepton masses and $\mu$ term phase. Using Eq.~\eqref{eq:interferences2}, and the fact that the higgses are uncharged under the horizontal symmetry, we find $\mathrm{Arg}(\mu) \sim \mathrm{Arg} \left( 1+ i \lambda^4 \right) \sim \lambda^4$. 
    \item Finally in the bottom right plot of Fig.~\ref{fig:horizontal+SCPV} we consider a scenario with the same charges as the second scenario, but this time we introduce a second flavon with $[S_2]=-2$ (similar to the third scenario) and assume the only source of CPV is through its interference with the first flavon. This is a typical scenario with an horizontal symmetry and spontaneous CPV that we will be studying in more details in this work.
\end{itemize}

Comparing the first scenario with the second scenario, we see the substantial effect of the horizontal symmetry in suppressing both LFV and eEDM. The order of magnitude of the contribution of these two scenarios to the eEDM and $\BRmutoe$ are given by Eqs.~\eqref{eq:anarchy_scaling}-\eqref{eq:scaling_horizontal}, respectively. The contribution in the eEDM of the second scenario is dominated by the $\mu$ term phase, which is kept constant in different trials; hence, the distribution in the eEDM has a much smaller spread. The third scenario with only spontaneous CPV, and no further flavor structure, also shows a great suppression in the eEDM thanks to the suppressed $\mu$ argument. Finally, the type of models we are studying in this work are represented by the fourth scenario. We find a significant suppression in at least one of the two observables compared to all the previous scenarios.

The spread in the distribution of the observables shows the importance of undetermined $\mathcal{O}(1)$ numbers in the UV theory. As mentioned earlier, we treat these undetermined coefficients as random numbers that change in different trials. The sampling of the random numbers is further explained in App.~\ref{app:matrices}.

Furthermore, this wide spread undermines the accuracy of naive order of magnitude estimations such as Eqs.~\eqref{eq:anarchy_scaling}-\eqref{eq:scaling_horizontal}. In the upcoming section, we develop a framework for interpreting the bounds on such models that turns this substantial effect of the undetermined UV parameters into a naturalness argument.

\subsection{A new measure of naturalness for different observables}
\label{subsec:tuning_def}

In the previous section, we highlighted the effects of a horizontal symmetry and spontaneous CPV in the MSSM that allow us to obtain the SM in the IR, while relaxing various LFV and eEDM bounds. However, a proper calculation of the contribution of these models to the IR operators is muddled owing to numerous undetermined UV theory coefficients in the mass matrices. As is  evident from Fig.~\ref{fig:horizontal+SCPV}, these coefficients can give rise to orders of magnitude spread in the predictions from various flavor observables. 

In particular, we argued that there is always some configuration of these random numbers that can completely suppress these observables. This suppression can happen for any scale of superpartner masses. As an extreme example, when all the off-diagonal random numbers in the slepton mass matrix of Eq.~\eqref{eq:slepton-mass-mtx} are zero, the LFV observables do not get any contribution from this model, regardless of the scale of the slepton masses. Nonetheless, as we lower the superpartner scale, the fraction of space of the undetermined UV numbers that can give rise to suppressed values of these observables decreases.

In this section, we develop a framework for interpreting the bounds from different observables on models with horizontal symmetries while remaining agnostic about the undetermined coefficients in the UV theory. We can model our ignorance of these coefficients by randomly sampling the space of all these coefficients and see how \textit{natural} it is for a given UV model to give rise to a viable set of IR observable values. One can view this from a Bayesian viewpoint: absent any strong prior reason to prefer one UV model to another, the models that are more likely (as we vary their parameters) to give rise to IR physics resembling what we see in our universe are more likely to be correct.

While, due to the effect of undetermined UV coefficients, we can not ever claim a $U(1)$-augmented MSSM is definitely inconsistent with the results of a given set of precision experiments, we can calculate the fraction of the space of all the random numbers in which these observables are suppressed enough such that the model is not constrained by the aforementioned experimental results. This motivates us to propose the following measure of fine tuning in these models. 

\begin{itemize}
    \item Let's assume we identify the space of all the undetermined UV $ \mathcal{O}(1)$ numbers that give rise to the right pattern of lepton masses and mixings as observed by the experiments and call its volume $V_{\rm SM}$. We then find the fraction of this space that further predicts the value of some observable $\mathbb{O}$ is in a certain range $\mathbb{I}$ and call its volume $V_{\mathbb{O}  \in \mathbb{I}}$. We take the ratio of these two volumes
    \begin{equation}
        \label{eq:t_def}
        t_{\mathbb{O} \in \mathbb{I}} \equiv \frac{V_{\mathbb{O} \in \mathbb{I}}}{V_{\rm SM}}
    \end{equation}
    to define a measure of fine tuning associated with observable $\mathbb{O}$. As an example, condition $\mathbb{I}$ could refer to the observable $\mathbb{O}$ being smaller than some fixed value $\hat{\mathbb{O}}$, i.e., $\mathbb{O} \leqslant \hat{\mathbb{O}}$.
\end{itemize}

Let us clarify this definition further through an example. Let's assume we fix all the charges and mass parameters of a $U(1)$-augmented MSSM model and generate 10,000 different mass matrices for fermions and sfermions, each with their own unique random numbers multiplying different matrix entries. Let's assume we find that 100 of these \textit{trials} give rise to the observed pattern of masses and mixings for leptons and that of these 100 ``good" trials, 25 of them give rise to $\mathbb{O} \leqslant \hat{\mathbb{O}}$ ($\mathbb{O}$ being some observable of interest). Here a trial refers to a set of fermion and sfermion mass matrices with its unique set of random numbers; see App.~\ref{app:matrices} for further details about distribution of these random numbers that we use in our study. Then $t_{\mathbb{O} \leqslant \hat{\mathbb{O}}}=25/100=25\%$, i.e., $25\%$ of the ``good" trials gives rise to $\mathbb{O} \leqslant \hat{\mathbb{O}}$.

The study of mass matrix models always suffers from the large number of free parameters. This complicates a rigorous study of the bounds on such models from IR observables, since the bounds can change significantly depending on what values are used for the undetermined free parameters. Our notion of naturalness enables us to sift through the entire parameter space of a model, a hitherto overlooked task that allows us to systematically update our priors on what models are preferred in light of experimental results. 

Notice that this definition relies on a definition of a ``good" trial. In our work we assume a trial is ``good" if the following conditions are satisfied. 
\begin{itemize}
    \item The charged lepton masses it predicts are within a factor of $\lambda$ of the observed lepton masses,
    \begin{equation}
    \lambda \leqslant \frac{m_l}{m_l^{\mathrm{obs}}} \leqslant \frac{1}{\lambda} \, ,
    \label{eq:good-condition-f-mass}
    \end{equation}
    \item After we rescale all the neutrino masses to get the observed total neutrino masses~\cite{Aghanim:2018eyx} $\sum_\nu m_\nu = 0.12$~eV, we find the neutrino mass differences to be within a $\lambda$ of the observed values:
    \begin{equation}
    \lambda \leqslant \left( \frac{\Delta m_{32}^2}{\Delta (m_{32}^2)^{\mathrm{obs}}} \right) \leqslant \frac{1}{\lambda},~~~\lambda \leqslant \left( \frac{\Delta m_{21}^2}{\Delta (m_{21}^2)^{\mathrm{obs}}} \right) \leqslant \frac{1}{\lambda} \, ,
    \label{eq:good-condition-nu-mass}
    \end{equation}
    with $(m_{32}^2)^{\mathrm{obs}}=2.453 \times 10^{-3}~\mathrm{eV}^2$ and $(m_{21}^2)^{\mathrm{obs}}=7.53 \times 10^{-5}~\mathrm{eV}^2$~\cite{Zyla:2020zbs}.
    \item The absolute value of the PMNS matrix entries are in the following range 
\begin{equation}
     \left(\begin{matrix} 
     0.7 & 0.4 & 0.0 \\ 
     0.1 & 0.3 & 0.5 \\ 
     0.1 & 0.3 & 0.5 
     \end{matrix}  \right) \leqslant  \Big| V^{\rm PMNS} \Big| \leqslant  \left(\begin{matrix} 
     0.95 & 0.7 & 0.3 \\ 
     0.6 & 0.8 & 0.9 \\ 
     0.7 & 0.8 & 0.9 
     \end{matrix}  \right)\, .
    \label{eq:PMNS_range_good}
\end{equation}
\end{itemize}

In the last equation, the acceptable range of PMNS entries are comparable to the uncertainties in the measured value of each entry~\cite{Giganti:2017fhf}.

It should be noted that the ranges in Eqs.~\eqref{eq:good-condition-f-mass}-\eqref{eq:PMNS_range_good} are simply one possible choice; we could use similar criteria with slightly different acceptable ranges for these parameters. The exact numerical value we find for the tuning depends on these criteria for a good trial (and on the range we draw our random numbers from). However, by taking the ratio in Eq.~\eqref{eq:t_def}, we suppress this sensitivity in the tuning calculation. For instance, one might have defined a ``good'' trial as one reproducing the measured masses and mixings within experimental uncertainties, but this is much less computationally tractable (and more of a time-dependent goal) and we expect that it would not substantially change our final results.\footnote{Another measure for the fine tuning is similar to Eq.~\eqref{eq:t_def} but with $V_{\rm SM}$ replaced with the volume of the entire space of random numbers. This definition not only is far more sensitive to our definition of a ``good" trial, but also muddles the intuition about the effect of the observable $\mathbb{O}$ since now it also includes the tuning required to merely get the right pattern of masses and mixings.} We have explicitly checked that changing the criteria for good experiments or the range from which we draw the random numbers at most gives rise to an $O(1)$ change in the final value of the tuning.

In studying the bounds on our setup, we use the tuning measure $t_{\mathbb{O} \leqslant \mathbb{O}^{lim}}$, where now $\mathbb{O}$ is (the absolute value  of) any of the observables from Table~\ref{tabs:obslist} and $\mathbb{O}^{lim}$ is the experimental limit on it. 
Notice that $t_{\mathbb{O} \leqslant \hat{\mathbb{O}}}$ is a non-decreasing function of $\hat{\mathbb{O}}$. 
As a result of this, we can always go to lower values of $\hat{\mathbb{O}}$ by introducing more fine tuning. This suggests that we can always push down the sleptons to lighter masses without violating the existing experimental bounds from observables such as the eEDM or $\BRmutoe$ but the required tuning is worse.

In practice, we approximate the volume of these subspaces by sampling them multiple times and use the ratio of total number of trials in each sample to calculate $t_{\mathbb{O} \leqslant \hat{\mathbb{O}}}$. To probe very small tuning values, we need to generate a large enough sample size of trials. Larger sample sizes not only allow us to probe lower-tuning subspaces, but also reduce the uncertainty in estimation of the tuning values.

A similar definition of fine tuning can be used in the context of any other $U(1)$-augmented model and for other flavor observables, e.g., the quark sector observables. Similarly to how we measure the \textit{naturalness} of a given model from the perspective of the higgs mass hierarchy problem using some fine-tuning measure, we can use the fine-tuning measure of Eq.~\eqref{eq:t_def} to see how \textit{natural} different models are from the perspective of various flavor physics observables.

It is also interesting to compare our naturalness measure to commonly-used naturalness measures of the higgs mass, such as the Barbieri-Giudice measure~\cite{Barbieri:1987fn}. While in the latter the tuning measure quantifies the sensitivity of an IR observable (e.g., the higgs mass) to a certain UV model parameter {\em at a given point in the parameter space} of the model, our tuning measure from Eq.~\eqref{eq:t_def} quantifies how naturally the experimental results, here the experimental results on a given observable $\mathbb{O}$, emerge from the undetermined UV parameters of the {\em model as a whole}. An analogue of the Barbieri-Giudice measure might be used when we fix all the undetermined UV parameters to some value and study the sensitivity of the observable at that specific point in the parameter space. We, however, want to study the entire parameter space of a model and see how typically the SM and other IR observables emerge from that setup. We aim to compare different models, rather than individual points in the parameter space of a single model. 

It will be interesting to apply our analysis to any BSM model to explain the SM flavor structure and see how much tuning should be introduced in order to satisfy various flavor physics constraints at a given mass scale of that model.

\subsection{Fine tuning and flavor bounds on the MSSM}
\label{subsec:tuning_calc}

In this section, we apply our tuning measure to the $U(1)$-augmented setup introduced in Sec.~\ref{subsec:model} to see how natural the light superpartners are given the current and future bounds on the observables of Table~\ref{tabs:obslist}. The specific question we want to answer is: \textit{for a given $U(1)$-augmented MSSM model (with fixed charges under the horizontal symmetry and mass scales), what tuning values should be tolerated in order to avoid the existing constraints on the LFV observables and the eEDM?}

We now consider a handful of different $U(1)$-augmented MSSM models and, for various amounts of tuning, calculate the bounds on their parameter space from LFV observables and the eEDM. We scanned over different charge assignments (assuming integer charges, and $|[f]|\leqslant 4$ for any superfield $f$) and $\tan \beta$ values. For any reasonable $\tan \beta$, we found on the order of $\sim 1000$ different charge assignments that can give rise to the right pattern of lepton masses and mixings (with the criteria in Eqs.~\eqref{eq:good-condition-f-mass}-\eqref{eq:PMNS_range_good}) at least once from among 1000 trials. 
We use the same assumptions on the mass spectrum as outlined in Sec.~\ref{subsec:model}. To review, we assume that all the $A$-terms and the sleptons masses and $|\mu|$ are proportional to the same scale $\tilde{m}$, $M_2=2M_1$, and the $\mu$ phase is given by Eq.~\eqref{eq:muphase}.

In this work we go a step beyond the common practice in the literature. Instead of studying an arbitrary single charge assignment, we study five (out of many) charge assignments which span a wide range of $t_\beta$.  
We show these charge assignments in Table~\ref{tabs:charges_less_tuned}. It should be noted that these are merely sample charges and it is possible that by scanning over the entire set of charges, we could find charge assignments that are less constrained. 
Our study is an intermediate step towards an exhaustive study of large number of charge assignments, which requires the development of novel scanning methods.\footnote{See Refs.~\cite{Harvey:2021oue,Hollingsworth:2021sii} for recent proposals for efficient methods of finding all potentially suitable charge assignments.}
The study of a large number of charge assignments emphasizes which features are a universal outcome of horizontal symmetries and which features are unique to a specific charge assignment. 

For the charges we study in Table~\ref{tabs:charges_less_tuned}, the charges in the first and fourth rows of the table obey the exact relations predicted by Eq.~\eqref{eq:charges_conditions}.\footnote{Notice the third row is an anarchy scenario in the neutrino mass matrix \cite{Hall:1999sn}, favored by the observations from various neutrino experiments. }
In the second, third, and fifth rows, the relation between the LH fermion charges slightly deviate from Eq.~\eqref{eq:charges_conditions}.

\begin{table}
\resizebox{\columnwidth}{!}{
\scalebox{0.05}[0.05]{
\begin{tabular}{|c|c|c|c|c|c|c|c|c|}
\hline 
$t_\beta$ & $[S_2]$ & $[L_1]$ & $[L_2]$ & $[L_3]$ & $[\bar{e}_1]$ & $[\bar{e}_2]$ & $[\bar{e}_3]$ & ``Good" Trial Ratio  \\ 
\hline 
25& -3 & 4 & 3 & 3 & 2 & 0 & -2 & $\sim 0.51\%$ \\
\hline 
15& -4 & 4 & 2 & 2 & 2 & 1 & -1 & $\sim 0.06\%$ \\
\hline 
10& -3 & 3 & 3 & 3 & 4 & 0 & -1 & $\sim 0.16\%$ \\
\hline 
5& -2 & 3 & 2 & 2 & 4 & 1 & 0 & $\sim 1.32\%$ \\
\hline 
2& -4 & 4 & 3 & 2 & 4 & 1 & 1 & $\sim 0.08\%$ \\
\hline
\end{tabular} 
}
}
\caption{ A few sample charge assignments that give rise to SM in the IR quite often. A total of $5 \times 10^5$ trials are generated for each charge assignment. In the last column we indicate the fraction of these trials that give rise to the observed pattern of masses and mixings among the leptons, i.e., ``good" trials as defined in Sec.~\ref{subsec:tuning_def}. We do find some flexibility in the conditions from Eq.~\eqref{eq:charges_conditions} on the charge assignments; yet, by and large, these conditions give the right ballpark of the charges needed to get a non-negligible number of good trials. }
\label{tabs:charges_less_tuned}
\end{table}

For each setup, we scan over the common slepton mass $\tilde{m}$ as well as gaugino masses $M_2=2M_1$.\footnote{Our calculation can straightforwardly be repeated for other MSSM spectra with different relationships between various mass scales.} For each point in the mass plane, we generate $N_{\rm tot}=5 \times 10^5$ different trials. In the last column of Table~\ref{tabs:charges_less_tuned} we report the fraction of trials from each setup that gives rise to the right pattern of lepton masses and mixings as reviewed in Sec.~\ref{sec:setup} and in Sec.~\ref{subsec:tuning_def}. 
We find that the first and fourth row's charge assignments, which follow the relations in Eq.~\eqref{eq:charges_conditions}, give rise to a slightly better ``good" trial ratio than others. 
Yet, other charge assignments with some deviation from these relations still give rise to SM-like pattern of masses and mixings in the IR with comparable frequency. It is worth noting that getting the SM masses and mixings in the IR is itself quite rare; we only find a sub-percent fraction of the trials that give rise to the SM masses and mixings in the IR. Nonetheless, the $U(1)$-augmented models give rise to SM-like masses and mixings far more often than a completely anarchic model would, where merely getting a sufficiently small electron mass (let alone all the other masses and mixings) would require a $\sim 10^{-5}$ tuning cost.

We then calculate the eEDM and the LFV observables of Table~\ref{tabs:obslist} for each of these trials. With these numbers, we can calculate $t_{\mathbb{O} \leqslant \hat{\mathbb{O}}}$ for different values of $\hat{\mathbb{O}}$ for different observables. This, combined with the experimental bounds from Table~\ref{tabs:obslist}, allows us to find the bounds on the slepton-gaugino mass plane for different fine tunings.

\begin{figure}
\centering
\resizebox{0.8\columnwidth}{!}{
	\includegraphics[scale=1]{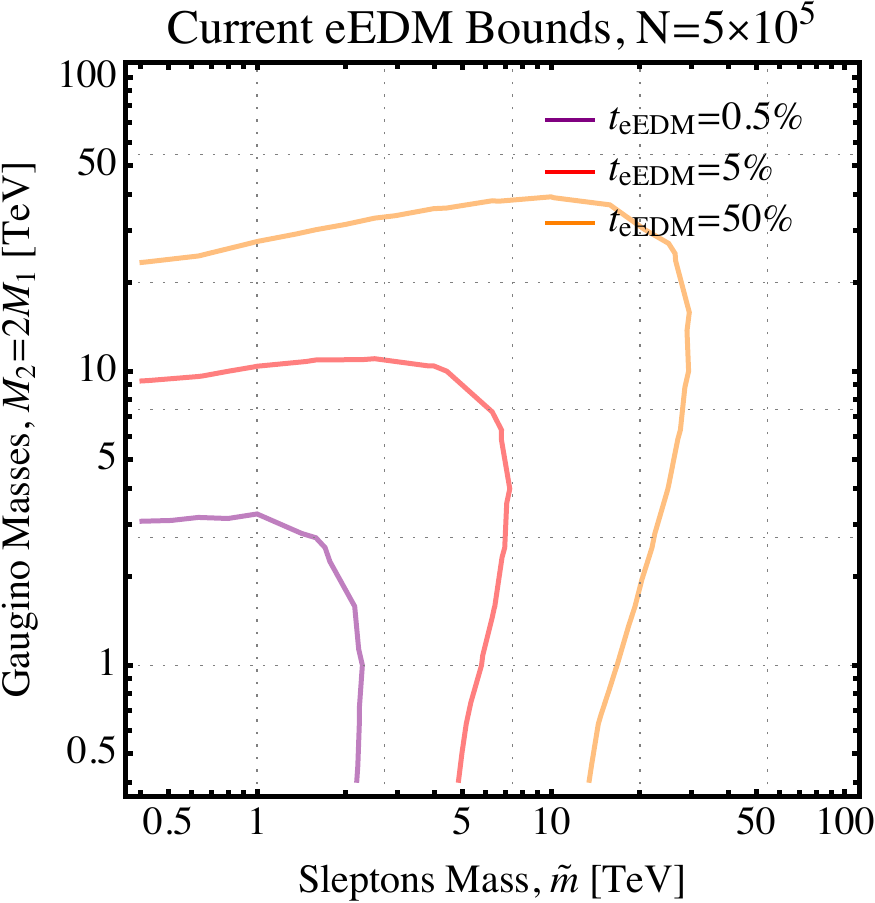}
	\hspace{1.5cm}
	\includegraphics[scale=1]{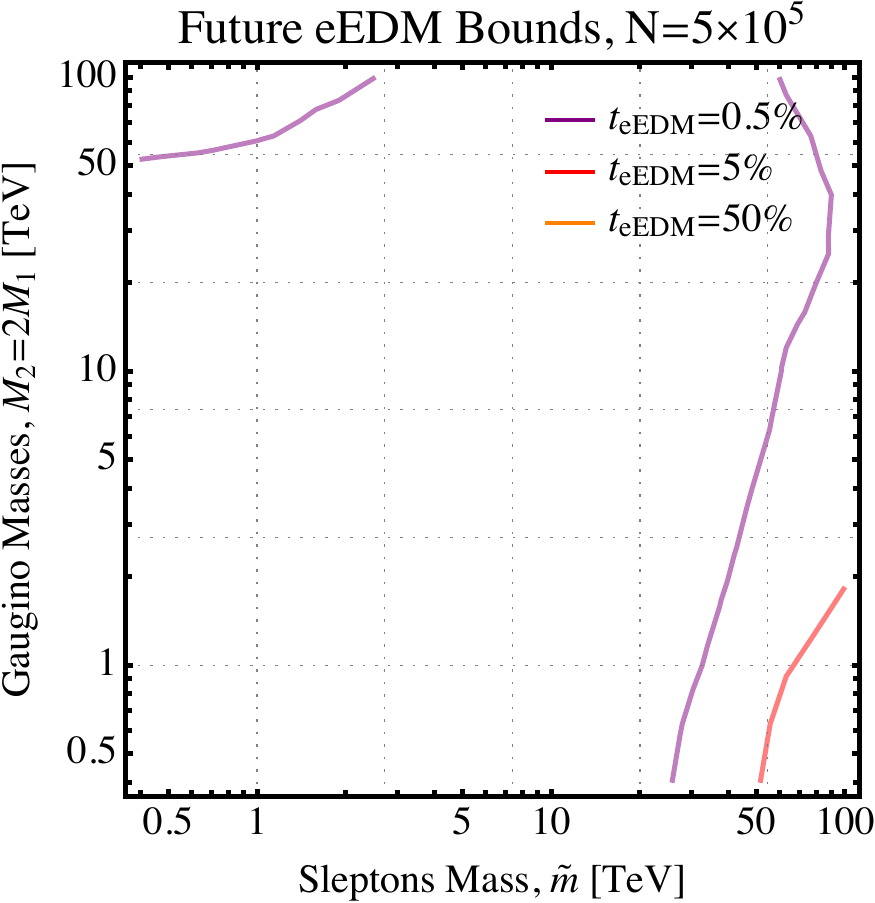}
}\\
\vspace{0.45cm}	
\resizebox{0.8\columnwidth}{!}{
	\includegraphics[scale=1]{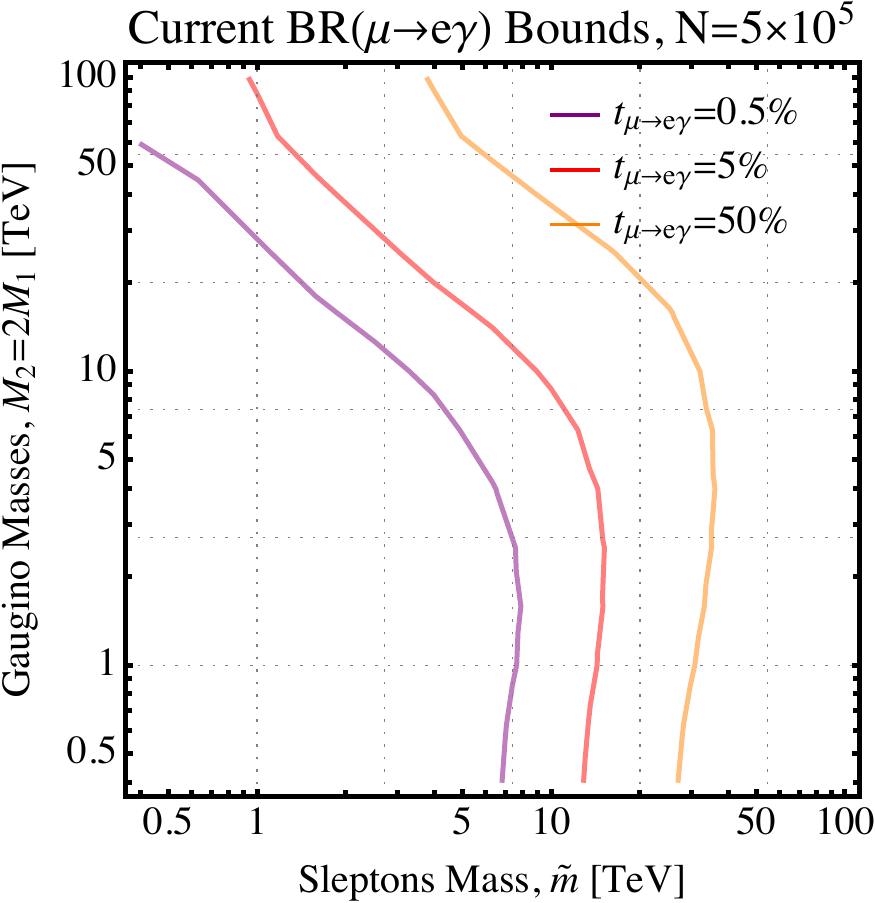}
	\hspace{1.5cm}
	\includegraphics[scale=1]{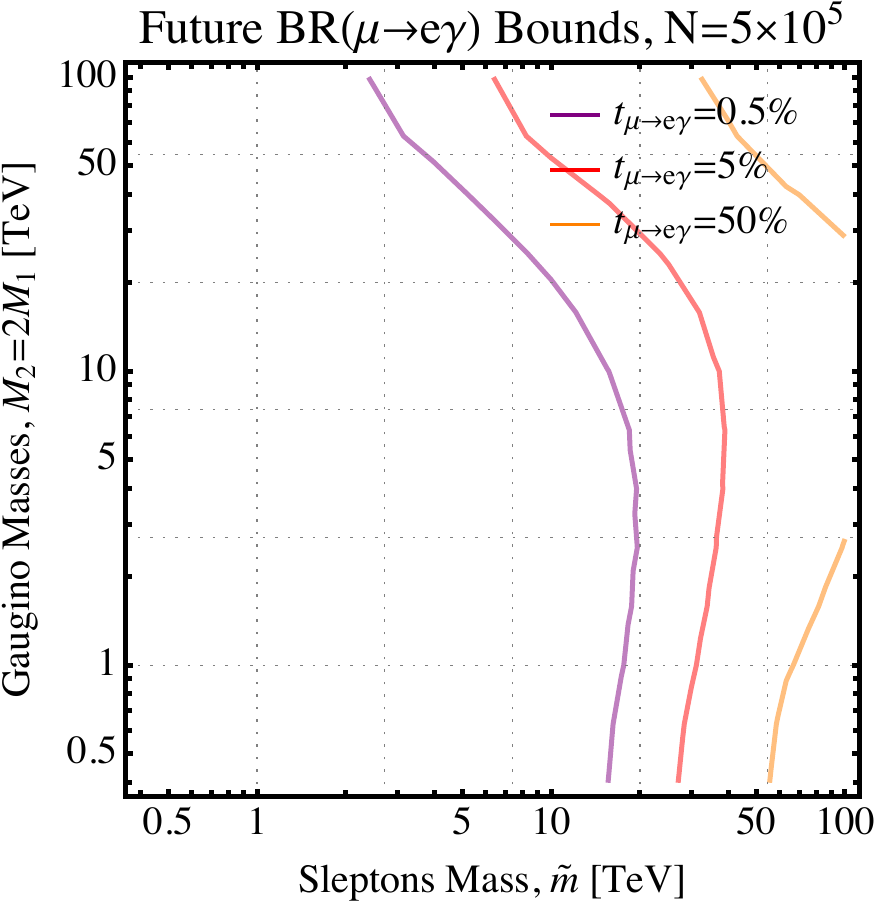}
}\\
\vspace{0.45cm}
\resizebox{0.8\columnwidth}{!}{
	\includegraphics[scale=1]{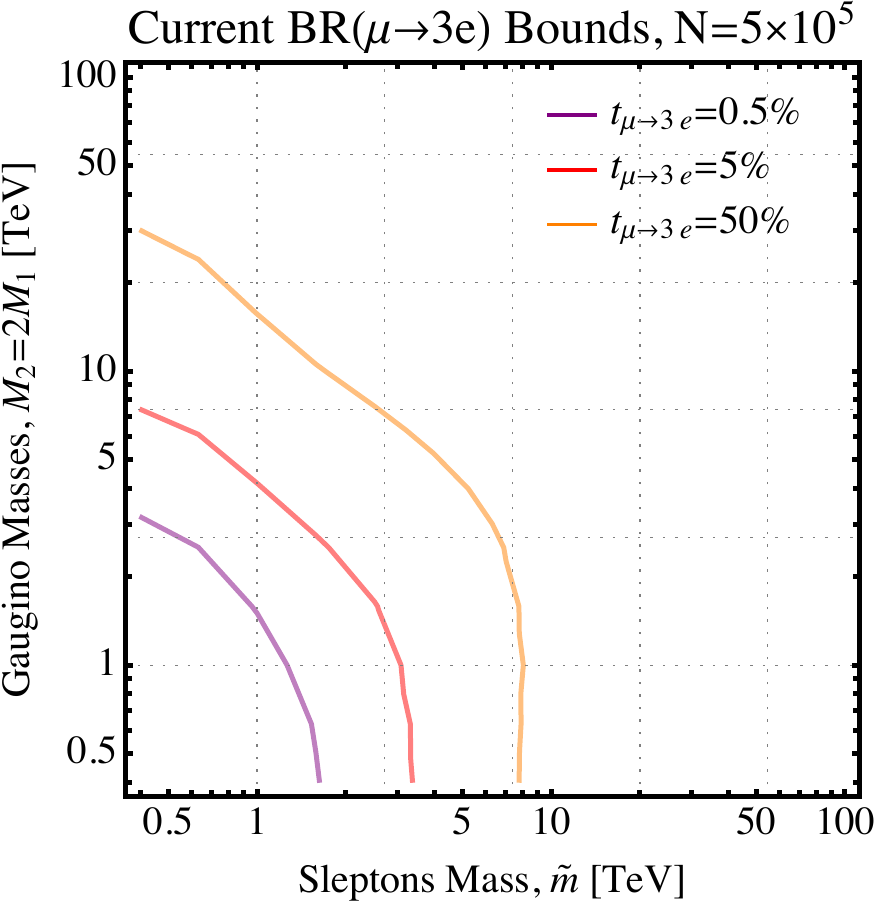}
	\hspace{1.5cm}
	\includegraphics[scale=1]{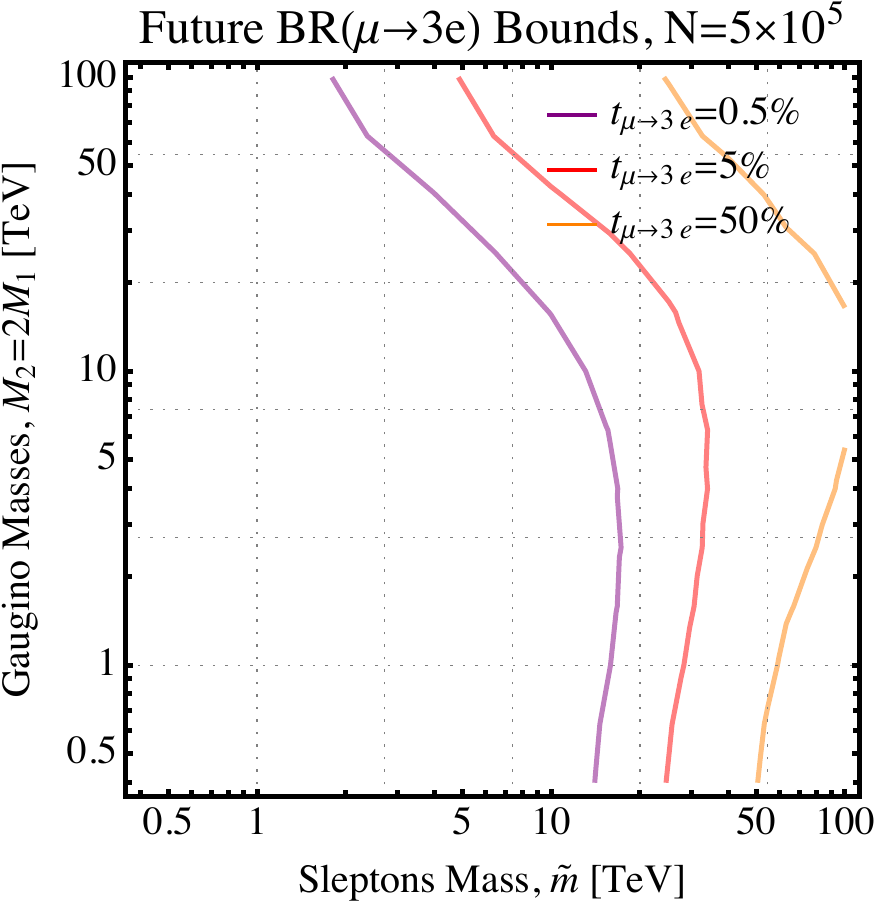}
}\\
\caption{The bounds on the charge assignment $[S_2]=-3$ and $\{[L_1],~[L_2],~[L_3],~[\bar{e}_1],~[\bar{e}_2],~[\bar{e}_3] \} = \{4,~ 3,~3,~2,~0,~-2\}$ (the first row of Table~\ref{tabs:obslist}) from eEDM (\textbf{top}), $\BRmutoe$ (\textbf{middle}), and $\BRmutoeee$ (\textbf{bottom}), with different amount of tunings. A total of $N=5 \times 10^5$ trials are generated and only the good trials are used in calculating the bounds. The \textbf{left (right) column} shows the current (future) bounds from each observable. We find that for this model to be natural, i.e., tuning of $\sim 50\%$ for each observable, the superpartners should be heavier than $\sim 10-50$~TeV; around percent-level tuning for each observable is required to realize superpartners of mass $\mathcal{O}(1)$~TeV.
}
\label{fig:mass-plane-bound-25}
\end{figure}

\begin{figure}
\centering
\resizebox{0.8\columnwidth}{!}{
	\includegraphics[scale=1]{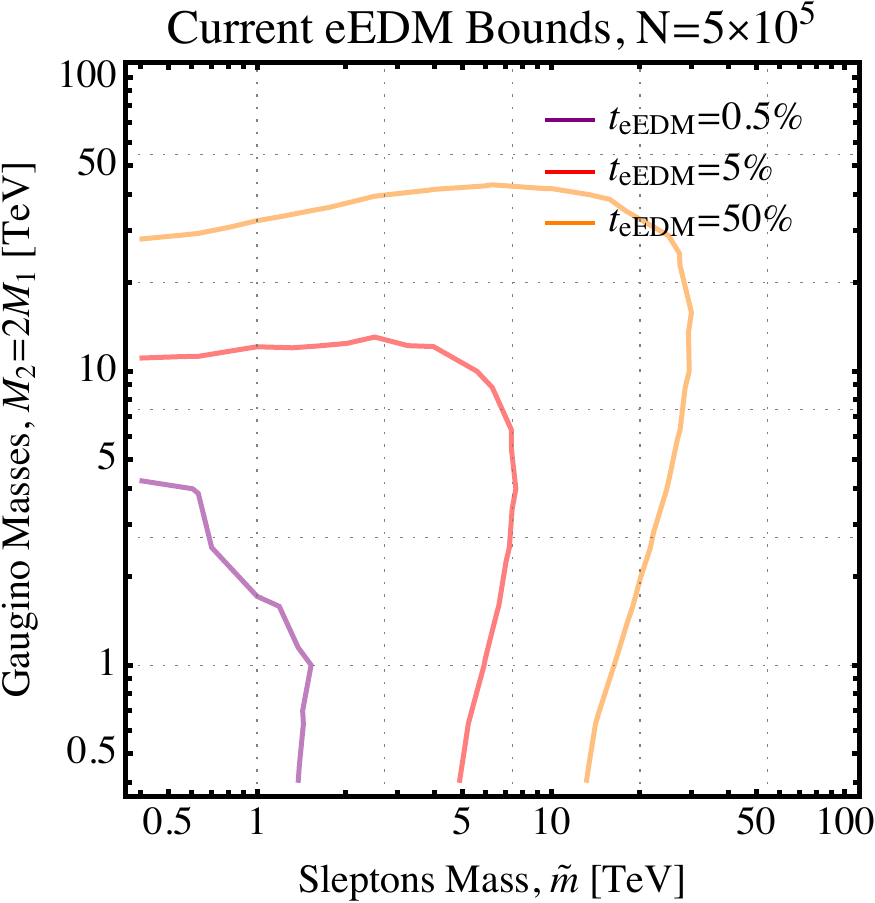}
	\hspace{1.5cm}
	\includegraphics[scale=1]{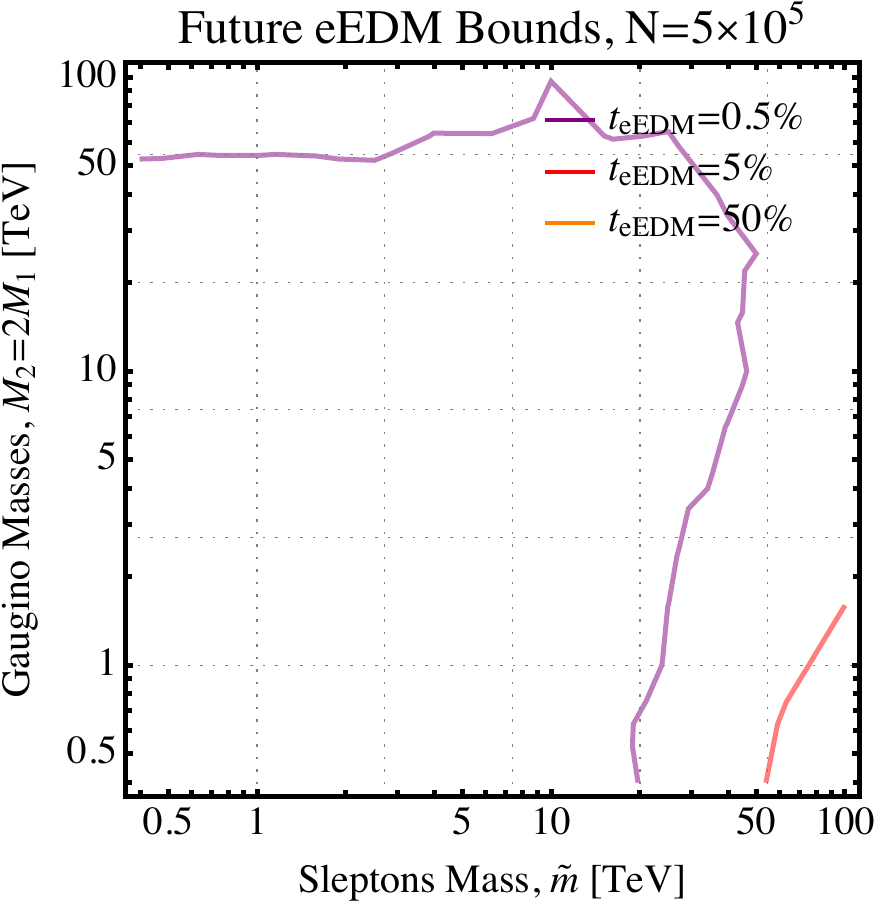}
}\\
\vspace{0.45cm}
\resizebox{0.8\columnwidth}{!}{
	\includegraphics[scale=1]{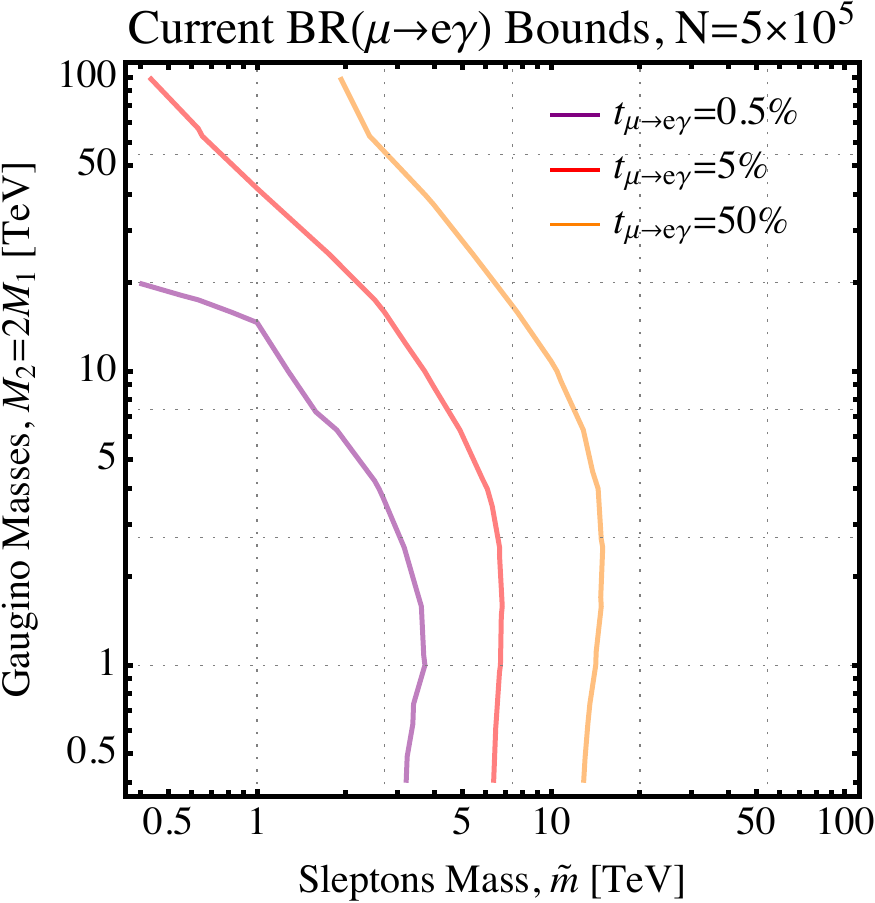}
	\hspace{1.5cm}
	\includegraphics[scale=1]{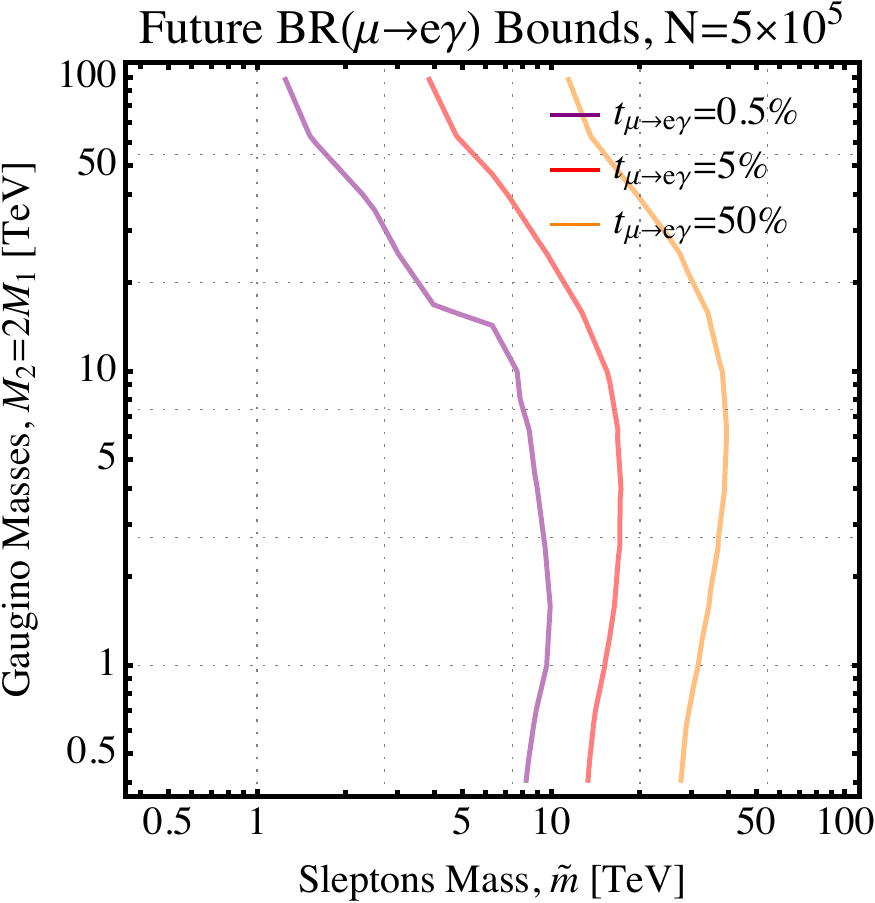}
}\\
\vspace{0.45cm}
\vspace{0.5cm}
\resizebox{0.8\columnwidth}{!}{
	\includegraphics[scale=1]{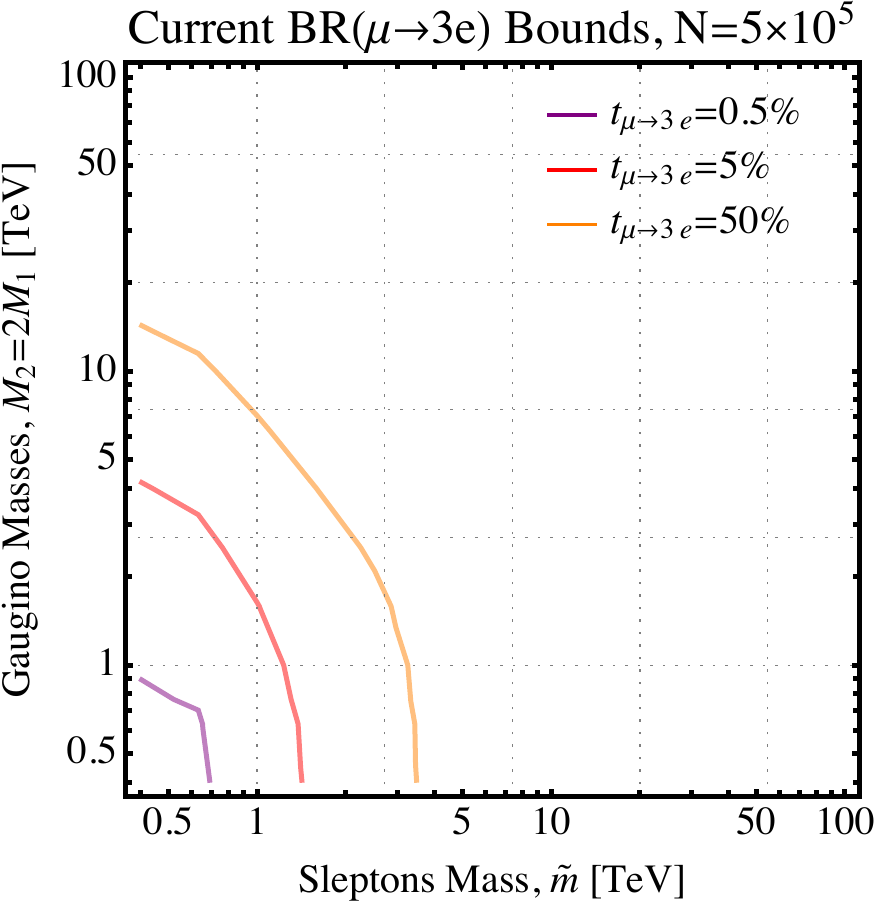}
	\hspace{1.5cm}
	\includegraphics[scale=1]{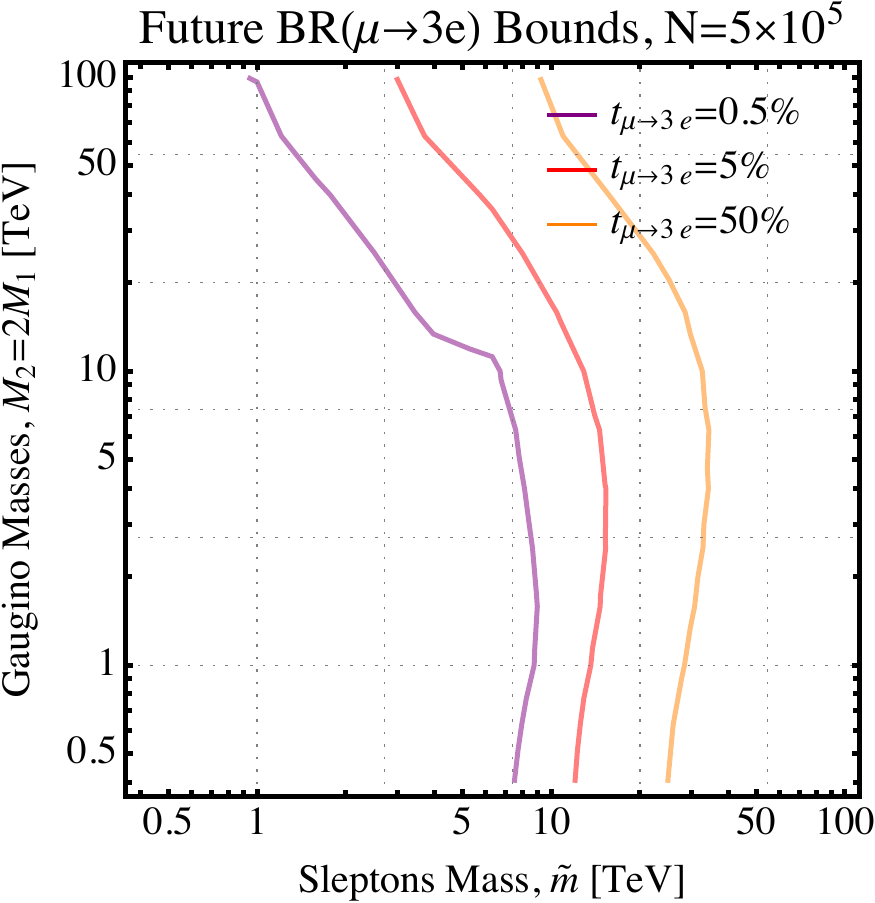}
}
\caption{Similar to Fig.~\ref{fig:mass-plane-bound-25} but for the charge assignment $[S_2]=-4$ and $\{ [L_1],~[L_2],~[L_3],~[\bar{e}_1],~[\bar{e}_2],~[\bar{e}_3] \} = \{4,~ 2,~2,~2,~1,~-1\}$ (the second row of Table~\ref{tabs:obslist}).
}
\label{fig:mass-plane-bound-15}
\end{figure}

\begin{figure}
\centering
\resizebox{0.8\columnwidth}{!}{
	\includegraphics[scale=1]{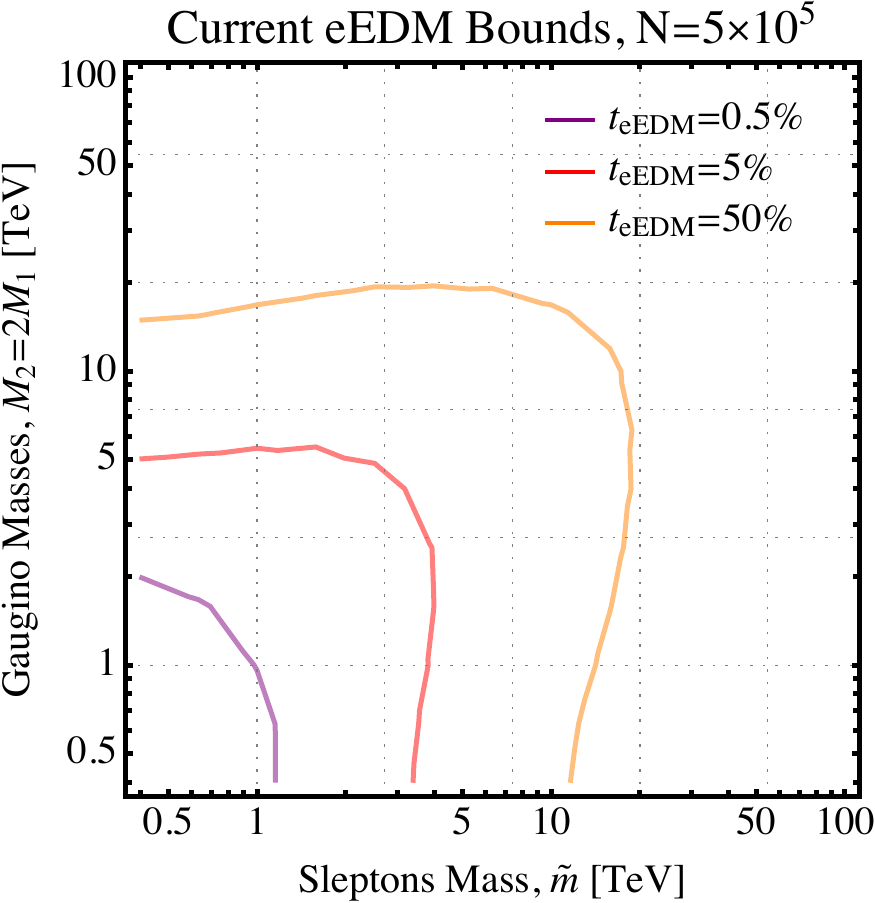}
	\hspace{1.5cm}
	\includegraphics[scale=1]{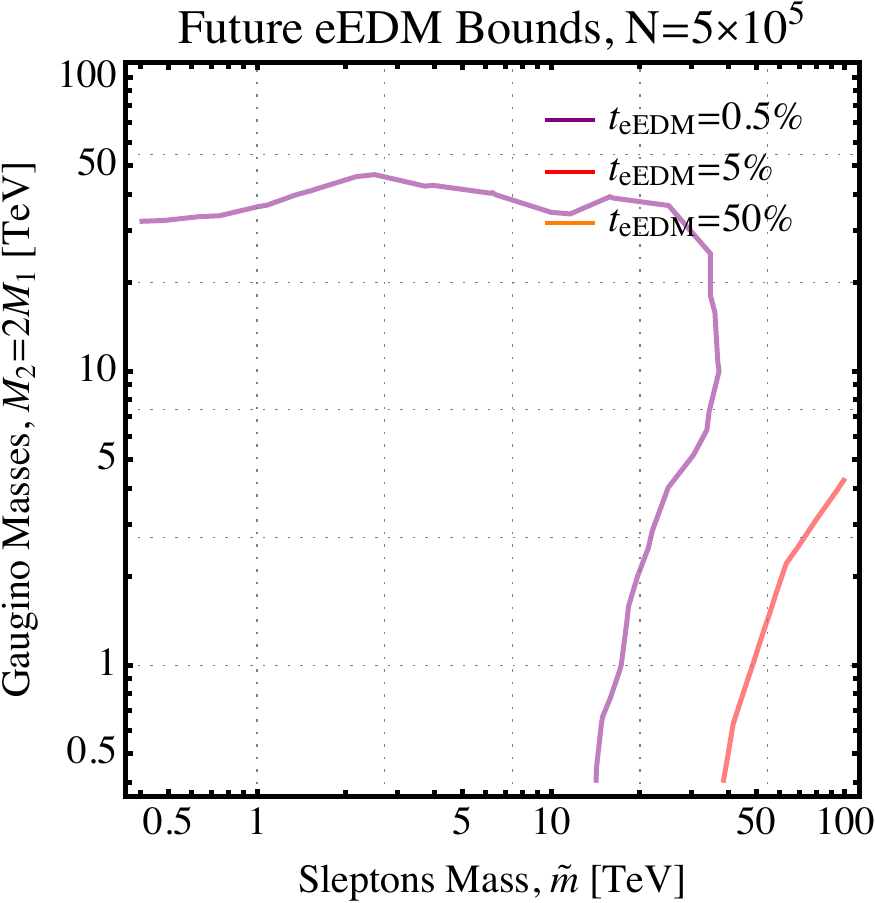}
}\\
\vspace{0.45cm}
\resizebox{0.8\columnwidth}{!}{
	\includegraphics[scale=1]{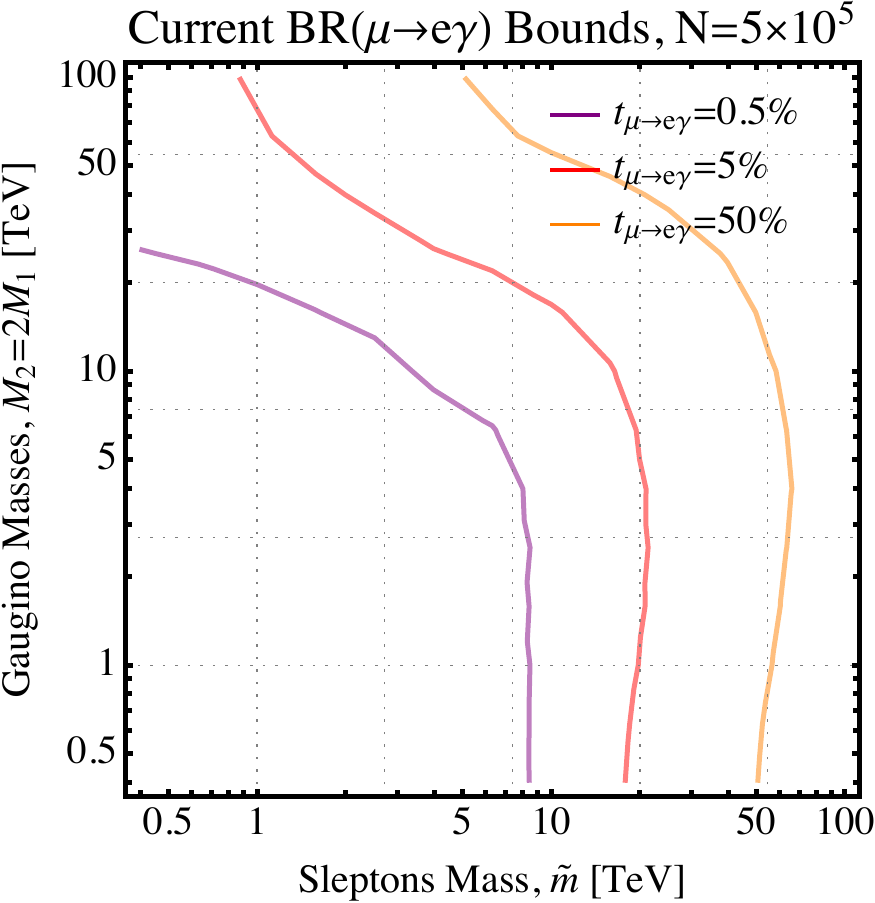}
	\hspace{1.5cm}	
	\includegraphics[scale=1]{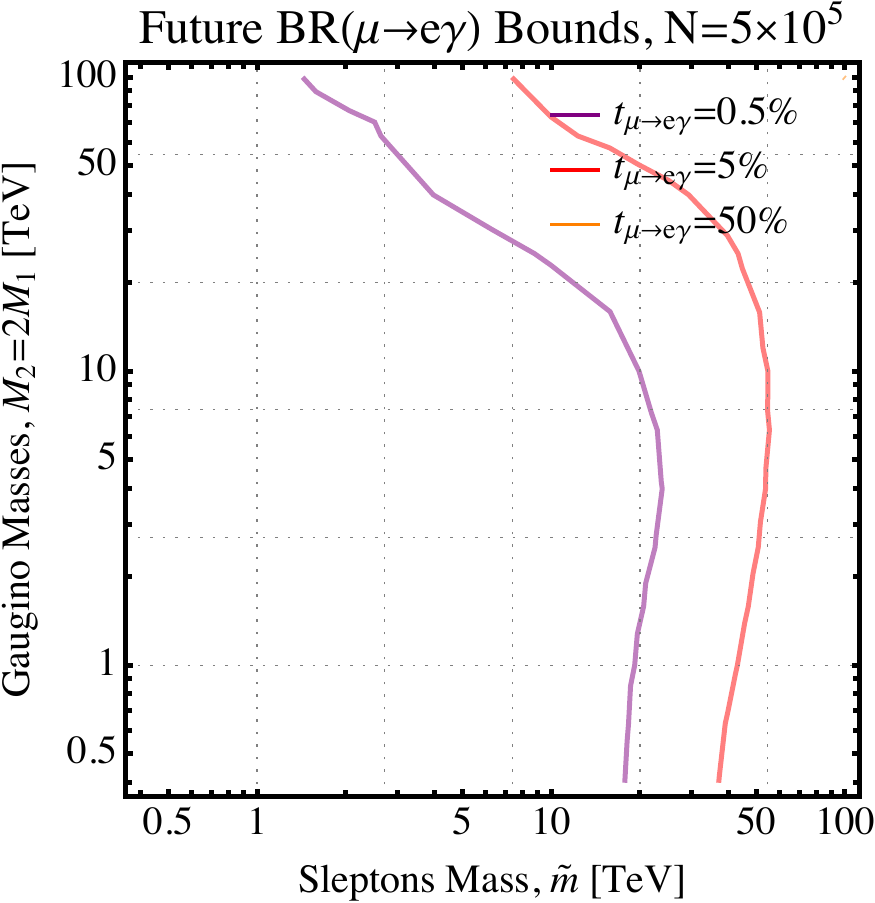}
}\\
\vspace{0.45cm}
\resizebox{0.8\columnwidth}{!}{
	\includegraphics[scale=1]{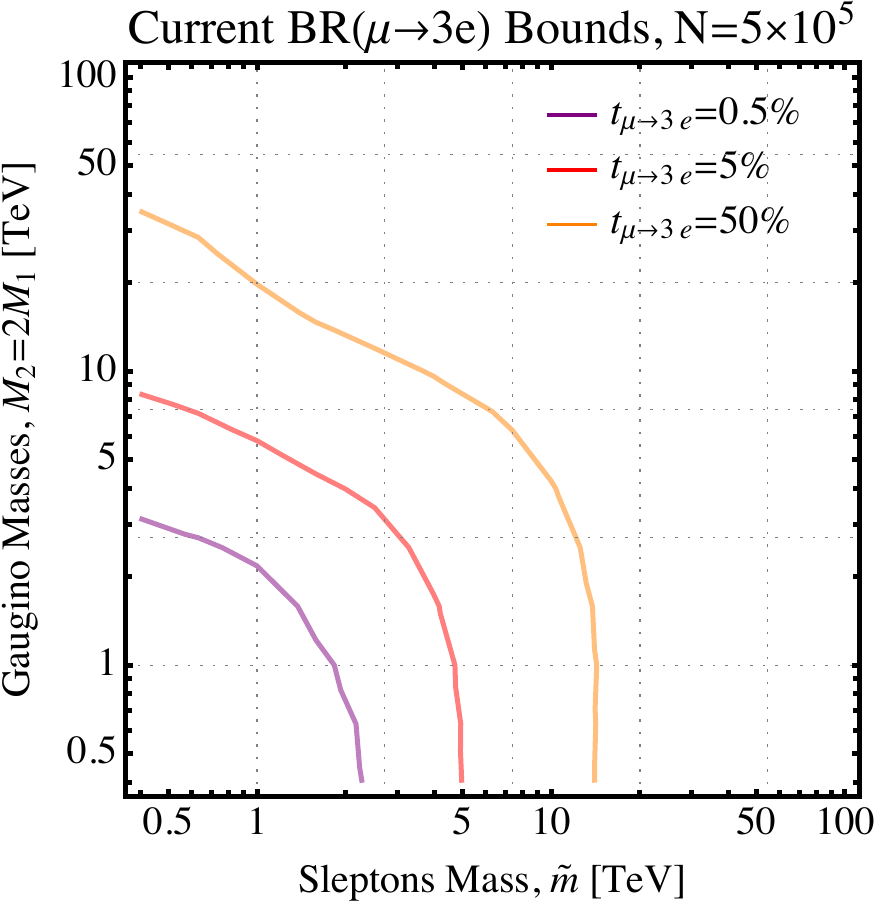}
	\hspace{1.5cm}
	\includegraphics[scale=1]{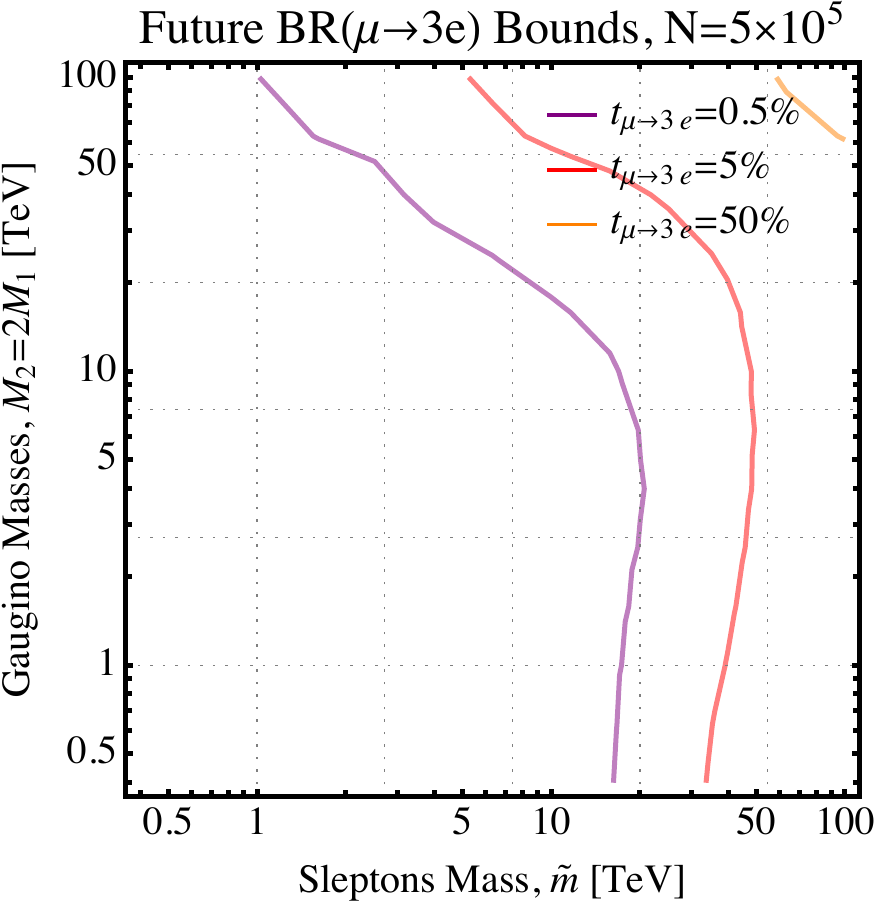}
}
\caption{Similar to Fig.~\ref{fig:mass-plane-bound-25} but for the charge assignment $[S_2]=-3$ and $\{ [L_1],~[L_2],~[L_3],~[\bar{e}_1],~[\bar{e}_2],~[\bar{e}_3] \} = \{3,~ 3,~3,~4,~0,~-1\}$ (the third row of Table~\ref{tabs:obslist}).
}
\label{fig:mass-plane-bound-10}
\end{figure}

\begin{figure}
\centering
\resizebox{0.8\columnwidth}{!}{
	\includegraphics[scale=1]{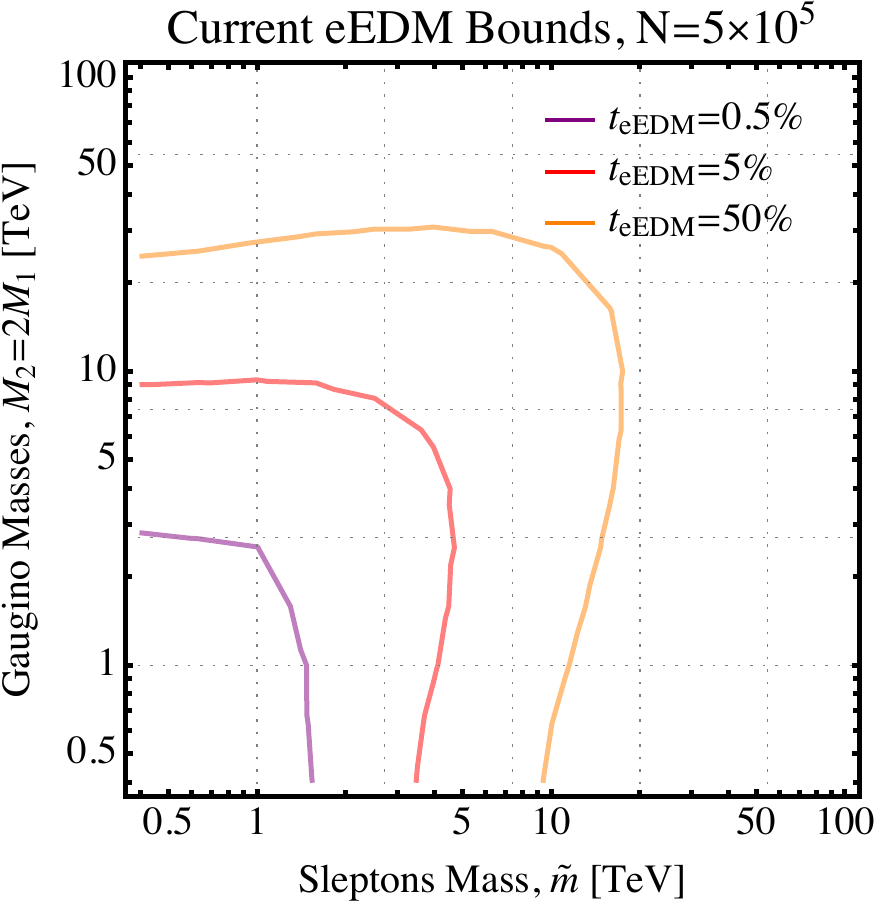}
	\hspace{1.5cm}
	\includegraphics[scale=1]{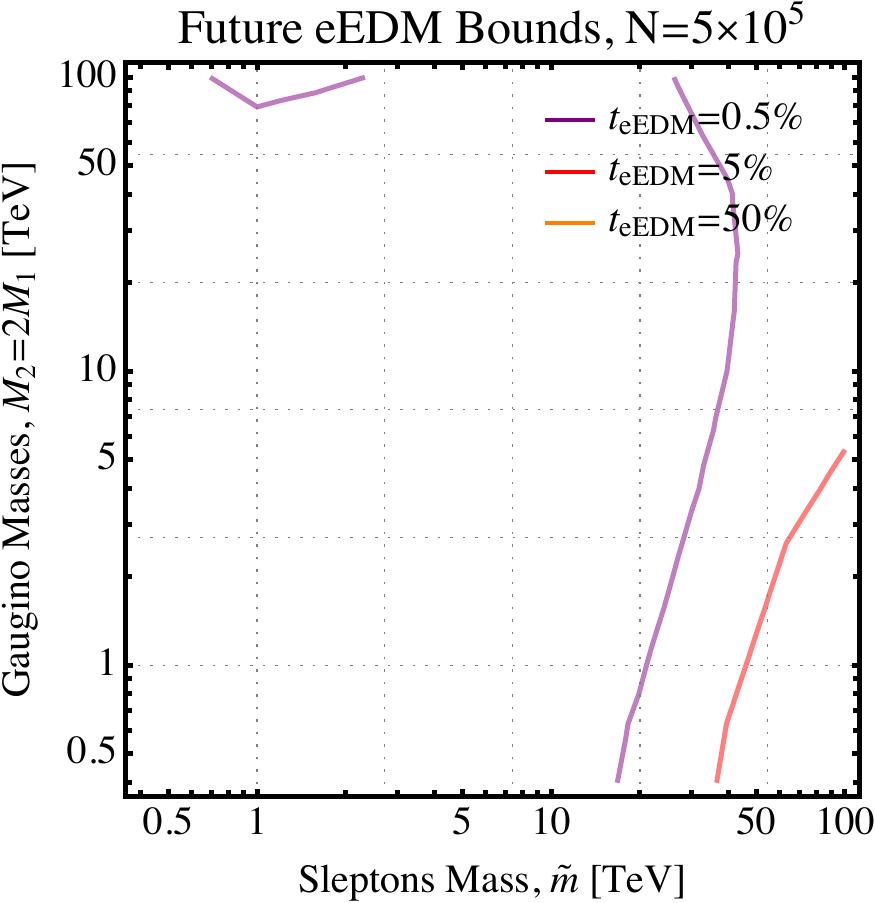}
}\\
\vspace{0.45cm}
\resizebox{0.8\columnwidth}{!}{
	\includegraphics[scale=1]{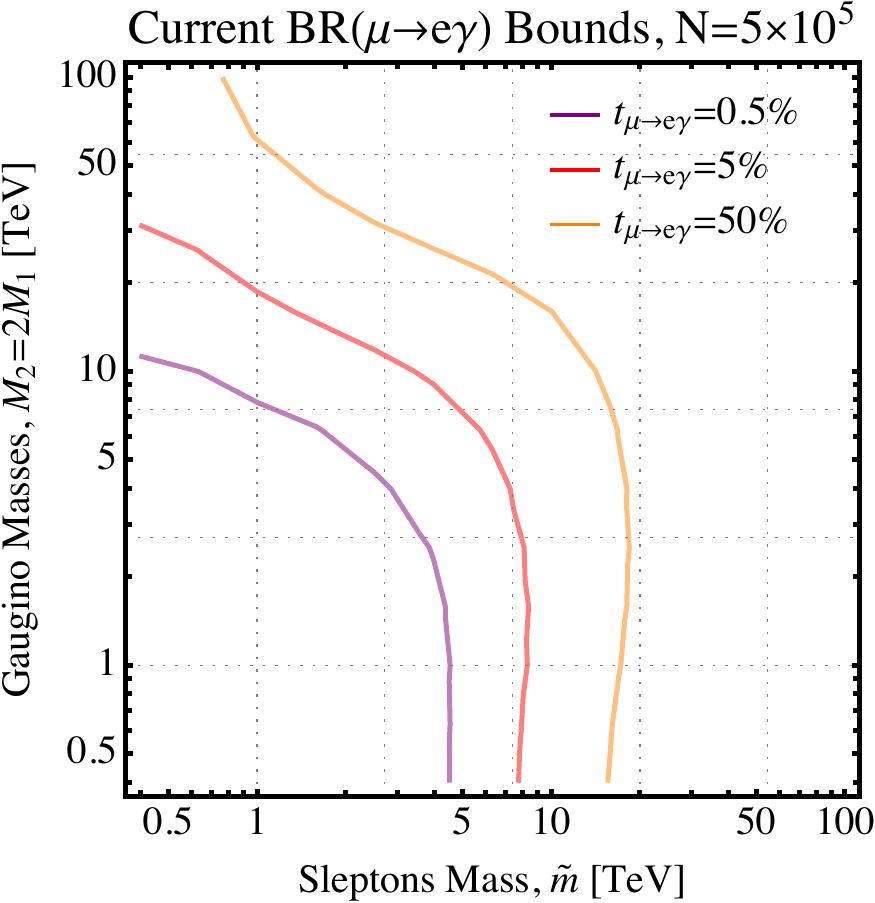}
	\hspace{1.5cm}	
	\includegraphics[scale=1]{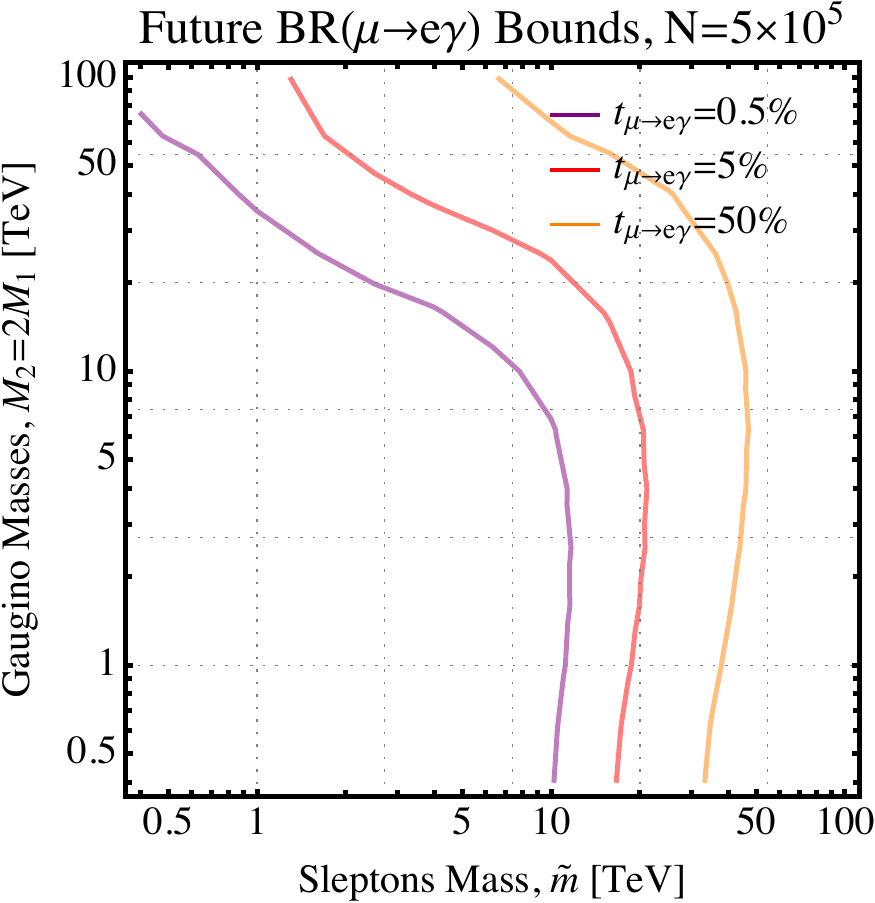}
}\\
\vspace{0.45cm}
\resizebox{0.8\columnwidth}{!}{
	\includegraphics[scale=1]{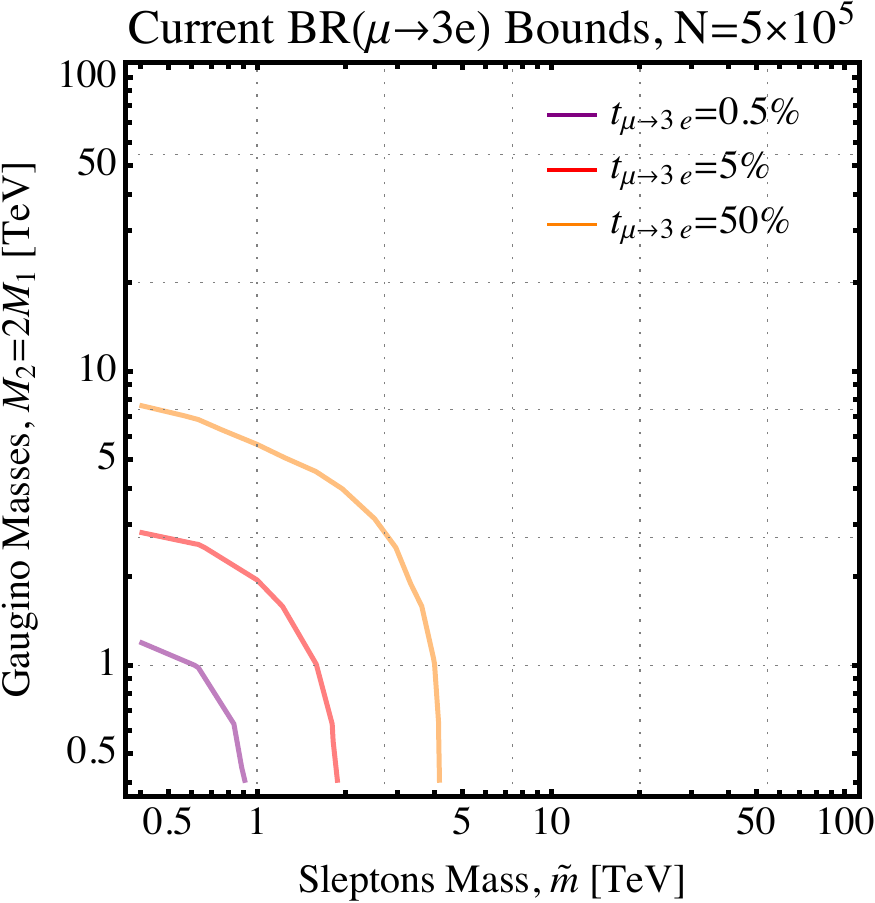}
	\hspace{1.5cm}
	\includegraphics[scale=1]{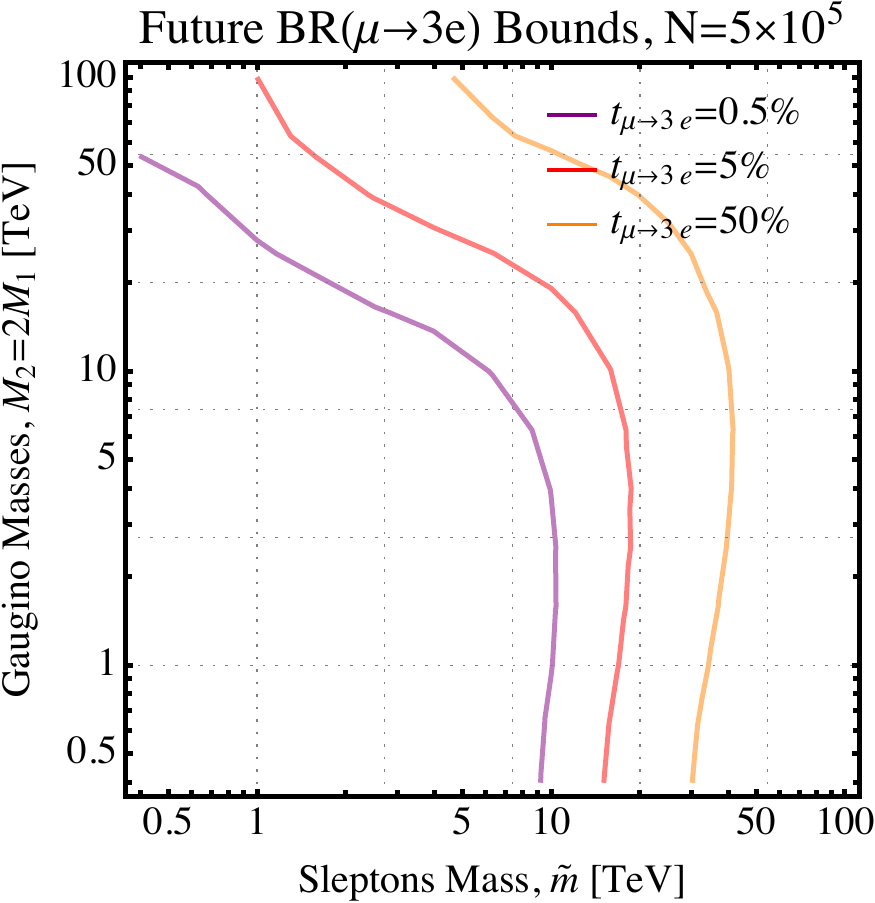}
}
\caption{Similar to Fig.~\ref{fig:mass-plane-bound-25} but for the charge assignment $[S_2]=-2$ and $\{ [L_1],~[L_2],~[L_3],~[\bar{e}_1],~[\bar{e}_2],~[\bar{e}_3] \} = \{3,~ 2,~2,~4,~1,~0\}$ (the fourth row of Table~\ref{tabs:obslist}).
}
\label{fig:mass-plane-bound-5}
\end{figure}

\begin{figure}
\centering
\resizebox{0.8\columnwidth}{!}{
	\includegraphics[scale=1]{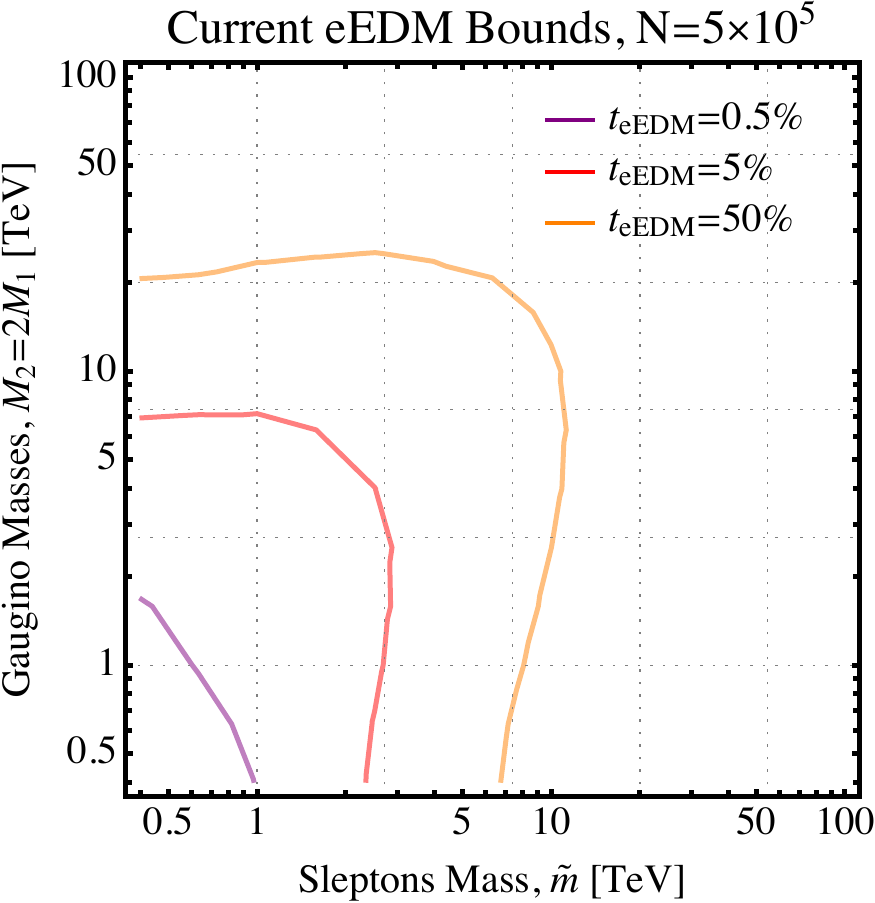}
	\hspace{1.5cm}
	\includegraphics[scale=1]{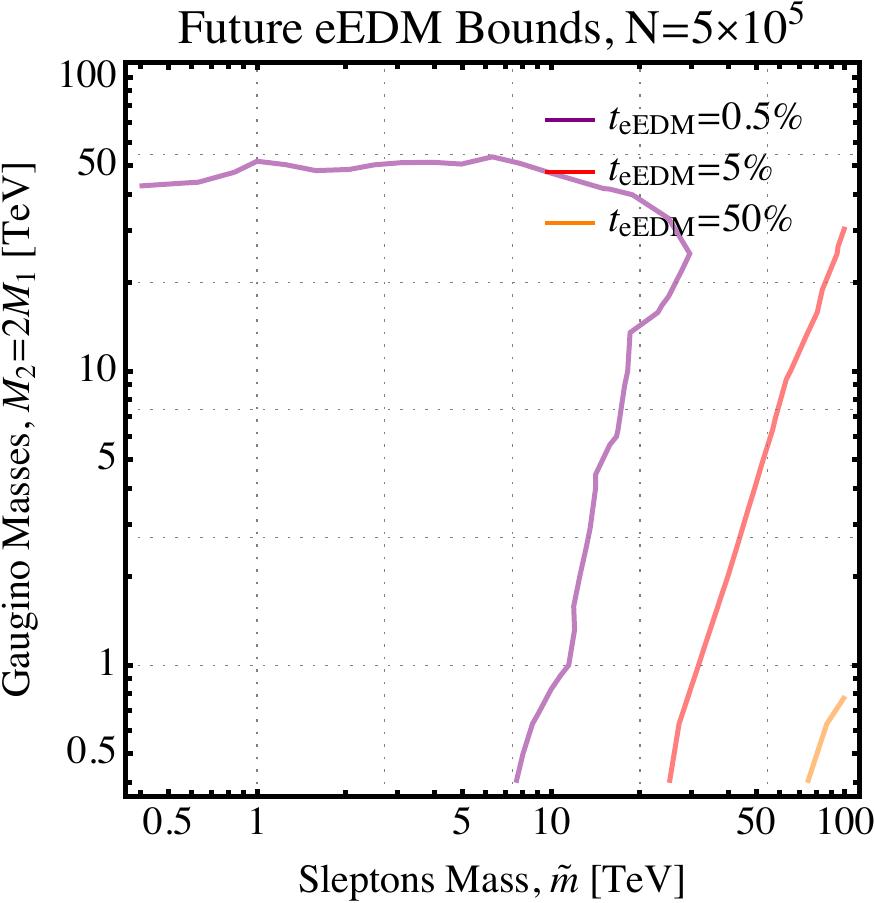}
}\\
\vspace{0.45cm}
\resizebox{0.8\columnwidth}{!}{
	\includegraphics[scale=1]{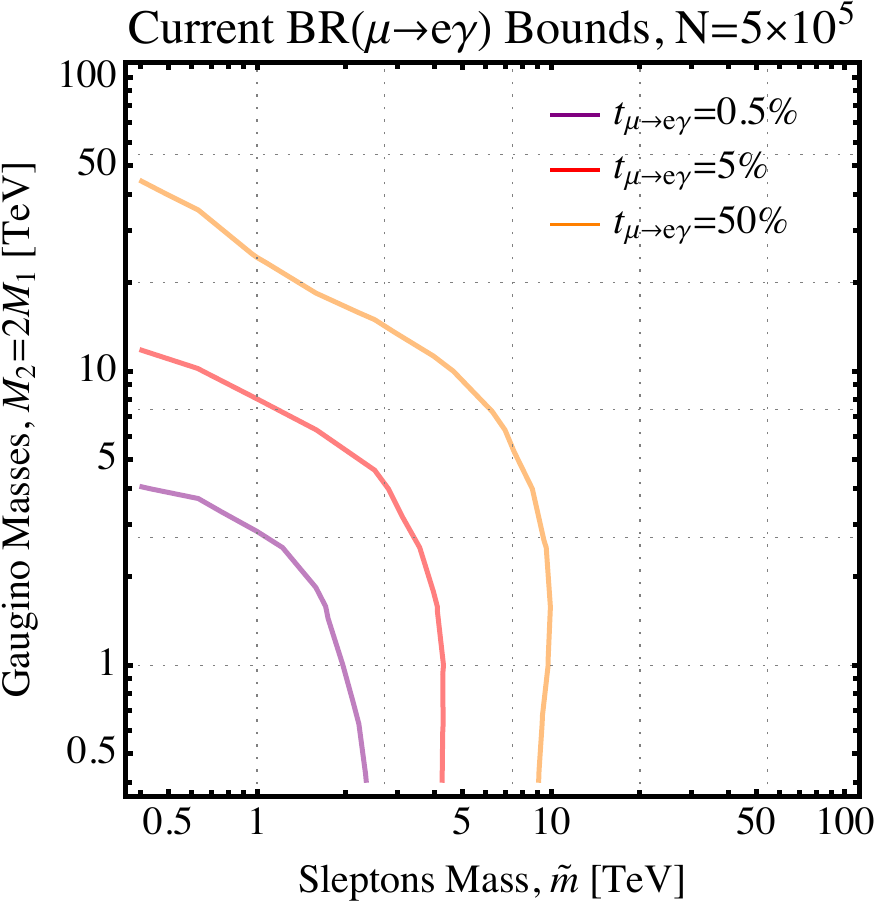}
	\hspace{1.5cm}	
	\includegraphics[scale=1]{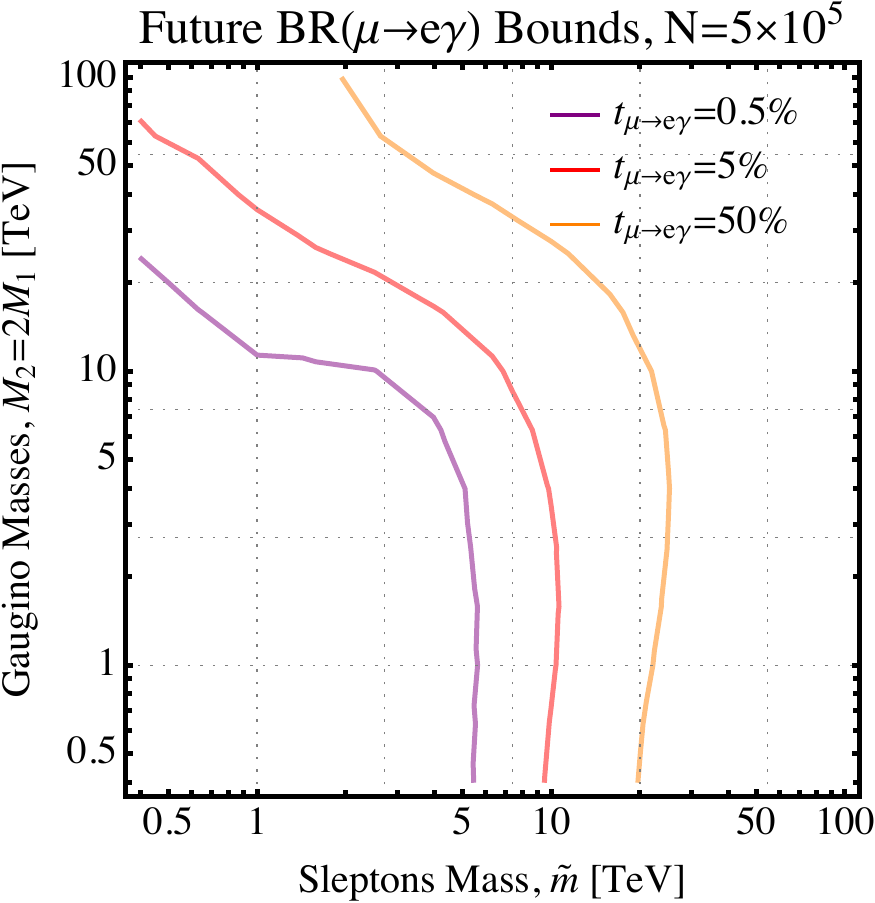}
}\\
\vspace{0.45cm}
\resizebox{0.8\columnwidth}{!}{
	\includegraphics[scale=1]{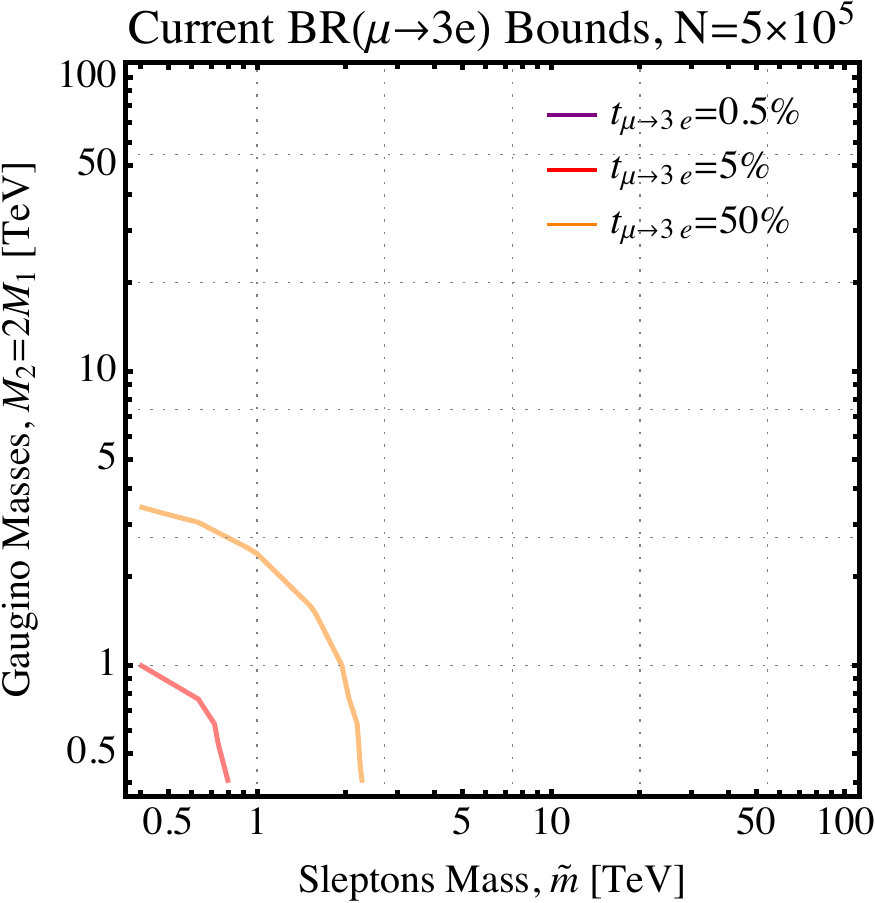}
	\hspace{1.5cm}
	\includegraphics[scale=1]{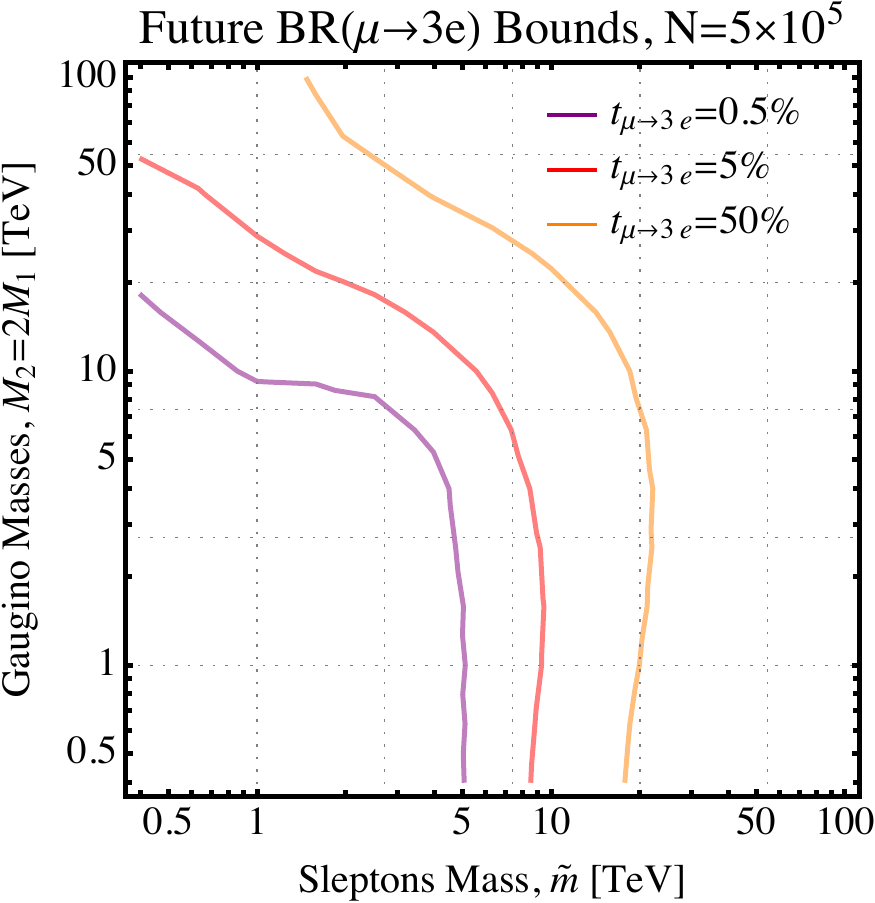}
}
\caption{Similar to Fig.~\ref{fig:mass-plane-bound-25} but for the charge assignment $[S_2]=-4$ and $\{ [L_1],~[L_2],~[L_3],~[\bar{e}_1],~[\bar{e}_2],~[\bar{e}_3] \} = \{4,~ 3,~2,~4,~1,~1\}$ (the fifth row of Table~\ref{tabs:obslist}). 
}
\label{fig:mass-plane-bound-2}
\end{figure}

We repeat this calculation for the five charge assignments in Table~\ref{tabs:charges_less_tuned} and for different SUSY parameters $\tilde{m}$ and $M_2=2M_1$. The results are shown in Figs.~\ref{fig:mass-plane-bound-25}-\ref{fig:mass-plane-bound-2}.\footnote{It should be noted that the x-axis in these plots is the shared slepton scale $\tilde{m}$. The exact slepton mass eigenvalues are a factor of $\mathcal{O}(1)$ times this scale.} In the top (bottom) row of each figure we show the current (future) bounds on the model; see Table~\ref{tabs:obslist}. 
We report the bounds on the models with different values of tuning allowed. 
The conclusions derived about the model from Figs.~\ref{fig:mass-plane-bound-25}-\ref{fig:mass-plane-bound-2} are more or less similar, so we discuss the main takeways only from Fig.~\ref{fig:mass-plane-bound-25}; the discussion for other models would be closely analogous.

In the upper panels of Fig.~\ref{fig:mass-plane-bound-25}, we find that if we want to avoid any significant tuning in the model, i.e., $t \approx 50\%$, and if all the particles are comparable in mass, the current bounds on sleptons and charginos/neutralinos from different observables are in the 10 -- 50 TeV ballpark. 
For TeV scale superpartners to avoid the current bound from each observable of Table~\ref{tabs:obslist}, a sub-percent-level tuning is required in this setup, and an even higher level of tunning is needed in order  to push the superpartners masses all the way down to the current bounds from the LHC \cite{Sirunyan:2020eab,ATL-PHYS-PUB-2021-007}. 
Note however that we present the average slepton mass scale, from which the mass eigenstates might differ by an $\mathcal{O}(1)$ factor. 
Without any correlation, the values of tuning are multiplied
over all the observables of Table~\ref{tabs:obslist} to find the total degree of tuning required in the model with a given mass scale.

As discussed in Sec.~\ref{subsec:model}, the assumption of $c_2=n_2$ for the $S_2$ flavon charge and its spurion's $\lambda$ scaling guarantees that we have the maximal CPV possible in our setup. Even so, we still find that, with the current experimental results summarized in Table~\ref{tabs:obslist}, $\BRmutoe$ gives rise to slightly more stringent bounds on the parameter space than the eEDM, while the current $\BRmutoeee$ bounds are comparable to the eEDM bounds for this charge assignment.
With the projected bounds of Table~\ref{tabs:obslist}, we find that the bounds from all these observables are comparable.

Similar deductions are made for the other charge assignments studied in Figs.~\ref{fig:mass-plane-bound-15}-\ref{fig:mass-plane-bound-2}. While the bounds on superpartners' masses can change by $\mathcal{O}(1)$ numbers across different charge assignments, the final conclusions are similar for different charge assignments studied: that different models' superpartners are currently ruled out all the way up to $\mathcal{O}(10)$~TeV masses if the model is completely natural, and to realize $\mathcal{O}(1)$~TeV scale (or even lighter) superpartners requires fine-tuning.\footnote{It is worth reiterating  that we explicitly checked that changing the definition of ``good" trials or the range of random $\mathcal{O}(1)$ numbers merely gives rise to $\mathcal{O}(1)$ change in the tuning value for different contours in Fig.~\ref{fig:mass-plane-bound-25}-\ref{fig:mass-plane-bound-2}. } 
The similarity of the results emphasizes that the observed patterns are a general feature of horizontal symmetries and not of a unique charge assignment.
It raises the question of how one can study a more complete set of good charges.
Our tuning measure allows one to study different charge assignments and find explicit bounds on its parameter space from these flavor observables (as well as other observables neglected in this study). 
An exhaustive study of charge assignments might uncover a unique preferred set of charge assignments, as well as exposing universal features of horizontal symmetries. 

We should also reemphasize that a similar calculation can be done for any model with undetermined UV parameters and for other observables. In particular, it will be interesting to repeat this analysis for $U(1)$-augmented models (either supersymmetric or not) in the quark sector. We leave such a study for a future work.


In calculating the tuning in this section we studied one observable at a time and considered the fraction of the UV parameter space that suppresses that observable. Instead, we could look at the subspace that simultaneously gives rise to a small enough value for different observables, which enables us to take account of correlations between different observables. We will leave a study of such correlations for a future work.

\section{Conclusion}\label{sec:conclusion}

The MSSM remains an extremely appealing candidate for a UV completion of the SM. The minimal version of the theory predicts disastrous contributions of superpartners to various LFV observables and the eEDM. In this work we focused on a $U(1)$-augmented MSSM, and carried out a complete calculation of the contribution to these observables from different sleptons and charginos or neutralinos. We showed that the combination of a horizontal symmetry and spontaneous CPV can suppress the contributions to the full set of leptonic observables.

The existence of undetermined $\mathcal{O}(1)$ numbers in mass matrix models always muddles the interpretation of the various experimental bounds on these models. To properly interpret the bounds from various observables on such models, we developed an intuitive notion of naturalness. We treat the undetermined UV coefficients as random parameters in the mass matrices and define the tuning measure as the fractional volume of the space of all these possible coefficients that satisfy a given condition on an observable, while giving rise to the SM patterns of masses and mixings in the IR. 
In particular, for any flavor observable we can use the condition that the model gives rise to a suppressed contribution to that observable. 
This provides us with a measure of how natural various mass matrix extensions of the SM are, considering the existing experimental results on any specific observable. We applied this naturalness notion to our study of the $U(1)$-augmented MSSM.

In all the charge assignments we studied, for a fixed value of tuning, the current bounds from LFV observables were stronger than those from the eEDM.
We found that if we want to avoid any significant tuning in the model, the current bounds on sleptons and charginos/neutralinos from different observables are in the 10 -- 50 TeV ballpark.
It is conceivable that by introducing more structures in the UV model, the TeV scale superpartners can become viable with mild tuning. Furthermore, the supersymmetric setups studied in this work also contribute to the muon $g-2$ via the $c^R_{22}$ Wilson coefficient; this motivates a future repeat of our tuning analysis with muon $g-2$ included in the list of observables.

We studied a handful of sample charge assignments, in contrast to the common practice in the literature of studying a single charge assignment; nevertheless, there are many more charge assignments that can give rise to SM-like theories in IR with comparable frequency. We emphasize that the development of novel methods is called for, in order to efficiently study all such charge assignments.

We can repeat our analysis for different charge assignments or, more generally, for any mass matrix model. Our naturalness notion enables a new rigorous way for interpreting the bounds from various flavor physics observables on such models in general. 
It also puts various flavor physics observables on the same footing as the higgs mass and allows us to compare the naturalness of different flavorful BSM models. While we only focused on a handful of lepton flavor observables, our analysis can be repeated for the quark sector observables in any mass matrix model as well.

\section*{Acknowledgments}

We would like to thank Jason Evans for discussions. The work of DA is supported by DOE grant DE-SC0015845. The work of PA was supported by the DOE Grant Number DE-SC0012567 and the MIT Department of Physics. MR is supported in part by the DOE Grant DE-SC0013607 and the Alfred P.~Sloan Foundation Grant No.~G-2019-12504.

\appendix

\section{Generic UV Contribution to the Dipole Operator and Lepton Flavor Observables}
\label{app:dipole-calc}

\begin{figure}
    \centering
    \resizebox{0.5\columnwidth}{!}{
    \includegraphics{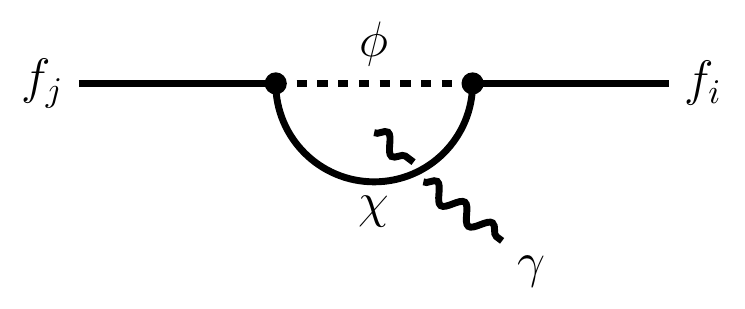}
    }
    \caption{The diagram generating the dipole operator of Eq.~\eqref{eq:L-dipole} from a generic UV theory with new heavy scalars ($\phi$) and fermions ($\chi$).  }
    \label{fig:dipole-diag-appx}
\end{figure}

The dipole operator in Eq.~\eqref{eq:L-dipole} can be generated by any generic flavorful BSM model. Let us consider a general Yukawa theory in the UV
\begin{equation}
    \mathcal{L} \supset y^R_{\phi \chi i} \phi \bar{\chi} P_R f_i + y^L_{\phi \chi i} \phi \bar{\chi} P_L f_i + \mathrm{h.c.},
    \label{eq:UV-yukawa}
\end{equation}
where $\phi$ and $\chi$ are, respectively, new heavy scalars and fermions, and $f$ denotes SM fermions. Such a theory can generate the dipole operator of Eq.~\eqref{eq:L-dipole} through the diagram in Fig.~\ref{fig:dipole-diag-appx}. 

After integrating out the new heavy particles and matching the UV diagram onto the dipole operator, we find 
\begin{eqnarray}
\label{eq:cR-UV-yukawa}
c^R_{ij} = \sum_{\phi \chi}  \frac{e}{64 \pi^2 m_\phi^2} \!\!\!\!\!\!\!\!\! && \left[ 	m_\chi y_{\phi \chi i}^{L *} y_{\phi\chi j}^{ R } \left( Q_\phi A(r) - Q_\chi B(r)	\right)	 \right. \\
&& + \left.  \left(	m_i 	y_{\phi\chi i}^{ R *} y_{\phi\chi j}^{ R } + m_j 	y_{\phi\chi i }^{ L *} y_{\phi\chi j}^{ L }	\right)   \left( Q_\phi \bar{A}(r) - Q_\chi \bar{B}(r)		\right)    \right], \nonumber
\end{eqnarray}
where $m_\phi$ ($m_\chi$) is the heavy scalar (fermion) mass while $Q_\phi$ ($Q_\chi$) is its electric charge, and $r=m_\chi^2/m_\phi^2$. The loop functions are given by 
\begin{eqnarray}
\label{eq:loop-funcs}
A(r) &=& \frac{r^2-1-2r \log (r)}{(r-1)^3}, \\
B(r) &=& \frac{r^2-4r+3+2 \log (r)}{(r-1)^3}, \\
\bar{A}(r) &=& \frac{2r^3+3r^2-6r+1-6r^2 \log (r)}{6(r-1)^4}, \\
\bar{B}(r) &=& \frac{r^3-6r^2+3r+2+6r \log (r)}{6(r-1)^4}.
\end{eqnarray}
We have explicitly checked that our formulas are in agreement with the corresponding equations from Refs.~\cite{Hisano:1995cp,Ellis:2008zy,Crivellin:2018qmi}. As evident from Eq.~\eqref{eq:cR-UV-yukawa}, the $A(r)$ and $B(r)$ loop functions correspond to diagrams with a heavy fermion mass insertion in the loop, while $\bar{A}(r)$ and $\bar{B}(r)$ correspond to diagrams with a mass insertion on the external fermion legs. The terms with $A(r)$ and $\bar{A}(r)$ ($B(r)$ and $\bar{B}(r)$) come from diagrams with the photon emitted from the scalar (fermion) in the loop. Evidently, only the term proportional to $y^Ly^R$ can contribute to the CP-odd observables.

\section{Diagrams Generating the Dipole Operator in the Mass Insertion Approximation}
\label{app:all_contributions}

As mentioned in Sec.~\ref{sec:setup}, there are many diagrams in the MSSM that can contribute to the dipole operator of Fig.~\ref{fig:dipole-diag}. In the absence of RPV, we only have diagrams with sleptons and higgsinos or gauginos running in the loop. 
To properly calculate the contribution of all these diagrams, one should go to the mass eigenbasis of these particles and use the result of the previous appendix with the new physical states and their couplings (which can be found in Ref.~\cite{Hisano:1995cp}). 
We use this basis to carry out all the calculations in deriving our numerical results. In this appendix, we provide a better analytic understanding of these contributions to $c^R_{ij}$ using the mass insertion approximation. 
We have checked explicitly that these approximations are in great agreement with our full numerical calculation as well as those obtained from publicly available code {\sf susy\_flavor\_v2.5}~\cite{Rosiek:2010ug,Crivellin:2012jv,Rosiek:2014sia}.

We choose a common scale $\tilde m$ for all the sleptons, such that their mass eigenvalues are comparable. We keep the original non-diagonal form of the slepton mass matrices and treat the $\delta$s from Eq.~\eqref{eq:slepton-mass-mtx} as perturbations on the propagators. Since the $\delta$s all depend on the horizontal charges of the superfields, this calculation allows us to study the effect of varying different charges. 

The gauginos and the higgsinos, on the other hand, can have vastly different masses. To keep track of their contributions, here we go to their mass eigenbasis. This gives rise to small modifications of their couplings to leptons and sleptons, proportional to the mixing parameter between the gauginos and the higgsinos. This mixing parameter can be treated as a perturbative parameter in our parameter space. We denote this parameter by $\varepsilon_n$ ($\varepsilon_c$) in case of neutralinos (charginos). Its smallness suggests that the mass eigenstates and the gauge eigenstates overlap substantially.

We show the leading order contributions to $c^R_{ij}$ from various particles in the loop in this basis in Tables~\ref{tab:eedm_diags}-\ref{tab:lfv_RR_diags}. 
We consider all possible combinations of fermions and sleptons in the loop. In each row we show the fermions in the loops and the SM coupling that appears in each vertex. (Other factors of $\mathcal{O}(1)$ at the vertices are neglected; we also denote both $g_1$ and $g_2$ gauge couplings of the SM by $g$ for simplicity.) The charge and the chirality of the slepton running in the loop can be determined from this information. 

There are three different perturbative parameters that help us identify the most dominant diagrams. These are (i) the SM fermion yukawa couplings $y_\mu$ and $y_e$, (ii) the neutralinos (chargino) mixing parameters $\varepsilon_n$ ($\varepsilon_n$), and (iii) the inverse of the common slepton soft mass $1/\tilde{m}^2$, which can be introduced through the loop functions, as well as through $\delta^{LR}$s (see Eq.~\eqref{eq:deltaLR}). Thus, we interchange this last expansion parameter with $\delta^{LR}$. For simplicity, we denote the mixing between all neutralinos (either the two gauginos with each other or one gaugino and one higgsino) by $\varepsilon_n$. In each row we consider the diagram with the least number of mass insertion $\delta$ and at most one $\delta^{LR}$ entry insertion. 
These $\delta$s are the source of CPV (in Table~\ref{tab:eedm_diags}) and LFV (in Tables~\ref{tab:lfv_RL_diags}-\ref{tab:lfv_RR_diags})\footnote{Notice that the hermitian property of $\delta^{LL}$ and $\delta^{RR}$ forces us to have at least three $\delta$ insertions in contributions to the eEDM in Table~\ref{tab:eedm_diags} when the slepton chirality is preserved in the loop.}. We emphasize that, depending on the flavor model, there are cases in which more $\delta$ insertions are actually the leading term. As our tables capture all terms with the correct flavor and chirality structure, it easy to obtain all higher order terms  by using a replacement of the sort:
\begin{align}
    \delta^{AB}_{\alpha\beta}\to
    \delta^{AC}_{\alpha\gamma}\delta^{CB}_{\gamma\beta}~.
\end{align}

\begin{table}[]
    \centering
    \resizebox{\columnwidth}{!}{
    \begin{tabular}{|c|c|c|c|c|}
    \hline
        The Diagram & $\chi$ & $c_1$ & $c_2$ & $\delta$ \\
        \hline
        \hline
        \multirow{14}{*}{\includegraphics[scale=0.8]{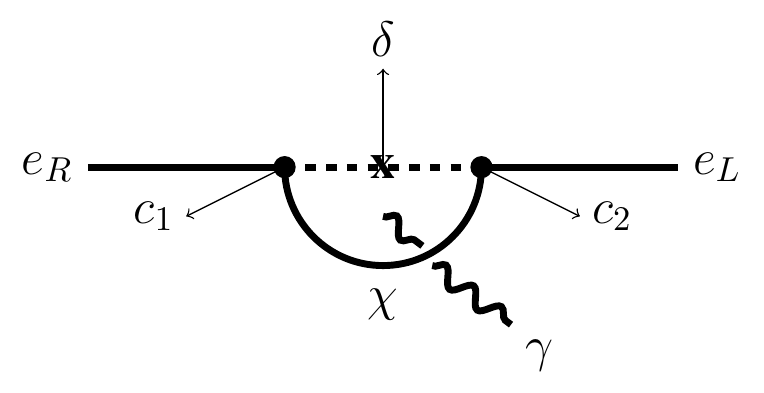}}  
        & $\tilde{H}^\pm$  & $y_e$  &  $\varepsilon_c g$  & $\delta^{LL}_{ef} \times \delta^{LL}_{ff'} \times \delta^{LL}_{f'e}$ \\
        \cline{2-5}
        & $\tilde{W}^\pm$   &  $\varepsilon_c y_e$  &  $g$  & $\delta^{LL}_{ef} \times \delta^{LL}_{ff'} \times \delta^{LL}_{f'e}$ \\
        \cline{2-5}
        &$\tilde{H}^0$ &  $y_e$  &  $y_e$  & $\delta^{LR}_{ee}$ \\
        \cline{2-5}
        & $\tilde{H}^0$   &  $y_e$  &  $\varepsilon_n g$  & $\delta^{LL}_{ef} \times \delta^{LL}_{ff'} \times \delta^{LL}_{f'e}$ \\
        \cline{2-5}
        & $\tilde{H}^0$   &  $\varepsilon_n g$  &  $\varepsilon_n g$  & $\delta^{LR}_{ee}$ \\
        \cline{2-5}
        & $\tilde{H}^0$   &  $\varepsilon_n g$  &  $y_e$  & $\delta^{RR}_{ef} \times \delta^{RR}_{ff'} \times \delta^{RR}_{f'e}$ \\
        \cline{2-5}
        & $\tilde{B}^0$   &  $g$  &  $g$  & $\delta^{LR}_{ee}$ \\
        \cline{2-5}
        & $\tilde{B}^0$   &  $g$  &  $\varepsilon_n y_e$  & $\delta^{RR}_{ef} \times \delta^{RR}_{ff'} \times \delta^{RR}_{f'e}$ \\
        \cline{2-5}
        & $\tilde{B}^0$   &  $\varepsilon_n y_e$  &  $\varepsilon_n y_e$  & $\delta^{LR}_{ee}$ \\
        \cline{2-5}
        & $\tilde{B}^0$  &  $\varepsilon_n y_e$  &  $g$  & $\delta^{LL}_{ef} \times \delta^{LL}_{ff'} \times \delta^{LL}_{f'e}$ \\
        \cline{2-5}
        & $\tilde{W}^0$  &  $\varepsilon_n g$  &  $g$  & $\delta^{LR}_{ee}$ \\
        \cline{2-5}
         & $\tilde{W}^0$  &  $\varepsilon_n g$  &  $\varepsilon_n y_e$  & $\delta^{RR}_{ef} \times \delta^{RR}_{ff'} \times \delta^{RR}_{f'e}$ \\
        \cline{2-5}
        & $\tilde{W}^0$  &  $\varepsilon_n y_e$  &  $\varepsilon_n y_e$ & $\delta^{LR}_{ee}$ \\
        \cline{2-5}
        & $\tilde{W}^0$  &   $\varepsilon_n y_e$  &  $g$  & $\delta^{LL}_{ef} \times \delta^{LL}_{ff'} \times \delta^{LL}_{f'e}$ \\
    \hline
    \end{tabular}
    }
    \caption{Diagrams with sleptons and gauginos/higgsinos in the loop contributing to the imaginary part of $c^R_{11}$. This Wilson coefficient contributes to the eEDM. The general topology of the diagram is shown on left. The second column shows the gaugino or the higgsino in the loop. The SM couplings appearing in each vertex, as well as $\varepsilon$s that characterize neutralino or chargino mixings, are shown in the next two columns. We consider all possible combinations of couplings with all possible fermions and sleptons in the loop. The last column includes the $\delta$ insertions on the slepton propagator. We are only including diagrams with the fewest number of $\delta$ insertions and at most one $\delta^{LR}$ entry. These $\delta$ insertions are the source of CPV in our setup. As evident from the table, when $\tilde{m}$ is not too large, i.e. $\delta^{LR}$ is not too small, the diagram with a Bino in the loop and a chirality flip on the propagator dominates, while other diagrams have more suppression thanks to fermion mixings $\varepsilon_{n/c}$ or the SM yukawas. }
    \label{tab:eedm_diags}
\end{table}

\begin{table}[]
    \centering
    \resizebox{\columnwidth}{!}{
    \begin{tabular}{|c|c|c|c|c|}
    \hline
        The Diagram & $\chi$ & $c_1$ & $c_2$ & $\delta$ \\
        \hline
        \hline
        \multirow{14}{*}{\includegraphics[scale=0.8]{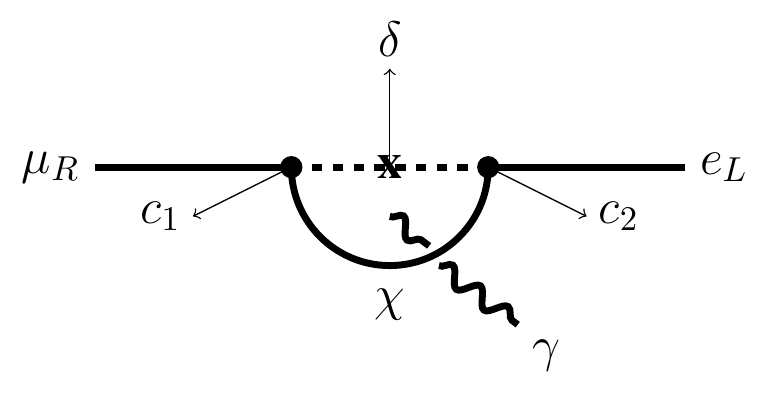}}  
        & $\tilde{H}^\pm$  & $y_\mu$  &  $\varepsilon_c g$  & $\delta^{LL}_{\mu e}$ \\
        \cline{2-5}
        & $\tilde{W}^\pm$   &  $\varepsilon_c y_\mu$  &  $g$  & $\delta^{LL}_{\mu e}$ \\
        \cline{2-5}
        &$\tilde{H}^0$ &  $y_\mu$  &  $y_e$  & $\delta^{LR}_{\mu e}$,~$\delta^{LR}_{\mu \mu} \times \delta^{RR}_{\mu e}$ \\
        \cline{2-5}
        & $\tilde{H}^0$   &  $y_\mu$  &  $\varepsilon_n g$  & $\delta^{LL}_{\mu e}$ \\
        \cline{2-5}
        & $\tilde{H}^0$   &  $\varepsilon_n g$  &  $\varepsilon_n g$  & $\delta^{LR}_{\mu e}$,~$\delta^{LR}_{\mu \mu} \times \delta^{LL}_{\mu e}$ \\
        \cline{2-5}
        & $\tilde{H}^0$   &  $\varepsilon_n g$  & $y_e$ & $\delta^{RR}_{\mu e}$ \\
        \cline{2-5}
        & $\tilde{B}^0$   &  $g$  &  $g$  & $\delta^{LR}_{\mu e}$,~$\delta^{LR}_{\mu \mu} \times \delta^{LL}_{\mu e}$ \\
        \cline{2-5}
        & $\tilde{B}^0$   &  $g$  &  $\varepsilon_n y_e$  & $\delta^{RR}_{\mu e}$ \\
        \cline{2-5}
        & $\tilde{B}^0$   &  $\varepsilon_n y_\mu$  &  $\varepsilon_n y_e$  & $\delta^{LR}_{\mu e}$,~$\delta^{LR}_{\mu \mu} \times \delta^{RR}_{\mu e}$ \\
        \cline{2-5}
        & $\tilde{B}^0$  &  $\varepsilon_n y_\mu$  &  $g$  & $\delta^{LL}_{\mu e}$ \\
        \cline{2-5}
        & $\tilde{W}^0$  &  $\varepsilon_n g$  &  $g$  & $\delta^{LR}_{\mu e}$,~$\delta^{LR}_{\mu \mu} \times \delta^{LL}_{\mu e}$ \\
        \cline{2-5}
         & $\tilde{W}^0$  &  $\varepsilon_n g$  &  $\varepsilon_n y_e$  & $\delta^{RR}_{\mu e}$ \\
        \cline{2-5}
        & $\tilde{W}^0$  &  $\varepsilon_n y_\mu$  &  $\varepsilon_n y_e$ & $\delta^{LR}_{\mu e}$,~$\delta^{LR}_{\mu \mu} \times \delta^{RR}_{\mu e}$ \\
        \cline{2-5}
        & $\tilde{W}^0$  &   $\varepsilon_n y_\mu$  &  $g$  & $\delta^{LL}_{\mu e}$ \\
    \hline
    \end{tabular}
    }
    \caption{Diagrams with sleptons and gauginos/higgsinos in the loop contributing to $c^R_{21}$ that come from the $y_Ly_R$ term in Eq.~\eqref{eq:cR-UV-yukawa}. This Wilson coefficient contributes to LFV observables such as $\BRmutoe$. The general topology of the diagram is shown on left. The second column shows the gaugino or the higgsino in the loop. The SM couplings appearing in each vertex, as well as $\varepsilon$s that characterize neutralino or chargino mixings, are shown in the next two columns. We consider all possible combinations of couplings with all possible fermions and sleptons in the loop. The last column includes the $\delta$ insertions on the slepton propagator. We are only including diagrams with the fewest number of $\delta$ insertions and at most one $\delta^{LR}$ entry. These $\delta$ insertions are the source of LFV in our setup. As evident from the table, when $\tilde{m}$ is not too large, i.e. $\delta^{LR}$ is not too small, the diagram with a Bino in the loop and a chirality flip on the propagator dominates, while other diagrams have more suppression thanks to fermion mixings $\varepsilon_{n/c}$ or the SM yukawas. }
    \label{tab:lfv_RL_diags}
\end{table}

\begin{table}[]
    \centering
    \resizebox{\columnwidth}{!}{
    \begin{tabular}{|c|c|c|c|c|}
    \hline
        The Diagram & $\chi$ & $c_1$ & $c_2$ & $\delta$ \\
        \hline
        \hline
        \multirow{14}{*}{\includegraphics[scale=0.8]{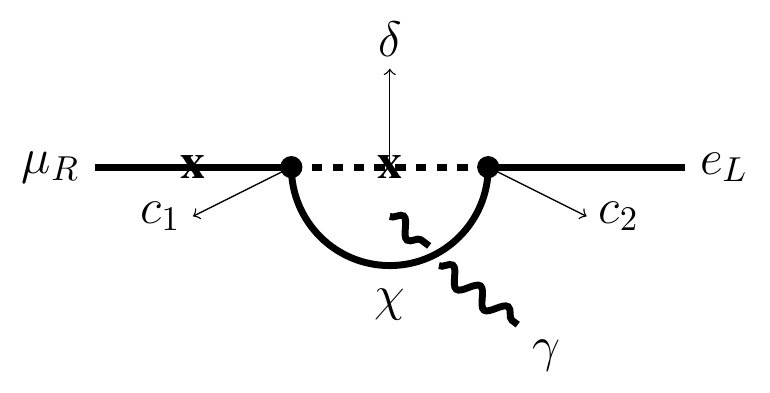}}  
        & $\tilde{H}^\pm$  & $\varepsilon_c g$  &  $\varepsilon_c g$  & $\delta^{LL}_{\mu e}$ \\
        \cline{2-5}
        & $\tilde{W}^\pm$   &  $g$  &  $g$  & $\delta^{LL}_{\mu e}$ \\
        \cline{2-5}
        &$\tilde{H}^0$ &  $y_\mu$  &  $y_e$  & $\delta^{RR}_{\mu e}$\\
        \cline{2-5}
        & $\tilde{H}^0$   &  $y_\mu$  &  $\varepsilon_n g$  & $\delta^{LR}_{\mu e}$,~$\delta^{LR}_{\mu \mu} \times \delta^{LL}_{\mu e}$ \\
        \cline{2-5}
        & $\tilde{H}^0$   &  $\varepsilon_n g$  &  $\varepsilon_n g$  & $\delta^{LL}_{\mu e}$ \\
        \cline{2-5}
        & $\tilde{H}^0$   &  $\varepsilon_n g$  &  $y_e$  & $\delta^{LR}_{\mu e}$,~$\delta^{LR}_{\mu \mu}  \times \delta^{RR}_{\mu e}$ \\
        \cline{2-5}
        & $\tilde{B}^0$   &  $g$  &  $g$  & $\delta^{LL}_{\mu e}$ \\
        \cline{2-5}
        & $\tilde{B}^0$   &  $g$  &  $\varepsilon_n y_e$  & $\delta^{LR}_{\mu e}$,~$\delta^{LR}_{\mu \mu}  \times \delta^{RR}_{\mu e}$ \\
        \cline{2-5}
        & $\tilde{B}^0$   &  $\varepsilon_n y_\mu$  &  $\varepsilon_n y_e$  & $\delta^{RR}_{\mu e}$ \\
        \cline{2-5}
        & $\tilde{B}^0$  &  $\varepsilon_n y_\mu$  &  $g$  & $\delta^{LR}_{\mu e}$,~$\delta^{LR}_{\mu \mu} \times  \delta^{LL}_{\mu e}$ \\
        \cline{2-5}
        & $\tilde{W}^0$  &  $g$  &  $g$  & $\delta^{LL}_{\mu e}$ \\
        \cline{2-5}
         & $\tilde{W}^0$  &  $g$  &  $\varepsilon_n y_e$  & $\delta^{LR}_{\mu e}$,~$\delta^{LR}_{\mu \mu}  \times \delta^{RR}_{\mu e}$ \\
        \cline{2-5}
        & $\tilde{W}^0$  &  $\varepsilon_n y_\mu$  &  $\varepsilon_n y_e$ & $\delta^{RR}_{\mu e}$ \\
        \cline{2-5}
        & $\tilde{W}^0$  &   $\varepsilon_n y_\mu$  &  $g$  & $\delta^{LR}_{\mu e}$,~$\delta^{LR}_{\mu \mu}  \times \delta^{LL}_{\mu e}$ \\
    \hline
    \end{tabular}
    }
    \caption{Diagrams with sleptons and gauginos/higgsinos in the loop contributing to $c^R_{21}$ that come from the $y_Ly_L$ term in Eq.~\eqref{eq:cR-UV-yukawa}. This Wilson coefficient contributes to LFV observables such as $\BRmutoe$. The general topology of the diagram is shown on left. The second column shows the gaugino or the higgsino in the loop. The SM couplings appearing in each vertex, as well as $\varepsilon$s that characterize neutralino or chargino mixings, are shown in the next two columns. We consider all possible combinations of couplings with all possible fermions and sleptons in the loop. The last column includes the $\delta$ insertions on the slepton propagator. We are only including diagrams with the fewest number of $\delta$ insertions and at most one $\delta^{LR}$ entry. These $\delta$ insertions are the source of LFV in our setup. As evident from the table, the diagrams with a Bino or a Wino in the loop and no chirality flip on the propagator dominate, while other diagrams have more suppression thanks to fermion mixings $\varepsilon_{n/c}$, the SM yukawas, or a $\delta^{LR}$ insertion on the slepton propagator. }
    \label{tab:lfv_LL_diags}
\end{table}

\begin{table}[]
    \centering
    \resizebox{1\columnwidth}{!}{
    \begin{tabular}{|c|c|c|c|c|}
    \hline
        The Diagram & $\chi$ & $c_1$ & $c_2$ & $\delta$ \\
        \hline
        \hline
        \multirow{14}{*}{\includegraphics[scale=0.8]{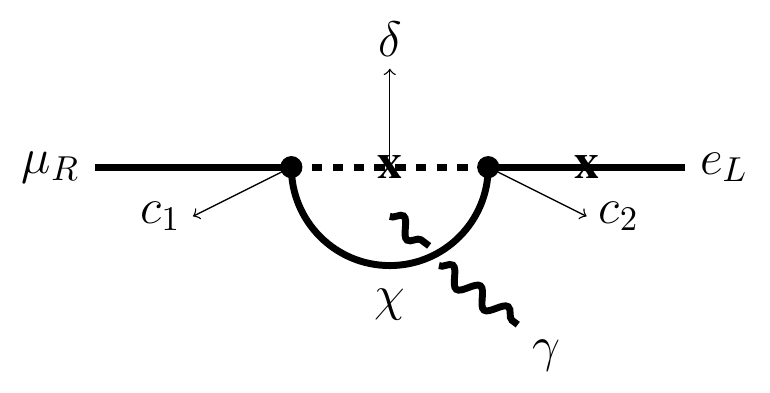}}  
        & $\tilde{H}^\pm$  & $y_\mu$  &  $y_\mu$  & $\delta^{LL}_{\mu e}$ \\
        \cline{2-5}
        & $\tilde{W}^\pm$   &  $\varepsilon_c y_\mu$  &  $\varepsilon_c y_\mu$  & $\delta^{LL}_{\mu e}$ \\
        \cline{2-5}
        &$\tilde{H}^0$ &  $y_\mu$  &  $y_e$  & $\delta^{LL}_{\mu e}$ \\
        \cline{2-5}
        & $\tilde{H}^0$   &  $y_\mu$  &  $\varepsilon_n g$  & $\delta^{LR}_{\mu e}$,~$\delta^{LR}_{\mu \mu} \times  \delta^{RR}_{\mu e}$ \\
        \cline{2-5}
        & $\tilde{H}^0$   &  $\varepsilon_n g$  &  $\varepsilon_n g$  & $\delta^{RR}_{\mu e}$ \\
        \cline{2-5}
        & $\tilde{H}^0$   &  $\varepsilon_n g$  &  $y_e$  & $\delta^{LR}_{\mu e}$,~$\delta^{LR}_{\mu \mu}  \times \delta^{LL}_{\mu e}$ \\
        \cline{2-5}
        & $\tilde{B}^0$   &  $g$  &  $g$  & $\delta^{RR}_{\mu e}$ \\
        \cline{2-5}
        & $\tilde{B}^0$   &  $g$  &  $\varepsilon_n y_e$  & $\delta^{LR}_{\mu e}$,~$\delta^{LR}_{\mu \mu}  \times \delta^{LL}_{\mu e}$ \\
        \cline{2-5}
        & $\tilde{B}^0$   &  $\varepsilon_n y_\mu$  &  $\varepsilon_n y_e$  & $\delta^{LL}_{\mu e}$ \\
        \cline{2-5}
        & $\tilde{B}^0$  &  $\varepsilon_n y_\mu$  &  $g$  & $\delta^{LR}_{\mu e}$,~$\delta^{LR}_{\mu \mu}  \times \delta^{RR}_{\mu e}$ \\
        \cline{2-5}
        & $\tilde{W}^0$  &  $\varepsilon_n g$  &  $\varepsilon_n g$  & $\delta^{RR}_{\mu e}$ \\
        \cline{2-5}
         & $\tilde{W}^0$  &  $\varepsilon_n g$  &  $\varepsilon_n y_e$  & $\delta^{LR}_{\mu e}$,~$\delta^{LR}_{\mu \mu} \times  \delta^{LL}_{\mu e}$ \\
        \cline{2-5}
        & $\tilde{W}^0$  &  $\varepsilon_n y_\mu$  &  $\varepsilon_n y_e$ & $\delta^{LL}_{\mu e}$ \\
        \cline{2-5}
        & $\tilde{W}^0$  &   $\varepsilon_n y_\mu$  &  $ \varepsilon_n g$  & $\delta^{LR}_{\mu e}$,~$\delta^{LR}_{\mu \mu}  \times \delta^{RR}_{\mu e}$ \\
    \hline
    \end{tabular}
    }
    \caption{Diagrams with sleptons and gauginos/higgsinos in the loop contributing to $c^R_{21}$ that come from the $y_Ry_R$ term in Eq.~\eqref{eq:cR-UV-yukawa}. This Wilson coefficient contributes to LFV observables such as $\BRmutoe$. The general topology of the diagram is shown on left. The second column shows the gaugino or the higgsino in the loop. The SM couplings appearing in each vertex, as well as $\varepsilon$s that characterize neutralino or chargino mixings, are shown in the next two columns. We consider all possible combinations of couplings with all possible fermions and sleptons in the loop. The last column includes the $\delta$ insertions on the slepton propagator. We are only including diagrams with the fewest number of $\delta$ insertions and at most one $\delta^{LR}$ entry. These $\delta$ insertions are the source of LFV in our setup. As evident from the table, the diagrams with a Bino in the loop and no chirality flip on the propagator dominates, while other diagrams have more suppression thanks to fermion mixings $\varepsilon_{n/c}$, the SM yukawas, or a $\delta^{LR}$ insertion on the slepton propagator. }
    \label{tab:lfv_RR_diags}
\end{table}

In Table~\ref{tab:eedm_diags} we show the diagrams contributing to the imaginary part of $c^R_{11}$, which is relevant for calculating the eEDM. As a result, only the couplings of the form $y_R y_L$ from Eq.~\eqref{eq:cR-UV-yukawa} are relevant. 
In Tables~\ref{tab:lfv_RL_diags}-\ref{tab:lfv_RR_diags} we show the contributions to $c^R_{21}$ that respectively come from the terms proportional to $y_R y_L$, $y_L y_L$, and $y_R y_R$ from Eq.~\eqref{eq:cR-UV-yukawa}. 
Results similar to those shown for $c^R_{21}$ in Tables~\ref{tab:lfv_RL_diags}-\ref{tab:lfv_RR_diags} can be obtained for any $c^R_{ff'}$ with trivial replacements.\footnote{It should be noted that then the $c^L_{ij}$ coefficients can be calculated using Eq.~\eqref{eq:WC-PL}.}

In Tables~\ref{tab:lfv_RL_diags}-\ref{tab:lfv_RR_diags}, we include two different $\delta$ insertions in the rows with $\delta^{LR}$. The first (second) insertion denotes the contribution of the $A$-terms (the $\mu$ term) from Eq.~\eqref{eq:deltaA_def}.\footnote{We should keep in mind that the $A$-term contributions to $\delta^{LR}$ can be non-diagonal in the SM fermions mass basis, while the contribution from the first term in Eq.~\eqref{eq:deltaLRwithC} will be proportional to $M_f$, i.e. diagonal in the SM fermions mass basis.} For these terms, our organizing principle (including the diagrams with the fewest number of $\delta$ insertions) could actually miss the dominant contributions to the LFV observables (for a given $\chi,~c_1,$ and $c_2$). For instance, in our setup, in the contribution from the third row of Table~\ref{tab:lfv_RL_diags}, the term with $\delta^{LR}_{\mu \mu} \times \delta^{RR}_{\mu e}$ can dominate the $\delta^{LR}_{\mu e}$ contribution thanks to the factor of $\tan \beta$ in Eq.~\eqref{eq:deltaLRwithC}.

\newpage
\section{Generating Random Mass Matrices}
\label{app:matrices}

As explained in Sec.~\ref{sec:minimal}, we study the effect of undetermined coefficients in the mass matrices by generating the mass matrices multiple times, with random $\mathcal{O}(1)$ numbers replacing these coefficients. The distribution of these random numbers gives rise to a spread in the predictions for various observables across different trials. In this appendix, we elaborate more on generating these random numbers and how we introduce them into the $\delta^{LL,RR,A}$ matrices. 
We emphasize that different prior distributions  give rise to different results. Therefore one has to make a choice of some reasonable distribution. In what follows we explain and motivate our choices. Our philosophy is to choose the simplest distribution that is close to a uniform distribution but also satisfies key physical constraints like hermiticity or positivity. 

We start by considering the four scenarios considered toward the end of Sec.~\ref{subsec:model}. This is the procedure used to prepare Fig.~\ref{fig:horizontal+SCPV}. 
In each scenario, we assumed that we start from the SM leptons' mass basis. It should be noted that in the notation of Sec.~\ref{subsec:model}, each entry of the $\delta$ matrices can get contributions from different interfering spurion insertions. Each of these interfering terms can be multiplied by their own random coefficients (hence the subscripts $a_1$ and $a_2$ in $C^{ij}_{a_1,a_2}$ in Eqs.~\eqref{eq:mass_mtx_fermions}-\eqref{eq:mass_mtx_neutrinos} and Eqs.~\eqref{eq:conditions}-\eqref{eq:deltaLRwithC}). 

As a result, each $\delta$ matrix can be broken into two pieces: (i) a $3\times 3$ matrix with random entries denoted by $C^{L,R,A}$, and (ii) a piece containing all interfering spurion insertions. (The latter only appears in the scenarios when we do have horizontal symmetries.) We then multiply these two matrices entry-by-entry. This is repeated for each of the interfering terms separately. We finally add up these interfering matrices to make the final $\delta^{LL}$, $\delta^{RR}$, and $\delta^{A}$ matrices. 

\begin{figure}
	\centering
	\resizebox{1\columnwidth}{!}{
\includegraphics{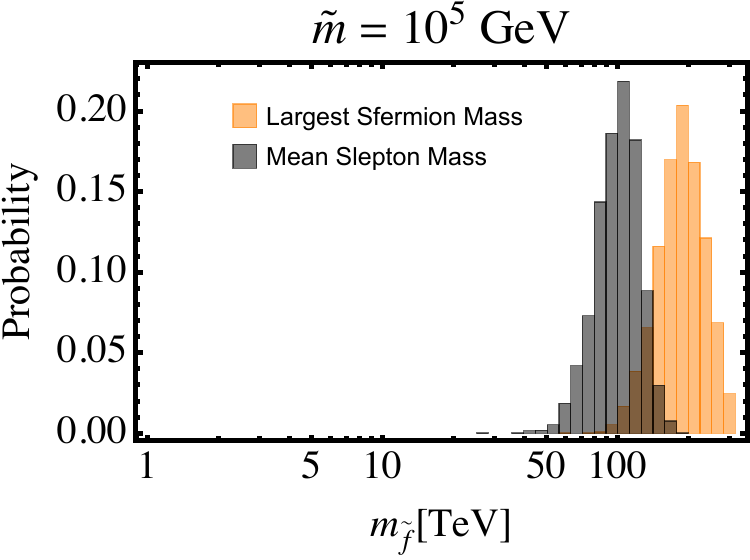}
\includegraphics{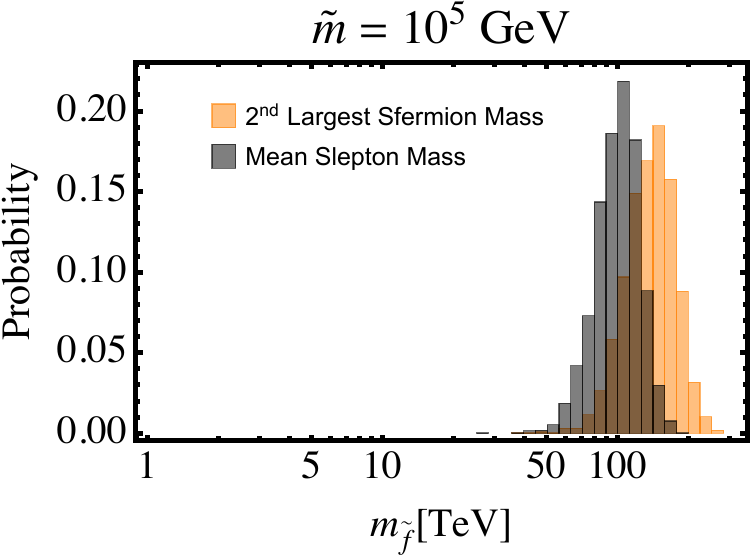}
\includegraphics{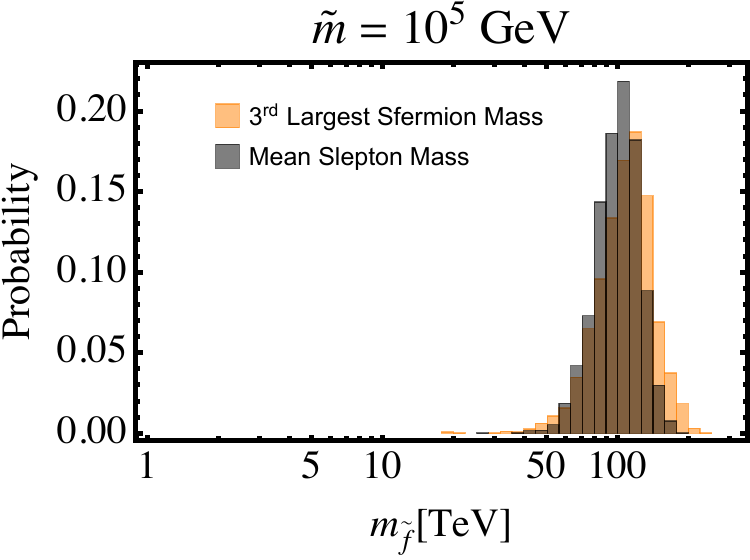}
	}\\
		\resizebox{1\columnwidth}{!}{
\includegraphics{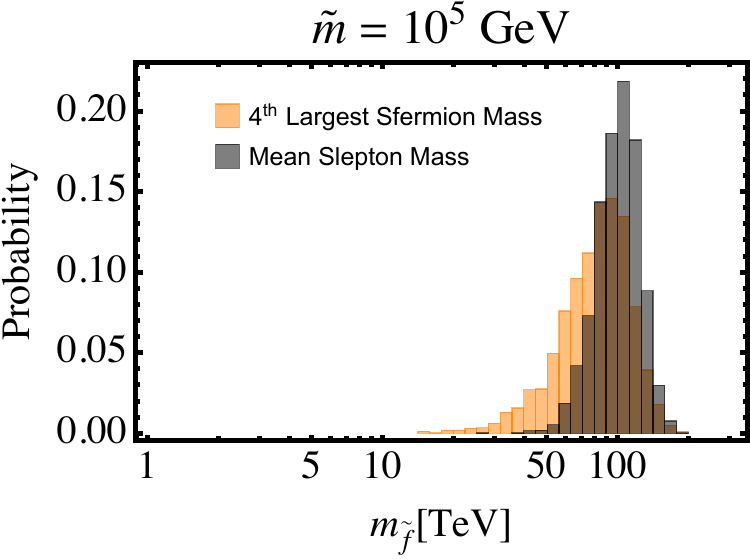}
\includegraphics{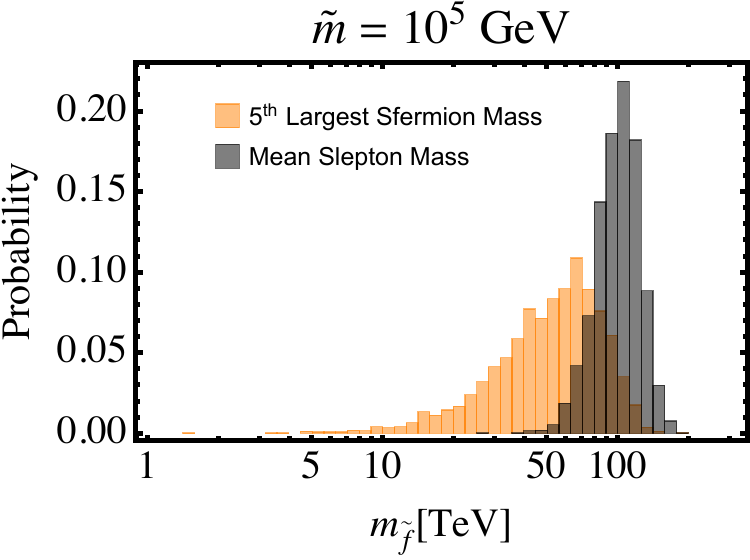}
\includegraphics{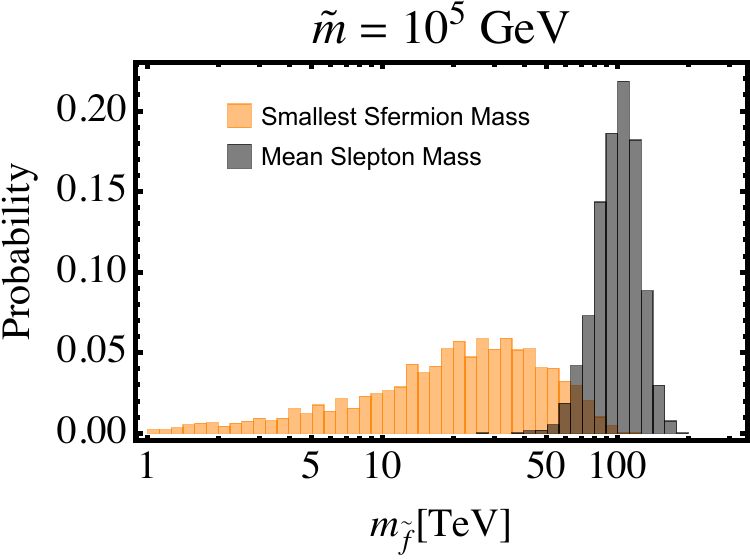}
	}\\
	\caption{The distribution of different slepton masses across different good trials for the charge assignment in the first row of Table~\ref{tabs:charges_less_tuned} with $\tilde{m}=100$~TeV. The orange histograms show the distribution of different mass eigenvalues, while the black histogram denotes the average slepton mass in different trials. Note that the average masses are sharply concentrated around $\tilde{m}$. While the heaviest eigenvalues are within an $\mathcal{O}(1)$ factor of $\tilde{m}$, the lightest eigenvalue for this charge assignment can be, in rare occasions, more than an order of magnitude smaller than the average slepton masses.  }
	\label{fig:slepton masses histograms}
\end{figure}

It should be noted that while $\delta^A$ can be a generic $3 \times 3$ matrix, $\delta^{LL}$ and $\delta^{RR}$ must be hermitian and positive definite. (The requirement that they be positive  definite follows from the phenomenological need to avoid tachyons that would break electromagnetism, not from theoretical consistency.) In the horizontal symmetry scenarios the spurion part of $\delta^{LL}$ and $\delta^{RR}$ are hermitian matrices that are very close to the identity matrix, thus we can guarantee the hermitian and positive definite properties of the overall $\delta^{LL}$ and $\delta^{RR}$ by making sure that $C^{L}$ and $C^{R}$ are hermitian and positive definite.\footnote{The $\delta^{LR}$ matrix could potentially spoil this property; however, its entries are suppressed compared to $\delta^{LL,RR}$ matrices by roughly $\langle H_d \rangle/\tilde{m}$, which can be a small number. As a result of this, we can guarantee the slepton mass matrices are positive definite and hermitian by only focusing on their diagonal blocks $\delta^{LL,RR}$.} To do that, in scenarios with spontaneous CPV and for each interfering term, we draw 9 random real numbers each from a uniform distribution between $(-1,~1)$ and call the resulting matrices $X^{L,R}$. Then we define $C^{L,R} = X^{L,R} \cdot \left(X^{L,R}\right)^\dagger$. 
(Note that while $X^{L,R}$ are uniformly distributed, $C^{L,R}$ are not. Nevertheless, this construction guarantees the physical constraints are satisfied by $C^{L,R}$.) 
We also generate 9 more random numbers uniformly distributed between $(-1,~1)$ and directly use them for each interfering term in $C^{A}$. 
In Fig.~\ref{fig:slepton masses histograms}  we show the distributions of the slepton mass eigenstates for an arbitrary point of the parameter space. We find that the average slepton mass agrees with $\tilde{m}$ at the percent level, while the individual mass eigenstates typically differ from the average by an $\mathcal{O}(1)$ number.

For the scenarios with explicit CPV, the only difference is that the random numbers will have a magnitude and phase that are drawn from a uniform distribution in the range $(0,~1)$ and $(-\pi,~\pi)$, respectively. 
In each scenario of Fig.~\ref{fig:horizontal+SCPV}, we repeat this process 1000 times. In each trial, we calculate the mass eigenstates and their contribution to the eEDM and $\BRmutoe$ observables.

A similar process is repeated for our calculations with the charge assignments of Table~\ref{tabs:charges_less_tuned} in Sec.~\ref{subsec:tuning_calc}, which were used to prepare Fig.~\ref{fig:mass-plane-bound-25}, \ref{fig:mass-plane-bound-15}, \ref{fig:mass-plane-bound-10},   \ref{fig:mass-plane-bound-5}, and \ref{fig:mass-plane-bound-2}. Unlike the scenarios in Sec.~\ref{subsec:model}, in these models we start from an arbitrary basis with off-diagonal entries in the SM fermion mass matrices. We generate the random matrices $X^{l,\nu}$ for the charged leptons and neutrinos mass matrices, as well as $X^{L,R,A}$ that will eventually appear in the corresponding $\delta$ matrices.

Physical consistency requires the neutrino mass matrix to be symmetric. 
Thus, we use $C^{\nu}=\frac{\left(X^\nu+\left(X^\nu\right)^\mathrm{T} \right)}{2}$ for the matrix of random numbers contributing to the neutrino masses. $C^{L}$ and $C^{R}$ are also made positive definite and hermitian as prescribed above. 
All the random numbers are still drawn from a uniform distribution between $(-1,~1)$. We then rotate to the SM fermion mass basis. Then, similar to the scenarios in Sec.~\ref{subsec:model}, we go to the mass basis of the new SUSY particles and calculate the dipole operator and the observables of Table~\ref{tabs:obslist} for each trial.

Note that depending on the distribution from which we draw the random numbers, the fraction of ``good" trials in Table~\ref{tabs:charges_less_tuned} can change. Using the uniform distribution reflects our complete lack of knowledge about the UV structure giving rise to the undetermined coefficients. While the good trial fraction can be affected by this change of distribution, it is natural to think the tuning ratio of Eq.~\eqref{eq:t_def} is far less sensitive to this distribution. This is why we choose to use this ratio as our measure of tuning. It not only captures the constraints from a specific observable (rather than getting the SM mass and mixing structure), but also partially cancels the sensitivity of the final tuning value to the distribution of these numbers in the numerator and the denominator.

\bibliography{ref}
\bibliographystyle{utphys}
\end{document}